\documentclass{aa}  
\usepackage{graphicx,amssymb}
\usepackage{natbib}
\usepackage{verbatim} 
\usepackage{txfonts}
\def\planck{\textit{Planck}}
\def\herschel{\textit{Herschel}}
\def\wise{\textit{WISE}}
\def\iras{\textit{IRAS}}
\def\spitzer{\textit{Spitzer}}

\def\lsun{\hbox{$\rm ~L_{\odot}$}}
\def\msun{\hbox{$\rm ~M_{\odot}$}}
\def\mism{\hbox{$\rm ~M_{\rm ISM}$}}
\def\mgas{\hbox{$\rm ~M_{\rm gas}$}}
\def\mstar{{M$_{*}$}}

\def\arcdeg{$^{\circ}$}
\def\H0{{\rm ~km~s^{-1}~Mpc^{-1}}}

\def\850{850$\mu$m}
\def\550{550$\mu$m}
\def\350{350$\mu$m}
\def\l850{L$_{\nu 850}$}
\def\lsubmm{L$_{submm}$}
\def\lpco{L$^{'}_{\rm CO}$}

\def\tdust{T$_{\rm dust}$}

\def\hi{\ion{H}{I}}

\def\hi{\ion{H}{I}}

\def\.25{0.25 keV\thinspace}

\def\arcsec{$^{\prime\prime}$}
\def\lir{\hbox{L$_{\rm IR}$}}
\def\lfir{\hbox{L$_{\rm FIR}$}}
\def\mdust{\hbox{M$_{\rm dust}$}}
\def\mstar{\hbox{M$_{ *}$}}
\def\mhi{\hbox{M$_{HI}$}}
\def\mh2{\hbox{M$_{H2}$}}

\def\mum{\hbox{$\mu$m}}
\def\tc{\hbox{T$_{\rm cold}$}}
\def\tw{\hbox{T$_{\rm warm}$}}
\def\hi{\hbox{H${\rm I}$}}
\def\h2{\hbox{H${2}$}}

\begin{document}
   \title{Molecular Gas, Dust and  Star Formation in Galaxies}
   \subtitle{I. Dust properties and scalings in $\sim$ 1600 nearby galaxies}

   \author{Orellana G.\inst{1,2}
          \and
          Nagar N. M.\inst{1}
          \and
          Elbaz D.\inst{3}
          \and
          Calder\'on-Castillo P.\inst{1}
          \and
          Leiton R.\inst{2}
          \and
          Ibar E.\inst{2}
          \and
          Magnelli B. \inst{4}
          \and
          Daddi E.\inst{3}
          \and 
          Messias H.\inst{5,6}
          \and 
          Cerulo P.\inst{1}
          \and
          Slater R.\inst{7}
          }
   \offprints{G. Orellana}

   \institute{Department of Astronomy, Universidad de Concepci\'on, Casilla 160-C Concepci\'on, Chile \\
              \email{gustavo.orellana@uv.cl,gorellana@udec.cl}
          \and
           Instituto de F\'isica y Astronom\'ia, Universidad de Valpara\'iso, Avda. Gran Breta\~na 
             1111, Valpara\'iso, Chile
          \and
           Laboratoire AIM-Paris-Saclay, CEA/DSM/Irfu - CNRS - Universit\'e Paris Diderot, 
            CE-Saclay, pt courrier 131, F-91191 Gif-sur-Yvette, France  
          \and
          Argelander-Institut f\"{u}r Astronomie, Universit\"{a}t Bonn, Auf dem H\"{u}gel 
             71, D-53121 Bonn, Germany 
          \and
           Joint ALMA Observatory, Alonso de C\'ordova 3107, Vitacura 763-0355, Santiago, Chile
          \and
            European Southern Observatory, Alonso de C\'ordova 3107, Vitacura, Casilla 19001, 19 Santiago, Chile
          \and
           Departamento de Ciencias Fisicas, Universidad Andres Bello, Sede Concepcion, 
             autopista Concepcion-Talcahuano 7100, Talcahuano, Chile
             }
   \date{Received 25 May 2016 ; Accepted 28 March 2017 }

  \abstract
   {Dust and its emission is being increasingly used to constrain the evolutionary stage of galaxies.
   A comprehensive characterization of dust, best achieved in nearby  
    bright galaxies, is thus a highly useful resource.}
   {We aim to characterize the relationship between dust properties (mass, luminosity
   and temperature) and their relationships with galaxy-wide properties (stellar, 
   atomic and molecular gas mass, and star formation mode). We also aim to provide
   equations to estimate accurate dust properties from limited observational datasets. 
   }
   {We assemble a sample of 1,630 nearby  (z $<$ 0.1) galaxies - over a large range of stellar masses (\mstar),
    star formation rates (SFR) and specific star formation rates (sSFR=SFR/\mstar) - for which comprehensive 
    and uniform multi-wavelength observations are available from \wise, \iras, \planck\ and/or SCUBA. 
    the characterization of dust emission comes from spectral energy distribution (SED) fitting using 
    Draine \& Li dust models, which we parametrize using two components  
    (warm at 45 - 70 K and cold at 18 - 31 K). 
    The subsample of these galaxies with global measurements of CO and/or \hi\ are used
    to explore the molecular and/or atomic gas content of the galaxies.}
   {The total infrared luminosity (\lir), dust mass (\mdust) and dust temperature 
   of the cold component (\tc) form a plane that we refer to as the \textit{dust plane}. 
   A galaxy's sSFR drives its position on the
   dust plane: starburst  (high sSFR) galaxies show higher \lir, \mdust\ and \tc\ compared to 
   Main Sequence (typical sSFR) and passive galaxies (low sSFR).
   Starburst galaxies also show higher specific dust masses (\mdust/\mstar) and specific gas masses (\mgas/\mstar).
   We confirm earlier findings of an anti-correlation between the dust to stellar mass ratio and \mstar.
    We also find different anti-correlations depending on sSFR; the anti-correlation becomes stronger 
   as the sSFR increases, with the spread due to different cold dust temperatures.
   The dust mass is more closely correlated with the total gas mass (atomic plus molecular) 
   than with the individual atomic and molecular gas masses.
      Our comprehensive multi wavelength data allows us to define several equations to 
   accurately estimate \lir, \mdust\ and \tc\ from one or two monochromatic luminosities in the infrared 
   and/or sub-millimeter.
} 
   { 
   It is possible to estimate the dust mass and infrared luminosity from a single monochromatic luminosity within 
   the Rayleigh-Jeans tail of the dust emission, with errors of 0.12 and 0.20 dex, respectively. 
   These errors are reduced to 0.05 and 0.10 dex, respectively, if the dust temperature of the cold component is used.
   The dust mass is better correlated with the total ISM mass (\mism\ $\propto$ \mdust$^{0.7}$).
   For galaxies with stellar masses 8.5 $<$ log(\mstar/\msun) $<$ 11.9, 
   the conversion factor between the single monochromatic luminosity at $850~\mum$ and the total 
   ISM mass ($\alpha_{850~\mum}$) shows a large scatter (rms = 0.29 dex) and a weak correlation with the \lir. 
   The star formation mode of a galaxy shows a correlation with both the gas mass and dust mass: 
   the dustiest (high \mdust/\mstar) galaxies are gas-rich and show the highest SFRs.
   
 }

   \maketitle
%
\section{Introduction}
\label{secintro}

Star formation occurs within dense ($\rm n(H_{2})\sim 10^{2}-10^{5} cm^{-3}$), 
massive ($10^{4}-10^{6}\msun$), and cold ($\rm T_{gas}\sim 10-50$ K) 
giant ($10\sim100$ pc) molecular clouds (GMC) \citep{keneva12}, 
where atomic gas, mainly atomic Hydrogen ($\rm HI$), 
is transformed into molecular gas (mainly $\rm H_{2}$) on dust grain surfaces
(e.g., \citealt{sco12}). Dust grains are formed within the cool, extended 
atmospheres of low mass ($\rm 1-4~M_{\sun}$) asymptotic giant branch (AGB) 
stars and are dispersed into the ISM via the strong AGB star winds
(\citealt{geh89}). In other words, the dust content is related to 
the star formation history of the galaxy.
Since much of our current knowledge of galaxy properties and evolution 
comes from studies of high temperature (T $>10^{3}$ K) regions, 
a global understanding of star formation requires a better knowledge 
of the role of cold gas and dust in the star formation process. 

Dust grains emit mainly in the far infrared (FIR; $40<\lambda<300~\mum$) 
and sub-millimeter (sub-mm; $300<\lambda<1000~\mum$). Early 
studies of dust content and emission have been done both from space 
(\iras, see \citealt{nauet84} and \textit{ISO}, see \citealt{keset96}) and from
the ground (SCUBA, see \citealt{holet99}, at the \textit{James Clerk Maxwell} Telescope
and MAMBO at the \textit{IRAM} 30 meter telescope). 
More recent missions - in the mid-infrared (e.g., \wise; \textit{Spitzer}), 
far-infrared (e.g., \textit{AKARI}; \textit{Herschel}), and sub-mm (e.g., \planck) 
- have revolutionized the field (e.g., \citealt{lutz14}).

The IR to sub-mm emission of dust has been characterized
in many samples (e.g., SLUGs by \citealt{dunet00}; HRS by \citealt{boet10};
KINGFISH/SINGS by \citealt{kenet03,dalet05}; SDSS-IRAS by \citealt{dacet10}
;ATLAS 3D by \citealt{capet11}; ERCSC by \citealt{neget13}) 
and at high-z (e.g., GOODS-Herschel by \citealt{maget10};
H-ATLAS by \citealt{ealet10}). Early studies modeled
the dust grain emission using grey-body emission from one or two dust
temperature components \citep[e.g.,][]{dunet00,dunet01}.
More complex and sophisticated dust emission models 
available today include the MAGPHYS \citep{dacet10} code -
which contains empirically-derived spectral energy density (SED) libraries 
from the ultraviolet (UV) to infrared (IR) -
and the model developed by \citet[DL07 hereafter]{drali07} which provides a more extensive
SED library covering the IR to sub-mm. The DL07 model has been successfully applied to the 
\textit{Spitzer} Nearby Galaxy Survey (SINGS) galaxies \citep{draet07}, and these authors note
that the presence of sub-mm photometry is crucial to constrain the mass of the cold
dust component.

The results on dust properties coming from many of the studies mentioned above are limited by 
poor statistics as a consequence of small samples and/or the limited sampling of the 
SED (especially at sub-mm wavelengths) which  decreases the reliability of the SED modeling.
Since dust properties are increasingly used at all redshifts to determine the evolutionary state of
a galaxy, and in general for galaxy evolution studies, it is crucial to fully characterize these
properties and their relationships and degeneracies in large samples of galaxies. 
Of specific interest is the degeneracy between dust temperatures and dust emissivity index,
the inter-relationships between dust mass, temperature, and luminosity, and the relationships
between these dust properties and other properties of the galaxy (e.g., stellar and gas masses,
SFR, specific star formation rate; sSFR=SFR/\mstar[yr$^{-1}$]).
 
The recent availability of \planck\ sub-mm (350~\mum\ to 850~\mum) fluxes for thousands of nearby
galaxies which are well studied in the optical to IR, allows, for the first time, comprehensive 
and accurate dust model fits to these. 
With a comprehensively modeled large sample of nearby galaxies in hand, one can test and
refine the many scaling relations and estimators now being used at all redshifts, e.g.,
estimating gas mass from a single flux measurement at 850~\mum\ \citep{sco12}, 
or estimating dust masses \citep{dunet00,dunet01} and/or luminosities (e.g., \citealt{sanmir96}, \citealt{elbet10}) 
from a few IR flux measurements.

The `starburstiness' of a galaxy is normally obtained from 
the ratio of the SFR and the stellar mass (\mstar). 
The SFR -\mstar\ plane shows that
while  most 'normal' star-forming galaxies follow a `main sequence' (MS)
of secular star formation \citep{elbet07}, a small fraction of galaxies show 
excessive SFR for a given \mstar: these galaxies are referred to as starburst (SB).  
The MS of galaxies is  observed over 
the redshift range z$\sim 0-4$ (e.g., \citealt{elbet07}, \citealt{magdet10},
\citealt{rodet11}, \citealt{elbet11}, \citealt{panet15}, \citealt{schet15})
and changes smoothly with redshift \citep{elbet11}.
In this work we use the MS proposed by \citet{elbet11}, at z=0: 
\begin{equation}
 \rm SFR=\frac{M_{*}}{4.0\times10^{9}}\left[\frac{\msun}{yr}\right]
\label{eq:SFR elbaz}
\end{equation}
This equation defines the specific star formation rate $\rm \left(sSFR\ [yr^{-1}] =SFR/\mstar\right)$
expected for MS galaxies (MS; $\rm -0.5 < ~sSFR ~[yr^{-1}]~ < 0.5$). We define SB galaxies as those
having $\rm sSFR\ [yr^{-1}] > 0.8$, and passive (PAS) galaxies as those at $\rm sSFR\ [yr^{-1}] < -0.8$. 
The two `transition' zones between the above three classifications, i.e. 
$\rm 0.5\leq\ SFR\ [\msun yr^{-1}] <0.8$ dex wrt the MS locus (an intermediate SB zone) and ;
$\rm-0.8\leq\ SFR\ [\msun yr^{-1}] <-0.5$ dex wrt to the MS locus (an intermediate passive zone) are
excluded in order to avoid contamination. 

In this paper, we capitalize on the recent availability of sub-mm (\planck) fluxes (for 
better dust model fits) and \wise\ fluxes (for stellar mass determinations)
to fit DL07 dust models to all nearby bright galaxies for which 
sufficient (for a reasonable fit to DL07 models) 
multi-wavelength uniform data are available from \wise, \iras, \planck, and/or SCUBA. 
The resulting model fits are used to explore the relationship between dust properties
(mass, luminosity, temperature) and their relationship with other galaxy-wide properties
(e.g., stellar and gas masses, sSFR). The comprehensive dust modeling also allows us
to refine estimations of total IR luminosity from one to a few IR to sub-mm fluxes,
the dust mass from a single sub-mm flux, and sSFR from IR to sub-mm colors. 

Throughout this paper we adopt a flat cosmology with $\rm \Omega_{m}=0.3$ and 
$\rm H_{0}=72 ~km~s^{-1}Mpc^{-1}$.

\section{Sample and Data }
\label{sec:sample}

We use two samples of nearby galaxies: 
(a) the sample of nearby galaxies with detections in \planck\
High Frequency Instruments (HFI) Second Data
Release catalog and global CO J:1-0 observations \citep{naget15}; and 
(b) all galaxies from the 2MASS Redshift 
Survey \citep[2MRS;][]{hucet12} which are listed as detections in the 
second \planck\ Catalog of Compact Sources at 350~\mum, 550~\mum, and 850~\mum\ 
\citep[PCCS2;][]{plaet15}. 

The \citet{naget15} sample is a compilation of $\sim$600 nearby galaxies 
($<$z$>$ = 0.06) with global CO J:1-0 observations, and sub-mm 
fluxes from \planck\ catalogs at 350~\mum, 550~\mum, and 850~\mum\ or 
SCUBA 850~\mum\ observations.
The names of the catalogs with the respective references are summarized in
Table \ref{tab:papers Neil}.

\begin{table}[ht]
\centering
\caption{Source samples in the Nagar et al. compilation}
\setlength{\tabcolsep}{4pt}
\begin{tabular}{c|l}
Sample               &  References               \\[2pt]
\hline\\
FCRAO                &  \citet{youet95}          \\[3pt]
SLUGS                &  \citet{dunet00,yaoet03}  \\[3pt] 
KINGFISH/            &  \citet{kenet03,dalet05}  \\
 ~~SINGS             &  \citet{draet07,mouet10}  \\
                     &  \citet{wilet12,dalet12}  \\[3pt]
ATLAS-3D             &  \citet{capet11,youet11}  \\[3pt]
Others               & \citet{sanet91,elfet96}   \\
                     & \citet{solet97,chuet09}   \\
                     & \citet{maoet10,papet12}   \\
                     & \citet{garet12,uedet14}
\end{tabular}
\label{tab:papers Neil}
\end{table}

The sample spans a range of morphological types - including spiral, 
elliptical and interacting galaxies - and luminosities from
normal to Ultra-Luminous Infrared Galaxies (ULIRGs).

The 2MRS sample consists of 44,599 nearby ($\rm < z >=0.03$) 2MASS \citep{sheet96} galaxies 
with K$_{\rm s} \geq$ 11.75 mag and Galactic latitude $|b|\geq5$ for which spectroscopic
redshifts have been obtained to 97.6\% completeness \citep{hucet12}.
We matched the 2MRS sample with the (PCCS2), using a maximum matching radius of 1 arcmin. 
The PCCS2 catalog contains only  galaxies 
with high reliabilities ($> 80\%$; signal to noise $> 5 $ in DETFLUX).
Sources with lower or unknown reliabilities, are listed in the equivalent excluded catalog (PCCS2E), 
which we have not used. 

The \planck\ satellite has a beam resolution of the order of 1 arcmin (e.g.\ 4.22 arcmin at 350~\mum, \citealt{plaet15}), 
and the reliability catalog contains $\sim$ 1000 sources (e.g.,  4,891 galaxies at 350~\mum, \citealt{plaet15}), 
with a density $<1$/ sources/deg$^2$ (e.g., 0.26 sources/deg$^2$ at 350~\mum, \citealt{plaet15}). 
This means that the resolutions of WISE and 2MASS ($\sim$ 1 arcsec) do not represent a problem for the match with 
the \planck\ source catalog, as we adopt a search radius of 1.0 arcmin. In order to remove any multiple match, 
we performed a visual inspection of all the matched objects. 
In some cases, we selected all the galaxies that have companions in the WISE, 2MASS and SDSS images and classified 
them as interacting systems. 
Furthermore, multiple detections in one \planck\ beam do not either represent a problem because the galaxies 
in our sample have a median diameter (parametrized as the D25 reported in HyperLeda \footnote{http://leda.univ-lyon1.fr}) 
of 1.5 arcmin, with only 291 having D25 $< 1.0$ arcmin ($\sim$ 18\% of the final sample).

\subsection{Flux densities and derived stellar mass}

The \planck\ collaboration \citep{plaet15} showed that, at 350 \mum, for sources with APERFLUX $>1.0$ Jy, 
the APERFLUXes reported in the Planck catalog are in agreement with those in the \emph{Herschel} Reference Survey.

\citet{naget15} compare the \planck\ observations at 850~\mum\ with SCUBA data (at 850~\mum ) 
in nearby galaxies, revealing that the APERFLUX and the DETFLUX from \planck\ show 
the existence of some systematic difference. However, \citet{naget15} also show that simple corrections
can solve this problem. They find that the observation of fluxes smaller than 
twice of the 90\% completeness limit (304 mJy at 850~\mum\ ) needs smaller corrections if the 
DETFLUX is used (the correction is $\rm f_{\planck~850~\mum}=f_{SCUBA~850~\mum}$+70), and for 
greater fluxes (greater than twice of the 90\% completeness limit) the fluxes are
more consistent with the APERFLUX (using the correction: $\rm f_{\planck~850~\mum}=f_{SCUBA~850~\mum}$+139).
Assuming a gray-body with a temperature of T=25 K and $\beta$=1.8, \citet{naget15} obtain similar corrections
for \planck\ observations at 350 and 550~\mum. For each wavelength, \citet{naget15} obtain three corrections:
one with free slope and intercept, a second with intercept zero and free slope, and a third with slope one, 
or the expected in the case of observations at 350 and 550 \mum. We use the last kind of correction in our work.
Additionally, to correct for the typical spectral shape of dust gray-body emission, we used correction factors of 
0.976, 0.903 and 0.887 at 350~\mum, 550~\mum\ and 850~\mum, respectively \citet{neget13}. 
Following \citet{naget15}, we assumed a 3\% contamination from the CO emission line at \planck\ 850~\mum\ and negligible 
CO emission line contamination at \planck\ 350 and 550~\mum. 
After these flux density corrections are applied, the limits obtained for the \planck-derived flux densities in our 
sample are 500 mJy, 315 mJy, and 175 mJy, at 350~\mum, 550~\mum, and 850 \mum, respectively.

Mid-infrared (MIR) fluxes  are obtained from the AllWISE Source Catalog `g' magnitudes
\footnote{wise2.ipac.caltech.edu/docs/release/allwise}.
These `g' magnitudes are calculated over apertures defined using 2MASS images, with additional 
corrections as described in \citet{jaret13}. 
The \wise\ (W1-W4) filters have a limiting sensitivity of 0.08, 0.11, 1 and 6 mJy, 
respectively. We select only sources with signal to noise (S/N) $\ge$ 5,
except for W4, where we consider a S/N $\ge$ 3.
We calculate the galaxy stellar mass (\mstar) using the \wise\ W1 filter (3.4~\mum) 
and the W1-W2 (4.6~\mum) color following the \citet{cluet14} calibration. 
The stellar mass ranges between $10^{9}$ and $10^{11}$ \msun\ for our sample. 
To test the consistency of our \wise-estimated stellar masses,  we compare our stellar masses to
those in three other catalogs based on SDSS-derived quantities (i.e. NASA-Sloan Atlas, Chang et al. 2015 
and MPA-JHU catalogs, see appendix \ref{apn:stellar mass}). We obtained a good agreement 
with the \mstar\ obtained in the NASA-Sloan Atlas (see appendix \ref{apn:stellar mass} for more details).

The infrared (IR) data comes from the Infrared Astronomical Satellite (\iras) 
at 12, 25, 60 and 100~\mum, obtained from the Galaxies and Quasars catalog 
\citep{fullon89}. We consider only sources with moderate or high quality fluxes
(no upper limits), with signal to noise $\ge$ 3.

\begin{figure}[tb]
\begin{center}
   \includegraphics[bb=83 222 281 347,width=0.5\textwidth,clip]{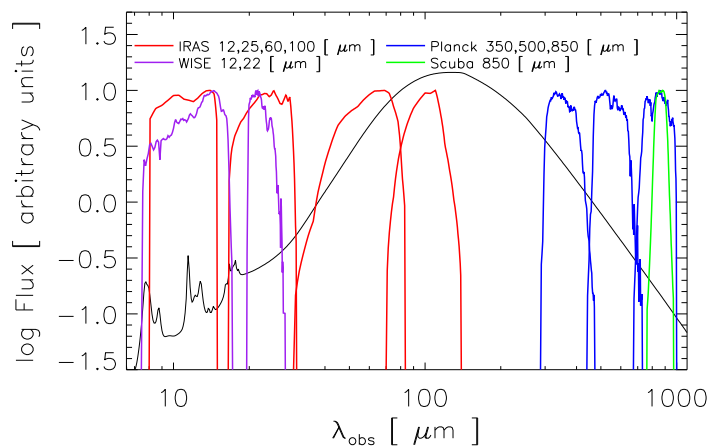}
   \caption{ An example dust emission template from \citet{drali07}, between 7 and 1100 $\mu m$.
   Overlaid are the transmission curves of all the filters used in this work: \wise\ W3 and W4 ; 
   \iras\ 12, 25, 60 and 100~\mum; \planck\ 350, 550 and 850~\mum\ and SCUBA 850~\mum. 
}  
\label{fig:bands}
\end{center}
\end{figure}

Figure \ref{fig:bands} shows the filter band-passes of all filters for which we compiled flux densities
which were then used to constrain the DL07 dust model fits: 
\wise\ W3 (12~\mum) and W4 (22~\mum) ; \iras\ 12, 25, 60 and 100~\mum\ ;  
\planck\ 350, 550 and 850~\mum\ and SCUBA 850~\mum. When multiple flux density
measurements at the same wavelength are available, we use \wise\ W3 and W4 
in preference to \iras\ 12 and 25~\mum, and \planck\ 850~\mum\ in preference to 
SCUBA 850~\mum.

\subsection{Gas masses and distances}

 For the comparison between the dust masses and the ISM content, we require HI data for
our galaxies. The integrated flux of \hi\ is obtained from different
surveys, detailed in table \ref{tab:hi surveys}

\begin{table}[ht]
\centering
\caption{\hi\ surveys used in this work}
\setlength{\tabcolsep}{4pt}
\begin{tabular}{c|l}
Sample         &  References \\
\hline\\
HiPASS         &  \citet{koret04,meyet04} \\
               & \citet{wonet06,doyet05}  \\[3pt]
HiJASS         &  \citet{lanet03}         \\ [3pt]
ALFALFA        &  \citet{hayet11,maret09} \\
               &  \citet{kenet08,saiet08} \\
               &  \citet{saiet07,gioet05}  \\[3pt]
Others         & \citet{spret05} 
\end{tabular}
\label{tab:hi surveys}
\end{table} 

The molecular gas mass ($\rm M_{mol}$) is calculated from global (non-interferometric) 
observations of the CO(J: 1-0) line (expressed in terms of the velocity-integrated flux 
or \lpco\ [K km s$^{-1}$ pc$^{2}$] \citealt{solet97} ) and the conversion factor 
$\rm \alpha_{CO}$ (\citealt{solval05}; \citealt{bolet13}). 
The correlation between the galaxy metallicity and the $\rm \alpha_{CO}$
value (\citealt{leret11}; \citealt{sandet13}) shows that the $\rm \alpha_{CO}$ takes values
$\sim2$ to $\sim20$ times the Galactic value for sources with metallicities (12+log(O/H)) 
smaller than 8.2. 
In our sample, the stellar mass ranges between $10^{9}$ and $10^{11}$ \msun. 
Over this stellar mass range, metallicities are expected to be between 8.4 and 9.1 \citep{treet04}: over
this limited metallicity range, it is valid to use a constant value of $\rm \alpha_{CO}$ \citep{leret11}. 
In our study, we use $\rm \alpha_{CO}=4.3$[\msun\ (K km s$^{-1}$ pc$^{2})^{-1}$] 
which includes a correction for heavy elements of 36\% \citep{bolet13}. 

Galaxy distances are derived from the redshift listed in 2MRS or the NASA/IPAC Extragalactic 
Database (NED)\footnote{https://ned.ipac.caltech.edu/} except for very nearby galaxies 
(z $<$ 0.045; $\rm D_{L} < 20$ Mpc) for which we use distances from 
the Extragalactic Distance Database (EDD) \footnote{http://edd.ifa.hawaii.edu/}.

\subsection{AGN contamination}

Since our study is focused on dust emission, it is crucial to discard galaxies in which the IR and sub-mm 
fluxes are highly contaminated by AGN emission.
We use the \citet{veceve10} catalog to identify and discard sources with AGN. 
This catalog contains 168,941 objects at redshifts between 0 and 6.43 (from which 5,569 are at z$<$0.1).  
We discarded all (345) sources in our sample which fall within 30\arcsec\ of any source 
in the \citet{veceve10} catalog. 
 Additionally, using the AGN selection criteria showed by \citet{cluet14} based on \wise\
colors (using filters W1,W2 and W3), we rejected 43 galaxies.
 Finally, we excluded all (73) sources with Rayleigh-Jeans (RJ) ($\lambda>300~\mum$)
spectral slope significantly lower than that expected from a gray-body  with $\beta=1.3$ and $\rm T$=15 K, 
since for these sources the emission in the RJ regime is likely contaminated by synchrotron emission.
In other words, these $\beta$ and T values imply the exclusion of 
all sources with colors: 
$\rm f_{350}/f_{550}<2.3$, $f_{350}/f_{850}<6.6$ and $f_{550}/f_{850}<2.9$ 
where $\rm f_{350}$, $\rm f_{550}$ and $\rm f_{850}$ are the \planck\ 
fluxes at 350, 550 and 850~\mum , respectively.

\subsection{Final sample}

Since our analysis requires accurate fitting of dust model SEDs from IR to sub-mm data, 
we restrict the two samples above to only those galaxies for which meaningful 
spectral fits are found (see Sect.~\ref{subsec:spectral fit}).
The final sample - with dust SED fits - comprises 1,630 galaxies, 
which all have reliable \mstar\ estimations. 
Of these galaxies, 136 are CO-detected and 1,230 have HI masses. 
From visual inspection of the SDSS and 2MASS images we classified 87 
galaxies as interacting in the sample.

\begin{figure}[tb]
\begin{center}
   \includegraphics[bb=85 80 279 285,width=0.24\textwidth,clip]{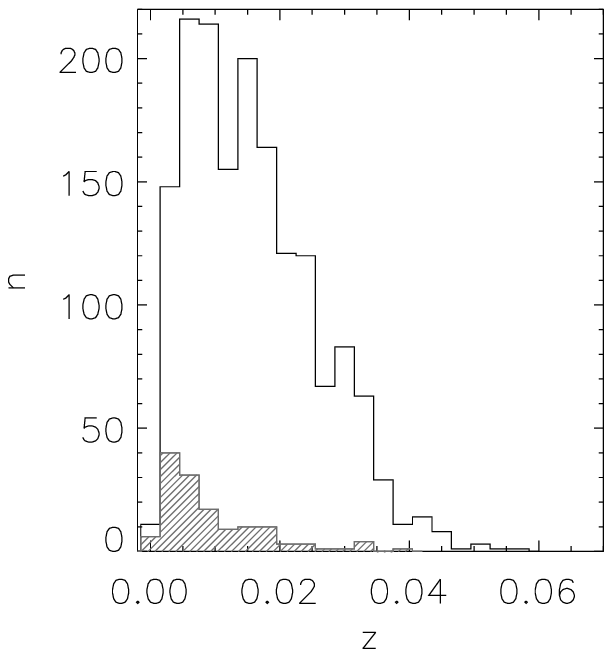}
   \includegraphics[bb=85 80 279 285,width=0.24\textwidth,clip]{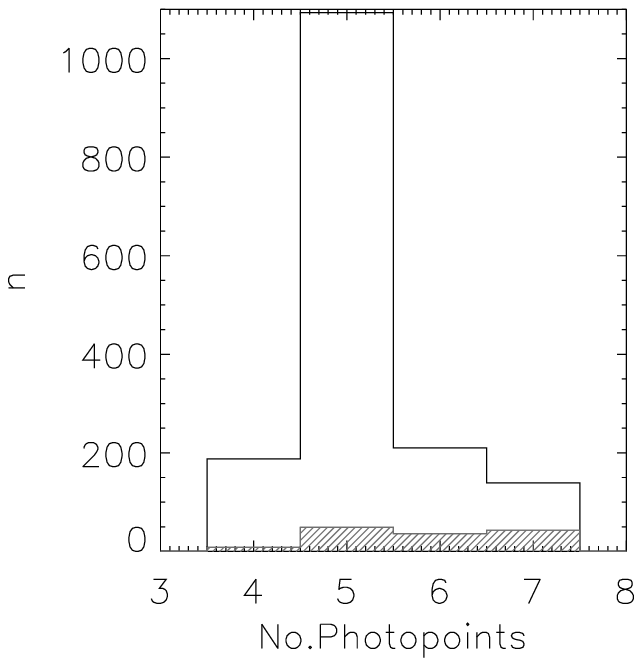}
   \caption{Distribution of redshift (left) and number of photometric points per galaxy (right) for the 1,630 objects 
   in our final sample. 
   The shaded histograms in both panels show the equivalent distributions for the subset of galaxies with CO detections.
   }  
\label{fig:sample z photo}
\end{center}
\end{figure}

The redshift distribution of the final sample is shown in the left panel of 
Fig.~\ref{fig:sample z photo}. The median redshift is $<$z$>$=0.015 for the entire
sample, with a mean of 0.012 and sigma = 0.011. 
For the subsample of galaxies with CO measurements, the median value is $<$z$>$=0.0066.
with mean of 0.0032 and sigma of 0.0042.
The distribution of the number of photometric data points 
(between 12~\mum\ and 850~\mum) per galaxy  is shown in the 
right panel of Fig. \ref{fig:sample z photo}: 
21\% of the sample have more than 6 photometric points,
67\% have five, and only 12\% have four photometric points. 
The subsample of
galaxies with CO observations has a median of 6 photometric 
points per galaxy; 59\% of these galaxies have $\geq$6 photometric 
data points and only $6\%$ have four photometric data points. 
In all cases we cover both sides of the emission peak
at $\sim$100~\mum; for the few galaxies with only 4 photometric data
points, these are distributed as 2 or 3 points at $\lambda<100~\mum$ 
and 1 or 2 at $\lambda>100~\mum$.

\section{Modeling Dust emission}
\label{sec:model dust emission}

The DL07 model describes the total galaxy spectrum by a linear combination 
of one stellar component, approximated by a black body with a specific color
temperature T$_{*}$, and two dust components.
One component with dust fraction  = (1 - $\gamma$)  is located in the diffuse 
interstellar medium (ISM) and  heated by a radiation field with constant 
intensity $\rm U=U_{min}$; the other component with dust fraction = $\gamma$ is exposed 
to a radiation field generated by photo-dissociation regions (PDRs) 
parametrized by a power-law $\rm U^{\alpha}$, over a range of intensities 
$\rm U_{min} < U < U_{max}$, with $\rm U_{max}\gg U_{min}$. 

Thus, the canonical model emission spectrum of a galaxy is:
\begin{equation}
\rm \emph{f}_{\nu}^{~model}=\Omega_{*}B_{\nu}\left(T_{*}\right)+\frac{M_{dust}}{4\pi D_{L}^{2}}\left[\left(1-\gamma\right)p_{\nu}^{(0)}+\gamma p_{\nu}\right]
\label{eq:model-flux}
\end{equation}
where $\Omega_{*}$ is the solid angle subtended by stellar photospheres, 
$\mdust$ is the dust mass, $\rm D_{L}$ is the luminosity distance, and  
$\rm p_{\nu}^{(0)}\left(q_{PAH},U_{min}\right)$ is the emitted power per 
unit frequency per unit dust mass from dust heated by a single starlight 
intensity $\rm U_{min}$.
The dust is a mixture of carbonaceous and amorphous silicate grains 
characterized by the polycyclic aromatic hydrocarbon (PAH) index, $\rm q_{PAH}$, 
defined as the percentage of the total grain mass contributed by PAHs 
with less than $10^{3}$ Carbon atoms. 
Finally, $\rm p_{\nu}\left(q_{PAH},U_{min},U_{max},\alpha\right)$ is similar 
to the previous term but for dust heated by a power law distribution of starlight 
intensities $\rm dM/dU\propto U^{-\alpha}$ extending from $\rm U_{min}$ to 
$\rm U_{max}$. 
For a known galaxy distance, the canonical dust model is thus characterized 
by eight free parameters: $\rm \Omega_{*},~T_{*},~q_{PAH},~U_{min},~U_{max},~\alpha,~\gamma$ 
and \mdust. 

The use of all eight free parameters in the DL07 model requires 
extensive observational datasets and it is computationally demanding. 
For the former reason, we limit the number and range of the free 
parameters as follows: (a) we use the Draine \& Lee model library 
(available on the web \footnote{www.astro.princeton.edu/$\sim$\ draine/dust/irem.html}). 
That library uses a limited parameter range for 
$\rm p_{\nu}^{(0)}\left(q_{PAH},U_{min}\right)$ and 
$\rm p_{\nu}\left(q_{PAH},U_{min},U_{max},\alpha\right)$ in 
Eq. \ref{eq:model-flux}; in which $\rm q_{PAH}$ takes 11 values between 
0.01 and 4.58, $\rm U_{min}$ takes 22 values between 
0.10 to 25.0 and $\rm U_{max}$ takes 5 values between $10^3$ to $10^7$
(as a reference, U = 1 corresponds to the starlight intensity estimate
for the local ISM) {and fixed the value $\alpha=2$};
(b) we follow \citet{draet07}, who show that the dust emission of 
the galaxies of the KINGFISH sample can be well fitted using DL07 models with 
fixed value of $\rm U_{max}=10^{6}$; 
(c) the stellar component, the first term in Eq. \ref{eq:model-flux}, 
is significant only at $\lambda\lesssim10~\mum$. 
Given that we use photometric data at $12~\mum\leq\lambda\leq 1000~\mum$, +
we do not require to use this stellar component. 
To test the influence of the stellar component in fluxes at 12 and 22 \mum, 
we extrapolate the power law obtained from fluxes at 3.4 and 4.6 \mum (W1 and W2, respectively), 
deriving an influence of 4\% and 1\% at 12 and 22 \mum, respectively. 
This means that the stellar component does not affect our dust emission results.
In summary: two parameters ($\rm \Omega_{*}$ and T$_{*}$) are not used
since we do not model the stellar component, 
two parameters ($\rm U_{max}$ and $\alpha$) are fixed to a single value, 
two parameters ($\rm q_{PAH}$ and $\rm U_{min}$) are limited in their range, and only
\mdust\ and $\gamma$ are allowed to  \emph{vary freely} (the \mdust\ is fixed after the minimization
described below in Eq. (\ref{eq:dustmass}) and $\gamma$ runs between 0.0 and 100 in steps of 0.1).  
With these restrictions we generate 24,200 template SEDs,  with luminosities per dust mass 
($\rm \nu L_{\nu}$/\mdust) in $\left[\lsun/\msun\right]$ and
wavelengths ($\lambda$) in $\left[\mum\right]$.
Each observed galaxy SED is fitted to each of the 24,200 SED templates solely 
by varying \mdust. The best fit value of \mdust\ is calculated by the 
minimization of $\chi^{2}$, where  
\begin{equation}
\rm  \chi^2\equiv\sum_i\frac{\left(F_{i}^{obs}-M_{dust}\left\langle \emph{f}_{\nu}^{~model}\right\rangle_{i}\right)^2}{\left(\sigma^{obs}_{i}\right)^2}
\label{eq:chi2}
\end{equation}
Here $\rm F_{i}^{obs}$ is the observed flux at the i$^{\rm th}$ band in Jy with an error $\rm \sigma_{i}^{obs}$ and 
$\rm ~\langle \emph{f}_{\nu}^{~model}\rangle_i$ is the DL07 template flux  per unit dust mass
in units of $\rm (Jy/\msun)$, convolved with the response function for the i$^{\rm th}$  band. 

The minimization of the Eq. \ref{eq:chi2} gives: 
\def\suma{\rm \sum_{i=0}^{N}\frac{F_{i}^{obs}\left\langle \emph{f}_{\nu}^{~model}\right\rangle_{i} }{\left(\sigma^{obs}_{i}\right)^{2}}}
\def\sumb{\rm \sum_{i=0}^{N}\frac{\left\langle \emph{f}_{\nu}^{~model}\right\rangle_{i}^{2}}{\left(\sigma^{obs}_{i}\right)^{2}}}
\begin{equation}
\rm M_{dust}\left[\msun\right]=\suma\left(\sumb\right)^{-1}
\label{eq:dustmass}
\end{equation}
The accuracy of the fit is parametrized by the reduced $\rm \chi^2$ 
$\rm \left(\chi^2_r\equiv \chi^2/dof~;~ dof=degrees~of~freedom\right)$ value.  

For this best fit value of \mdust~ (and for each of the 24,200 SED templates)
we calculate the template spectrum from Eq. (\ref{eq:model-flux}) and
obtain the total infrared [8 to 1000~\mum] luminosity (\lir) following:
\begin{equation}
\rm  L=\int^{\lambda_{max}}_{\lambda_{min}}L_{\nu}\left(\lambda\right)\times\frac{c}{\lambda^2}\left[\lsun\right]~d\lambda
\label{eq:lum integred}
\end{equation}

Instead of using only the final best fit template for a given galaxy, it is more robust to
use a final template fit (FTF) which is the weighted mean of all templates for which 
$\rm \chi^{2}_{r} \leq \hbox{min}\left(\chi^{2}_{r}\right)+1$.
Thus, the values of \mdust\ and \lir\ 
are calculated as the geometric mean, weighted by the individual 
$\rm \chi^{2}_{r}$, of all templates which satisfy our $\rm \chi^{2}_{r}$ criteria. 

Given that the dust in the DL07 models is distributed in two components (diffuse and PDR, each 
with a different radiation field intensity) a large range of dust temperatures is present. 
For several reasons - especially to search for systematic changes with other parameters - it
is useful to characterize the dust as having a single, or at most two, temperature(s). We use
two methods to characterize the effective temperature(s) of the FTF.
We calculate the luminosity weighted temperature (T$_{\rm weight}$) of the FTF, defined as:
\begin{equation}
\rm T_{weight}[K]=\sum_{i=0}^{N}\frac{b}{\lambda_i}~ L_{\lambda,i} \left(\sum_{i=0}^{N} L_{\lambda,i}\right)^{-1}
\label{eq:t weighted}
\end{equation}
where b is the Wien's displacement constant ($\sim$ 2,897 [$\rm \mum ~K$]),
and $\rm L_{\lambda,i}$ is the monochromatic luminosity at wavelength $\lambda_{i}$.
We also fit a two-temperature dust model to the FTF of the galaxy, using 
a cold dust component $\left(\rm T_{cold}\right)$ and a warm dust component 
$\left(\rm T_{warm}\right)$, each described by a gray-body spectrum: 
\begin{equation}
\rm  S_{tot}=A_1\nu^{\beta}B(\nu,T_{cold})+A_2\nu^{\beta}B(\nu,T_{warm})
\label{eq:two component grey bodies}
\end{equation}
where $\nu$ is the frequency, $\rm A_1$ and $\rm A_{2}$ are normalization factors for
each gray-body, $\beta$ is the dust emissivity index (assumed to be the same for
both components)\footnote{The spectral index (commonly referred as $\alpha$) 
of the dust emission in the Raleigh-Jeans (RJ) limit is thus 2+$\beta$.}, 
and $\rm B(\nu,T_{cold})$ and $\rm \left(B(\nu,T_{warm}\right)$ 
are the Planck functions for the cold and warm 
dust components, respectively. 
The fit was performed using the MPFIT code\footnote{www.physics.wisc.edu/$\sim$craigm/idl/fitting.html}, 
which uses a robust minimization routine to obtain the best fit. The two-temperature
dust model fits were performed over the wavelength range 22 - 1000~\mum; 
wavelengths shorter than 22~\mum\ were not used to avoid the complexity
of the PAH emission features. 

In this work we use dust mass (\mdust) obtained from the DL07 fits, i.e. from the FTF. However, for
comparison, we also calculate the dust mass implied by the two temperature dust model fit ($\rm M_{dust}^{2gb}$).
The total dust mass of the two temperature dust model fit is calculated as follows: 
(see \citealt{dunet01}):
\begin{equation}
 \rm M_{dust}^{2gb}=\frac{S_{850}D_L^{2}}{\kappa_{850}}\times\left[\frac{N_{cold}}{B(850,T_{cold})}+\frac{N_{warm}}{B(850,T_{warm})}\right]
 \label{eq:2gb Md}
\end{equation}
where $\rm S_{850}$, $\kappa_{850}$, and $\rm B(850,T)$ are the observed flux, the dust 
emissivity and the black body emission at 850~\mum, respectively, $\rm T_{cold}$  and  $\rm T_{warm}$
are the dust temperatures of the cold and warm components, and $\rm N_{cold}$
and $\rm N_{w}$ are the relative masses of the cold and warm dust components. 
Using the SLUGs sample, \cite{dunet00} obtained a dust emissivity value of 
$\rm\kappa_{850}=0.077 m^2~kg^{-1}$. However, more recent works support lower 
emissivity values at 850~\mum:  $\rm \kappa_{850}=0.0383~m^2kg^{-1}$ \citep{dra03}, i.e.
higher dust masses for a given observed flux. 
In our study, we use the latter value to calculate the dust mass using the two dust components.

\section{Results}
\label{sec:results}

\subsection{Spectral fits}
\label{subsec:spectral fit}

\begin{figure}[tb]
\begin{center}
   \includegraphics[bb=85 80 279 279,width=0.25\textwidth,clip]{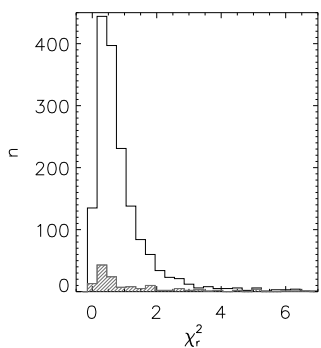}
   \caption{The histogram of the reduced $\chi^{2}$ $\rm \left(\chi^{2}_{r}\right)$ distribution 
   obtained for the DL07 model fits to our sample galaxes. The gray histogram shows the distribution for the subset 
   of sources with CO observations. 
}  
\label{fig:sample red chi}
\end{center}
\end{figure}

Using the procedure outlined in the previous Section, we were 
able to obtain a FTF for 1,630 galaxies. 
The distribution of the $<\rm \chi^2_r>$ obtained for these 
fits are shown in Fig. \ref{fig:sample red chi}: the median 
value is $<\rm \chi^2_r>=0.62$ and $92\%$ of the spectral 
fits satisfy $\rm \chi^2_r<2.0$.

For our sample, the DL07 dust model fits result in the following parameter ranges.
$\gamma$ ranges between 0.0 and 0.02 with a median value equal to 0.01;
75\% of the sample have $\rm U_{min}$ in the range between 0.2 to 3.0,
with a typical value equal to 1.5;
$\rm q_{PAH}$ shows a typical value 3.19, and 91\% of the sample are best fit with
templates based on Milky Way models
(see Appendix \ref{apn:parameters} for more details).

\begin{figure*}[!ptb]
\begin{center}
  \includegraphics[bb=116 95 417 234,width=0.9\textwidth,clip]{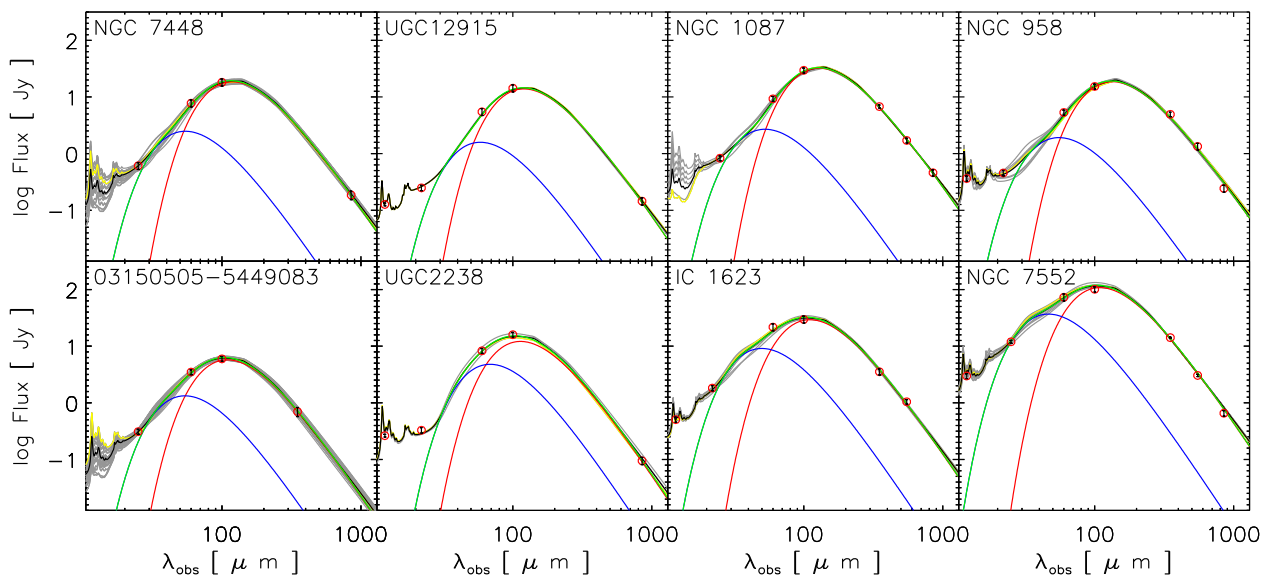}
         \caption{Examples of DL07 spectral energy distribution (SED) fits to galaxies in our sample, as named
         in each panel. Photometric data points are shown with red circles and corresponding 
         black error bars.
         Panels are organized (left to right) by the number of photometric data points (4 to 7
         points). The top (bottom) row shows 
         'normal' star forming  (starburst; SB) galaxies;
         (see Sect.~\ref{subsec:SFM}).  The single best-fit DL07 template SED is shown in yellow.
         All SEDs which satisfy the reduced $\chi^2$ criterion (see Sect.~\ref{sec:model dust emission})
         are shown in gray; the weighted geometric mean of these is used as our final
         template fit (FTF; shown in black).  The two temperature component model fits 
         (eqn.~\ref{eq:two component grey bodies}) are shown in red (cold dust component), 
         blue (warm dust component), and green (sum of cold and warm dust components). 
}  
\label{fig:SED}
\end{center}
\end{figure*}

\begin{figure*}[!pht]
\begin{center}
  \includegraphics[bb=95 79 285 283,width=0.33\textwidth,clip]{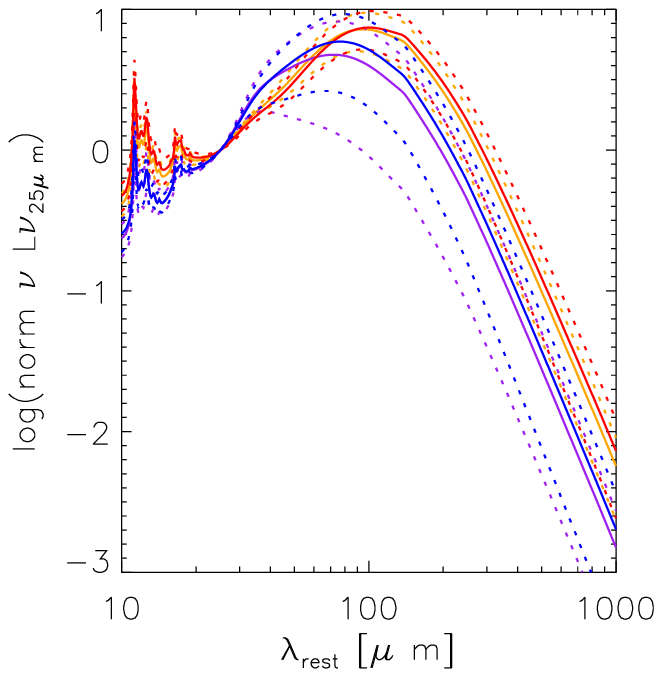}
  \includegraphics[bb=95 79 285 283,width=0.33\textwidth,clip]{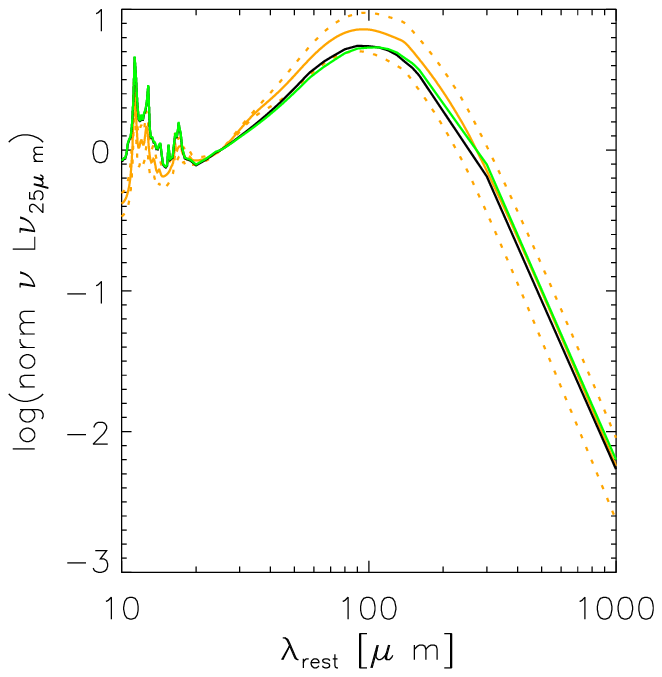}
  \includegraphics[bb=95 79 285 283,width=0.33\textwidth,clip]{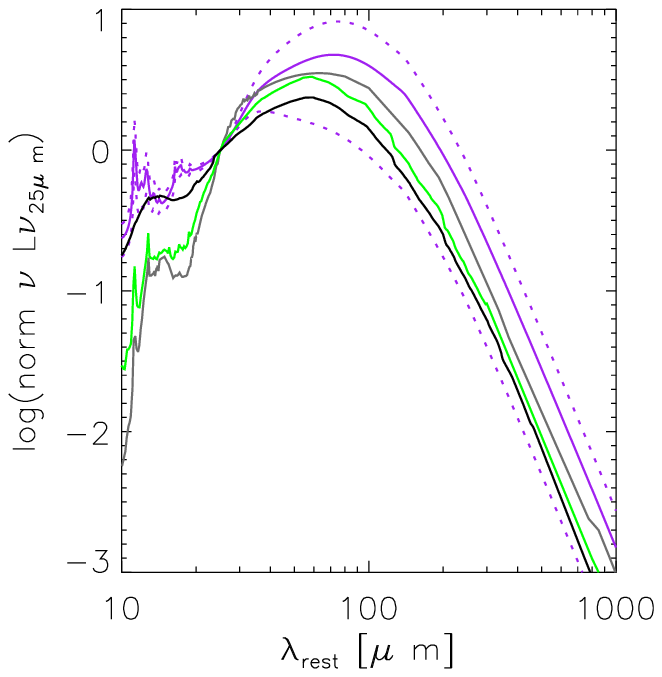}
         \caption{Left panel: Composite SEDs of galaxies with normal sSFR (875 normal 
         star forming galaxies; red) and high sSFR (26 SB galaxies; blue), 
         for wavelengths between 10 and 1000~\mum. The orange spectrum is from a 
         subsample (277 templates) of the normal galaxies closest to the Main Sequence 
         ($\rm -0.5 < sSFR [yr^{-1}]< 0.5$, see Section \ref{subsec:SFM}). 
         Similarly, the purple spectrum is a subsample (11 templates) of the 
         most extreme SB galaxies in our sample ($\rm sSFR [yr^{-1}]>$ 0.8).
         Middle panel: Comparison of composite SED of the subsample of galaxies closest to the Main Sequence (orange) 
         with templates of Sa and  Sc galaxies (black and green, respectively, obtained from the SWIRE template
         library, see text). 
         Right panel: comparison of composite SED of the subsample of the most extreme SB galaxies (purple) 
         with spectra of Arp220 (a starburst-ULIRG; gray), \iras\ 22491-1808 
         (a starburst-ULIRG; green), and \iras\ 19254-7245 South (a Seyfert 2-starburst-ULIRG; black).
         In all panels, the composite spectra are plotted using solid lines, and their corresponding 
         dispersions ($\pm 0.7\sigma$) are plotted using dotted lines of the same color. 
         All SEDs are normalized to the 
         25~\mum\ luminosity ($\rm\nu L_{\nu25~\mum}$) in order to best contrast changes in
         the SED peak ($\sim$100~\mum), sub-mm luminosities, and PAH emission at $\sim5$ to 11~\mum.
         Note that our fits use data between 12~\mum\ and 850~\mum. Further, 
         our composite spectra do not include a stellar component, and thus do not match the 
         comparison spectra below $\sim 10~\mum$.
        }  
\label{fig:mean SEDs}
\end{center}
\end{figure*}

Figure~\ref{fig:SED} shows eight example SED fits 
- both DL07 model fits and two temperature component fits - to galaxies in our sample. 
Clearly, when observed fluxes at $\leq$25~\mum~ are absent, a large number of DL07 templates
can be fitted: these templates show large differences at $\lambda<60~\mum~$, but are similar at
wavelengths in the Rayleigh-Jeans tail ($\lambda>300~\mum$). However, as shown by the robustness test 
(Section \ref{subsec:robustness} and Appendix \ref{apn:robutsness test}) for a two-temperature
component fit, the warm dust component (blue spectrum) is not affected by the absence 
of an observed flux at $\leq$25~\mum\ if we follow our fitting criteria (see Section \ref{sec:model dust emission}
and Appendix \ref{apn:robutsness test}). In a similar way, the cold dust component (red spectrum) does not
vary significantly between different templates, as long as the galaxy has at least one measurement
in the Rayleigh-Jeans tail. The figure also illustrates that the two temperature component model
(green spectrum) reproduces well both the best fit SED (yellow spectrum) and the 
final template fit used by us (FTF; black spectrum). 

To obtain the typical SEDs of galaxies with normal (MS) and high (SB) sSFR,
we use the geometric mean (weighted by $\chi^{2}$) of all MS and SB galaxies. 
The left panel of Figure \ref{fig:mean SEDs} shows the composite SED of all 875 MS galaxies (red spectrum) 
and all 26 SB galaxies (blue spectrum), where the SB galaxies have $\rm sSFR >0.8~[yr^{-1}]$.
We also show the composite SEDs of two `cleaner' sub-samples: those closest 
to the MS ($\rm 0.8 < sSFR [yr^{-1}] <1.2$, orange spectrum) 
and those with the highest SFR in our sample ($\rm sSFR  >1.6~[yr^{-1}]$, purple spectrum).

For each individual composite (MS and SB, Figure \ref{fig:mean SEDs} middle-panel) SED, 
the dispersions are small in the RJ tail: 
they are thus easily distinguishable from each other at wavelengths 
longer than $\lambda>200~\mum$, as long as a good short-wavelength ($\sim$25~\mum) point is available 
for relative normalization. The largest differences between the two template spectra are seen near the 
FIR peak: high sSFR (SB) galaxies have a more dominant warm dust component: thus their emission peak
is shifted to smaller wavelengths and the width of the peak is larger.

To compare our composite SEDs to those of galaxies with well characterized SEDs, we use the spectra 
available in the SWIRE template library 
\citep{polet07}\footnote{www.iasf-milano.inaf.it/$\sim$polletta/templates/swire\_templates.html}.
 The SWIRE templates, which are based on the GRASIL code \citep{silvet98}, contain SEDs for ellipticals,
spirals and starburst galaxies

The composite SED of our galaxies closest to the MS is compared to templates of Sa (black line) and Sc (green line) 
spiral galaxies (the Sb template is not shown as it is very similar to that of Sa galaxies)
in the middle panel of Fig.~\ref{fig:mean SEDs}. Clearly, there is a good agreement 
- within the 3$\sigma$ dispersion - for $\lambda>20~\mum$; at shorter
wavelengths the stellar component, present in the Sa and Sc templates but not in our 
MS composite, is the main reason for the observed differences. 
The composite SED of our highest sSFR sub-sample is compared to the spectrum of Arp220 (gray), 
\iras\ 22491-1808 (green) and \iras\ 19254-7245 South (black) in the right panel of Fig.~\ref{fig:mean SEDs}. 
The latter three spectra show a shift of the emission peak to shorter wavelengths compared to 
our SB composite SED, and \iras\ 22491-1808 and \iras\ 19254-7245 South 
show large absorption features at $\lambda\ <$ 25~\mum\ which are not seen in our composite SED or
indeed in any individual DL07 template. 

\subsection{Robustness of the SED fitting}
\label{subsec:robustness}

To test the robustness of our SED fitting, we examine 24 galaxies of our sample with 7 photometric
observations.
Then we explore how the \mdust,\lir, dust temperatures (cold and warm), quantity of SEDs, $\chi^2$ 
and reduced $\chi^2$ vary  with the amount of points and the rejection of specific points 
(e.g., how it is affected by the rejection of the flux at 12 and 22~\mum\ 
or at 350~\mum).
The result of this test reveals that our results are very robust, with the parameters showing factors 
of difference smaller than 0.1 (for \mdust, \lir\ , and dust temperatures). 
This means that the final results and relations obtained in our work, are robust and not affected 
by the amount of points or the distribution in wavelengths of them (following our SED fitting criteria, 
see Section \ref{sec:model dust emission}). 
For more details, see Appendix \ref{apn:robutsness test}.

\subsection{Star formation mode}
\label{subsec:SFM}

The SFR is often estimated from the IR (integrated between 8 to 1000~\mum) and 
ultraviolet ($\lambda=2700\AA$) luminosities (e.g., \citealt{muret11}; \citealt{sanet14}). 
\citet{sanet14} suggest that both luminosities together provide the best estimate
of SFR, but if only one is available, then \lir, rather than UV luminosity, is 
the more reliable. 
We estimate the SFR for our sample galaxies using \lir\ derived from our DL07 model fits
and the relationship in \citet{ken98b}, (assumes a Salpeter initial mass function)
which assumes a
\begin{equation}
\rm SFR~\left[\frac{\msun}{yr}\right]=1.78\times10^{-10}~\lir~\left[L_{\sun}\right].
\label{eq:sfr}
\end{equation}

\begin{figure}[ht]
 \begin{center}
    \includegraphics[bb=84 103 281 165,width=0.45\textwidth,clip]{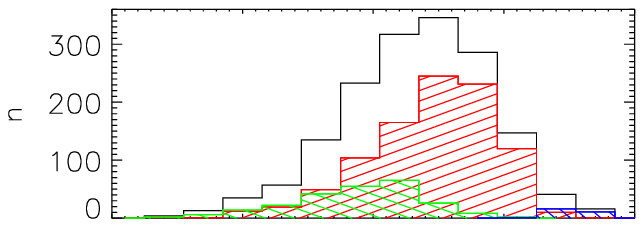}
    \includegraphics[bb=84 77 281 129,width=0.45\textwidth,clip]{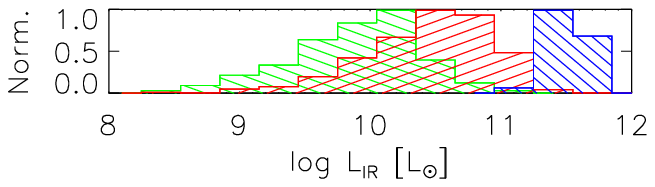}
    \includegraphics[bb=84 103 281 165,width=0.45\textwidth,clip]{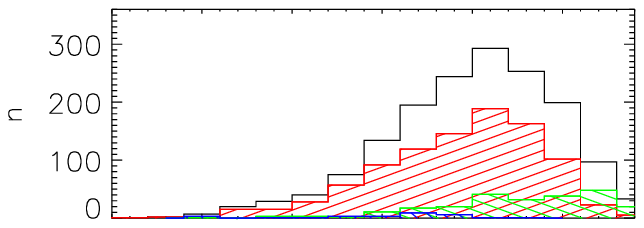}
   \includegraphics[bb=84 77 281 129,width=0.45\textwidth,clip]{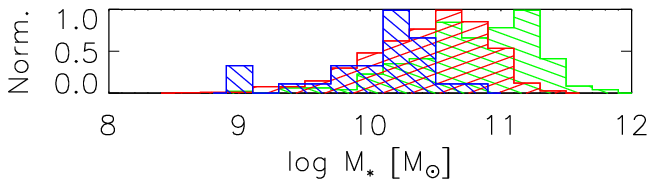}
         \caption{ Distribution of \lir\ (from the DL07-model-fits; top panels) and stellar mass
         (following \citet{cluet14}; bottom panels) for our sample.  
         Histograms for our full sample (black), all SBs (blue), all MS (red) and all passive galaxies (green)
         are shown.  The bottom sub-panels show the equivalent normalized histograms. 
          }  
  \label{fig:lir-mstar histo}
  \end{center}
\end{figure}
The top panels of Figure \ref{fig:lir-mstar histo} show the \lir\ distribution (black histogram)
for our full sample. Infrared luminosities are $3.9\times10^{6}$ to $7.0 \times10^{11}$ $\rm L_{\sun}$ 
with a typical error of 13\% and median value of $2.2\times10^{10}$ $\rm L_{\sun}$ . 
Comparing the \lir\ measured for the KINGFISH galaxies and our final sample with good SED fitting, 
we obtain a median ratio of 1.4.
The bottom panel in Figure \ref{fig:lir-mstar histo} shows the distribution of the stellar mass (\mstar),
which ranges between $3.5\times10^8$ and $7.9\times10^{11}$ $\rm M_{\sun}$ with 
a median value of $3.6\times10^{10} \rm M_{\sun}$ and typical error of 20\%. 

\begin{figure}[ht]
  \begin{center}
    \includegraphics[bb=43 145 520 645,width=0.45\textwidth,clip]{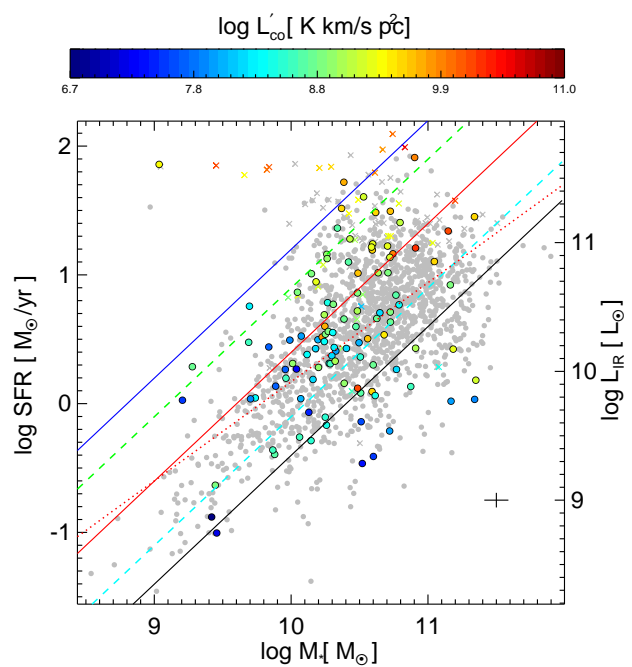}
         \caption{ The star formation rate - stellar mass plane. 
         Interacting galaxies are shown with Xs, and non-interacting 
         galaxies (or those which we were unable to classify due to lack of deep imaging) with
         circles. 
         Symbol colors indicate the CO J:1-0 luminosity, with gray denoting galaxies 
         without global CO J:1-0 measurements. 
         The red solid line shows the MS locus from \citet{elbet11},  
         and the red dotted line shows the MS locus from \citet{elbet07}. 
         The other lines delineate the limits used to classify our sample in starburstiness:
         SB galaxies lie above the blue solid line;  
         Intermediate-SB (InSB) galaxies lie between the blue solid and green dashed lines; 
         normal star forming galaxies (MS) lie between the green dashed and the cyan dashed lines; 
         and PAS galaxies lie below the black solid line. The typical error is shown in the bottom right
         corner. 
}  
  \label{fig:sSFR}
  \end{center}
\end{figure}

Figure \ref{fig:sSFR} shows all our sample galaxies in the SFR-\mstar\ plane.
The sample covers roughly three orders of magnitude in both SFR and stellar mass, 
and are distributed on both sides of the locus of the \citet{elbet11} MS line. 
Given the cutoffs in sSFR we use for SB, MS, and passive galaxies (see Sect.~\ref{secintro})
the percentages of these sub-groups in our sample are 2.0\%, 58.9\% and 15.\%, respectively.
In on equivalent maner, the \citet{elbet07} MS (dotted red line in figure
\ref{fig:sSFR}) is used, the same sSFR displacements from the MS are used to define MS, 
SB and PAS galaxies, we see no great changes in the number of MS galaxies.
However, the number of SBs increases (factor 1.8) and PAS galaxies decrease (factor 4.6). 

\subsection{Dust Masses, Temperatures, and emissivity index ($\beta$)}
\label{subsec:dust properties}

\begin{figure*}[ht]
  \begin{center}
  \includegraphics[bb=97 115 276 276,width=0.245\textwidth,clip]{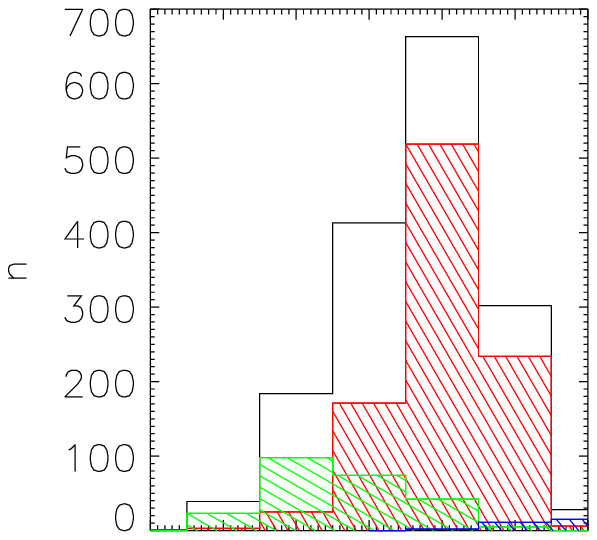}
  \includegraphics[bb=97 115 276 276,width=0.245\textwidth,clip]{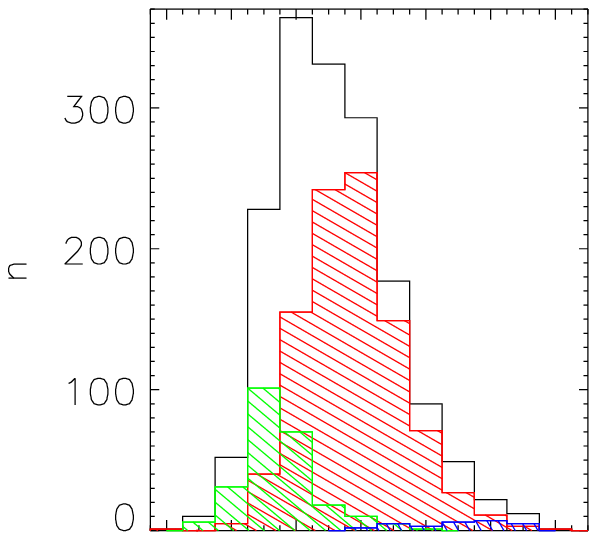}
  \includegraphics[bb=97 115 276 276,width=0.245\textwidth,clip]{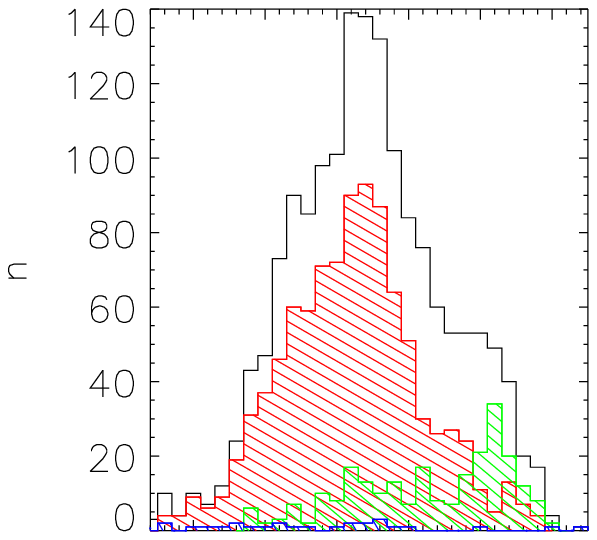}
  \includegraphics[bb=97 115 276 276,width=0.245\textwidth,clip]{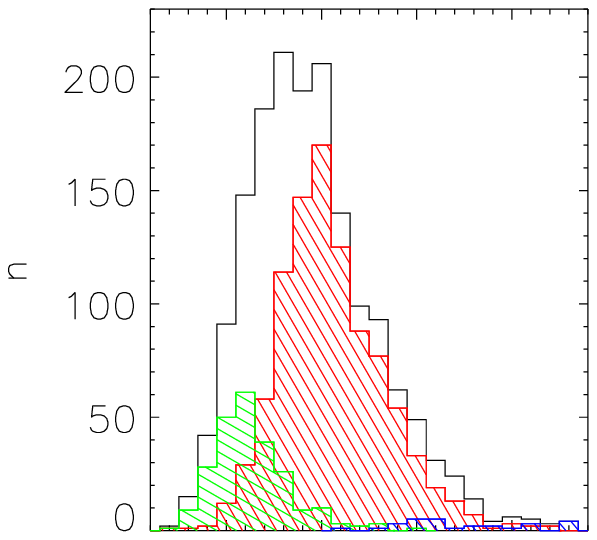}
  \includegraphics[bb=97 77 276 165,width=0.245\textwidth,clip]{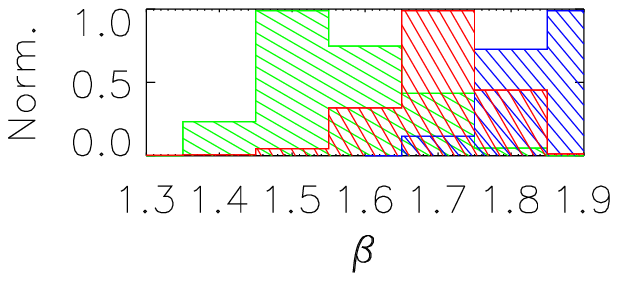}
  \includegraphics[bb=97 77 276 165,width=0.245\textwidth,clip]{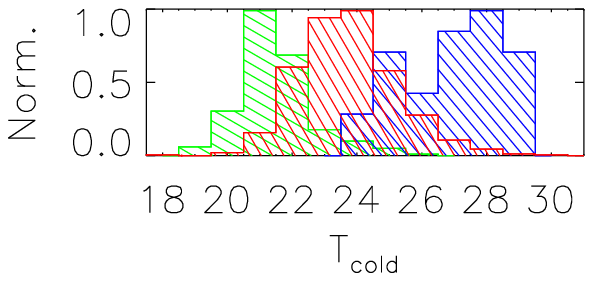}
  \includegraphics[bb=97 77 276 165,width=0.245\textwidth,clip]{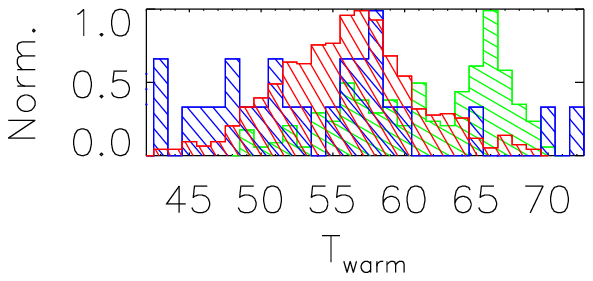}
  \includegraphics[bb=97 77 276 165,width=0.245\textwidth,clip]{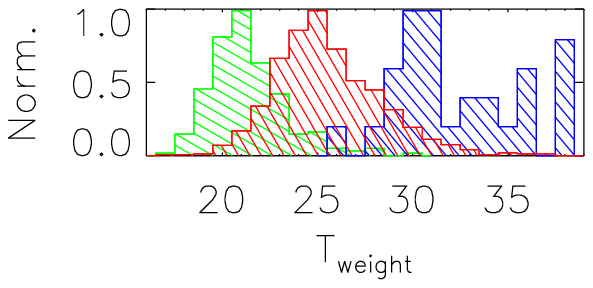}
         \caption{ From left to right: the distributions of 
         the emissivity index ($\beta$), the temperatures of the cold (\tc) and
         warm (\tw) dust components , and the luminosity-weighted 
         dust temperature ($\rm T_{weight}$).
         In all panels, the distribution of the full sample is shown in black, 
         SB galaxies in blue, MS galaxies in red, and PAS galaxies in green. 
         The lower panels show the equivalent normalized histograms.  
         }
  \label{fig:dust temp}
  \end{center}
\end{figure*}

The presence of multiple dust temperatures in the DL07 models 
(see Sect.~\ref{sec:model dust emission}) precludes the direct 
application of Wien's law to the model template (and thus our FTF) in order to obtain a dust temperature.
For this reason we use two temperature component model fits to the FTF to parameterize
the dust temperature (see Sect.~\ref{sec:model dust emission}). 
Figure \ref{fig:dust temp} shows the distributions of
 the gray-body emissivity index ($\beta$), the temperatures of the cold (\tc) 
and warm (\tw) dust components in the two temperature component fits, 
(eqn. \ref{eq:two component grey bodies}),
and the luminosity weighted dust temperature ($\rm T_{weight}$, eqn. \ref{eq:t weighted}). 
In agreement with previous results (e.g \citealt{dunet01}, \citealt{clemet10}, \citealt{clemet13})
for nearby galaxies, our fitted values of $\beta$ are distributed over 1.3 - 1.9, 
with a median of $\sim$1.7 (Fig.~\ref{fig:dust temp}). 
The $\beta$ distributions for the sSFR-classified sub-samples are significantly different: 
PAS galaxies show the lowest values ( median $\beta\sim1.4$), 
MS galaxies typically show values in the range $\beta$=1.3 - 1.9 with $<\beta>\sim1.7$, and
SB galaxies typically show values of $\beta$=1.7 to 1.9 with $<\beta>\sim1.9$.
Similar differences are seen in the distributions of, and median, temperatures of 
the cold dust component: the median temperature of the cold dust component
is 21.4~K for PAS galaxies, 23.6~K for MS galaxies, and
27.1 K for SB galaxies. 

For the full sample, the warm dust component (from the two component fit;
Fig. \ref{fig:dust temp}) shows a median value of \tw=57 K.
Unexpectedly, the PAS galaxies show the hottest warm components, though
the relative luminosity of this warm component is neglible w.r.t. the
luminosity of the cold component.
MS galaxies have warm component temperatures distributed relatively tightly around 
\tw=57 K while SB galaxies show a more uniform spread in the distribution of \tw. 
In relative luminosity, however, the SBs are more dominated by the warm dust component:
SB and MS galaxies show a median warm component luminosity to total luminosity 
(cold plus warm component) ratio of 0.01 and 0.14, respectively. 
If, instead, the FTF is characterized by the weighted dust 
temperature, the full sample shows a median weighted temperature of 24.1 K. 
PAS, MS and SB galaxies are clearly separated in T$_{\rm weight}$, 
with median values of 21.0, 25.2 and 31.1 K, respectively.

\begin{figure}[ht]
  \begin{center}
   \includegraphics[bb=84 103 281 165,width=0.45\textwidth,clip]{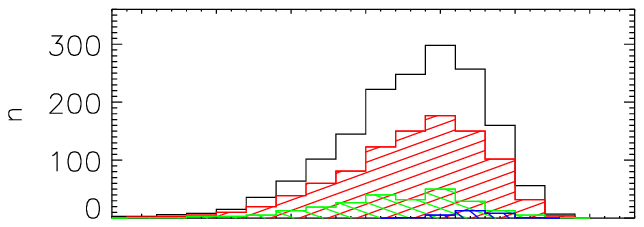}
   \includegraphics[bb=84 77 281 129,width=0.45\textwidth,clip]{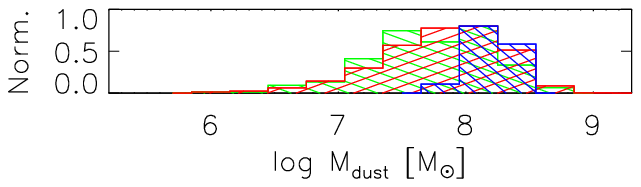}
         \caption{Distribution of the DL07-model-derived dust mass. 
         The full sample is shown in black, and the sub-samples of SB, MS, and PAS galaxies are
         shown with the same colors as in Fig. \ref{fig:dust temp}. The bottom panel shows the equivalent
         normalized histograms. 
          }  
  \label{fig:Md dl07}
  \end{center}
\end{figure}

Figure \ref{fig:Md dl07} shows the distribution of the dust masses, as derived from
the DL07 model fits (see Sect.~\ref{sec:model dust emission}), 
for the entire sample (black), and for the different sSFR sub-samples
(SB in blue, MS in red and PAS in green). 
In the full sample, dust masses range between 
$6.2\times10^{5}\msun$ and $8.6\times10^{8}\msun$, with a median value of $7.5\times10^{7}\msun$
and an estimated typical error of 20\%. This median value is similar (considering our errors) 
to that obtained by \citet[]{clemet13} ($7.8\times10^7\msun$), who used MAGPHYS modeling.
Note that they corrected their model results to an emissivity value of 
$\rm \kappa_{850}=0.0383~m^{2}kg^{-1}$, the same value assumed in our dust mass estimations from 
two dust components and in agreement with the results obtained with the DL07 templates (see below). 
Passive galaxies tend to have lower dust masses than MS and SB galaxies. 
The median dust mass for PAS, MS and SB galaxies are
$6.2\times10^7$, $7.6\times10^7$, $1.7\times10^{8}$ \msun, respectively.
For the two temperature component models, the cold dust 
component dominates the total dust mass: the median contri4bution of warm dust to
the total (warm plus cold) dust mass ($\rm M_w/M_{tot}$) is
0.2\%, with 97\% of the sample at $\rm M_w/M_{tot}$ $<$ 1\%. 
The highest values of $\rm M_w/M_{tot}$ (up to 4\%) are seen in SB galaxies.

\begin{figure}[ht]
  \begin{center}
   \includegraphics[bb=30 136 479 651,width=0.35\textwidth,clip]{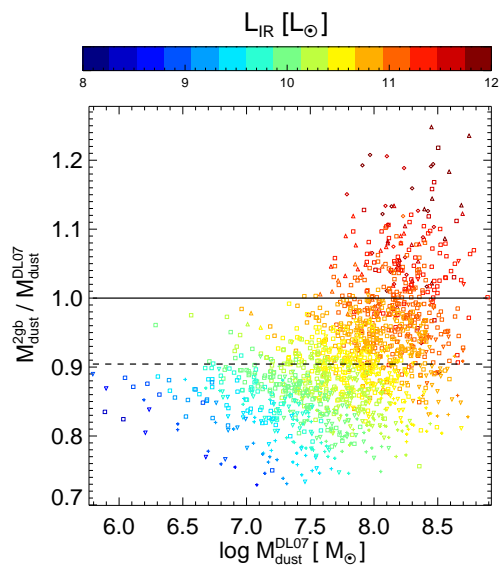}
         \caption{ The ratio of dust masses (two component fit to DL07 model fit) as a function of DL07 model fit dust masses. 
         SBs, MS, and PAS galaxies are plotted as diamonds, squares, and crosses, respectively. 
         Upward and downward triangles show intermediate SB and intermediate PAS sources.
         The line of equality (solid black line) and median ratio (dashed black line) are also shown. 
         Symbols are colored by IR luminosity following the color-bar.
          }  
  \label{fig:mass comp}
  \end{center}
\end{figure}

A comparison of the dust masses derived using DL07 models to the dust masses derived from
our two component $\rm \left(M^{2gb}_{dust};~ eqn.~ \ref{eq:2gb Md}\right)$ 
fits is shown in Figure \ref{fig:mass comp}. Clearly there is a systematic difference
in the two values. Recall that the two
component model was obtained via fits to the DL07 FTF fit rather than a fit to the 
individual photometric data points. The dust mass ratios 
$\rm M^{2gb}_{dust}/M^{DL07}_{dust}$ show a median of 0.91, and 
the best fit relating the two dust masses (see Fig.\ref{fig:mass comp}) is:
\begin{equation}
\rm \frac{M^{2gb}_{dust}}{[\msun]}=10^{-0.34\pm0.06} \left(\frac{M^{DL07}_{dust}}{[\msun]}\right)^{1.04\pm0.01}
 \end{equation}
The symbol colors (by IR luminosity) clearly reveal that the inconsistency in the DL07
and two-component derived dust masses is correlated to the IR luminosity:
sources with lower \lir\ show relatively higher DL07 model dust masses,
while sources with higher \lir\ have relatively higher two gray-body-fit dust masses.
Alternatively, the difference in the two masses is related to the dust temperature of the cold component,
showing an increment from the bottom left corner to the right top corner for the points in the figure. 
The difference in masses is likely a result of
the DL07 models using a more complex calculation of dust mass for a given dust
luminosity, i.e., different grain types and sizes related to the parameter $\rm q_{PAH}$
(see details in \citealt{draet07}), while the dust mass of the two temperature 
component fit is derived via a single emissivity index.
In any case, the difference in dust masses is less than 0.2 dex (factor of 1.58), so these differences 
are relatively unimportant in the correlations presented in the following sections (which
use the DL07-derived dust masses).

\subsection{ \lir,~\mdust, and ~\tdust\ Plane }
\label{subsec:fundamentalplane}

The relationship between dust mass, dust temperature, and dust
luminosity (\lir) is in general well understood (e.g., \citealt{drali07}, \citealt{sco12}): 
when dust grains absorb UV photons from young OB stars, they are heated and re-emit their energy at IR wavelengths.
\citet{clemet13} have shown a strong correlation 
between the SFR/\mdust\ ratio and the dust temperature of the cold dust component, 
especially for sources with T$_{\rm dust,cold}>18$ K. 
Our sample shows a similar correlation, and our larger sample size allows 
us to clearly demonstrate that all galaxies lie in a single plane, 
which we refer to as the \emph{dust plane}, in the \lir\
in the \lir\ (thus SFR), \mdust\ and \tdust\ phase space
(similar to the relation shown by \citealt{genet15}).

\begin{figure}[ht]
  \begin{center}
  \includegraphics[bb=50 150 480 650,width=0.4\textwidth,clip]{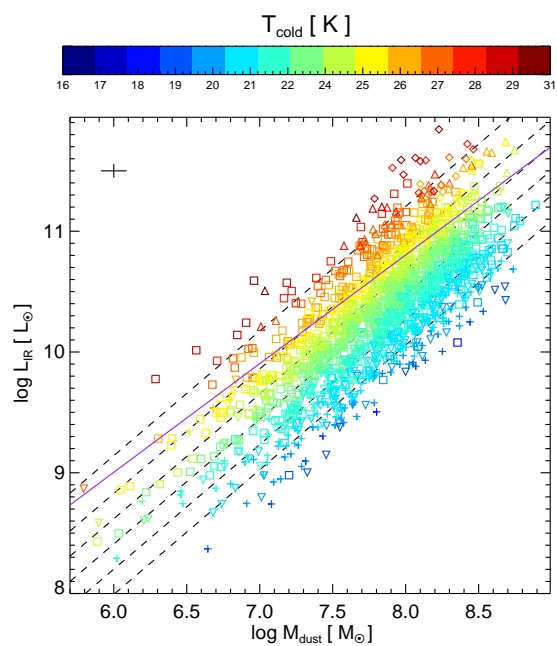}
         \caption{The relationship between IR luminosity (thus SFR) and dust
         mass for our sample, with the symbols colored by the temperature of the cold dust component.
         The figure is a projection of the dust plane we find between \lir, \mdust, and \tc. 
         Symbols - which denote the sSFR classification of the galaxies - are the
         same as in Fig.~\ref{fig:mass comp}.
         The black dashed lines are the fits to the \lir-\mdust\ plane for cold component
         dust temperature bins centered at (bottom to top) 20, 22, 24, 26 and 28~K.
         The purple line shows the relationship between \lir\ and \mdust\ obtained by 
         \citet{dacet10} where dust temperature is not taken into account. 
         The typical error is shown in the top right of the plot.
         }  
  \label{fig:plane1}
  \end{center}
\end{figure}

Figure \ref{fig:plane1} shows a projection of this dust plane in our sample: 
in this case we plot \lir\ vs. \mdust\ and color the symbols by the (cold component)
dust temperature.
The dashed black lines delineate the
\lir-\mdust\ relationships for different cold component dust temperature bins, and 
the purple line shows the relation between \lir\ and \mdust\ 
obtained by \citet{dacet10}.
Here the dust temperature used is that of the cold component of the two component fit; 
this is the dominant component, in mass and luminosity, for all galaxies in the sample. 
The best fit to this dust plane is:
\begin{equation}
 \rm \log \left(\frac{L_{IR}}{[\lsun]}\right) - 1.07 ~\log \left(\frac{M_{dust}}{[\msun]}\right)- 0.19 ~\left(\frac{T_{cold}}{[K]}\right) + 2.53 = 0
 \label{eq:plane}
\end{equation}
Clearly, the placement of a galaxy in the dust plane is most sensitive to the temperature of 
the cold component, rather than the values of \mdust\ and \lir. 
For example, if the dust temperature is constant and we change \mdust\ by a factor of 10, 
the \lir\ is required to change by a factor of 12. 
However, increasing \tc\ by 10 K, with constant \mdust, requires \lir\ to increase by a factor 80.
An increase in \lir\ is more easily achieved by increasing the dust temperature rather
than the dust mass. 
The data of \citet{clemet13} are consistent with a plane similar to that defined 
by Eq. \ref{eq:plane}. However, their data suggest higher values of
both \lir\ and \mdust\ for a given value of \tc. 

If the luminosity weighted dust temperature (T$_{\rm weight}$) is used instead of \tc,  the sample galaxies
still fall on a single dust plane, though there is a larger scatter. 
In this case, the dust plane is parametrized by:
\begin{equation}
 \rm \log \left(\frac{L_{IR}}{[\lsun]}\right) - 1.02 ~\log \left(\frac{M_{dust}}{[\msun]}\right)- 0.10 ~\left(\frac{T_{weight}}{[K]}\right) + 0.30 = 0
 \label{eq:plan1e2}
\end{equation}
If the temperature of the warm component (from the two components fit) 
of the dust is used instead of the temperature of the cold component, the sample galaxies no longer
fall on a single dust plane. 

The correspondence between interacting galaxies and SB galaxies is one-to-one 
only for extreme starbursts: of all interacting galaxies in the sample, only  
those with high cold-dust
component temperature ($\rm T_{cold}> 27$ K) and high SFR ($\rm log (L_{IR}/\lsun) > 11$, 
i.e. LIRGs) are classified as SBs.

\begin{figure*}[ht]
  \begin{center}
  \includegraphics[bb=9 121 475 654,width=0.33\textwidth,clip]{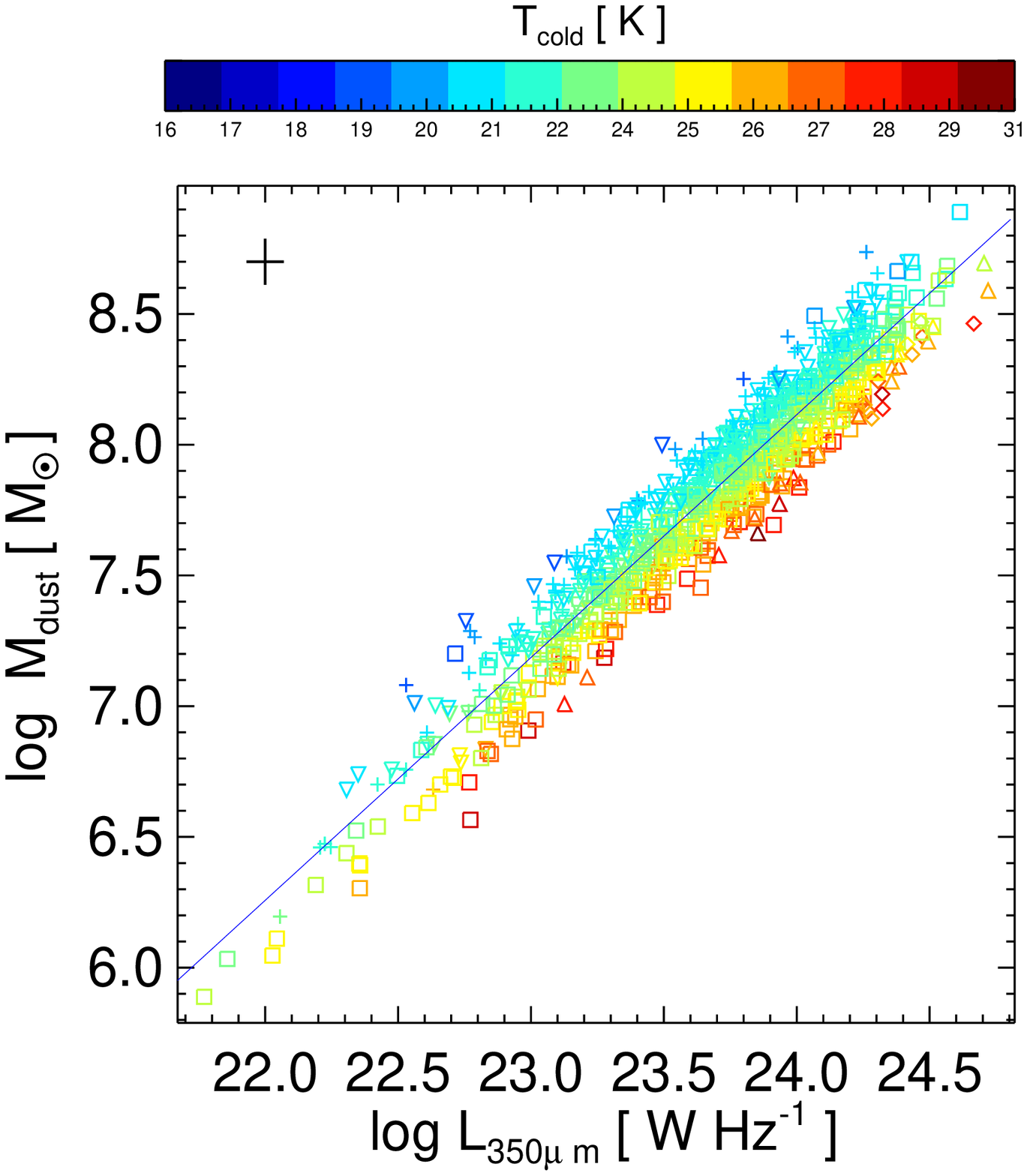}
  \includegraphics[bb=9 121 475 654,width=0.33\textwidth,clip]{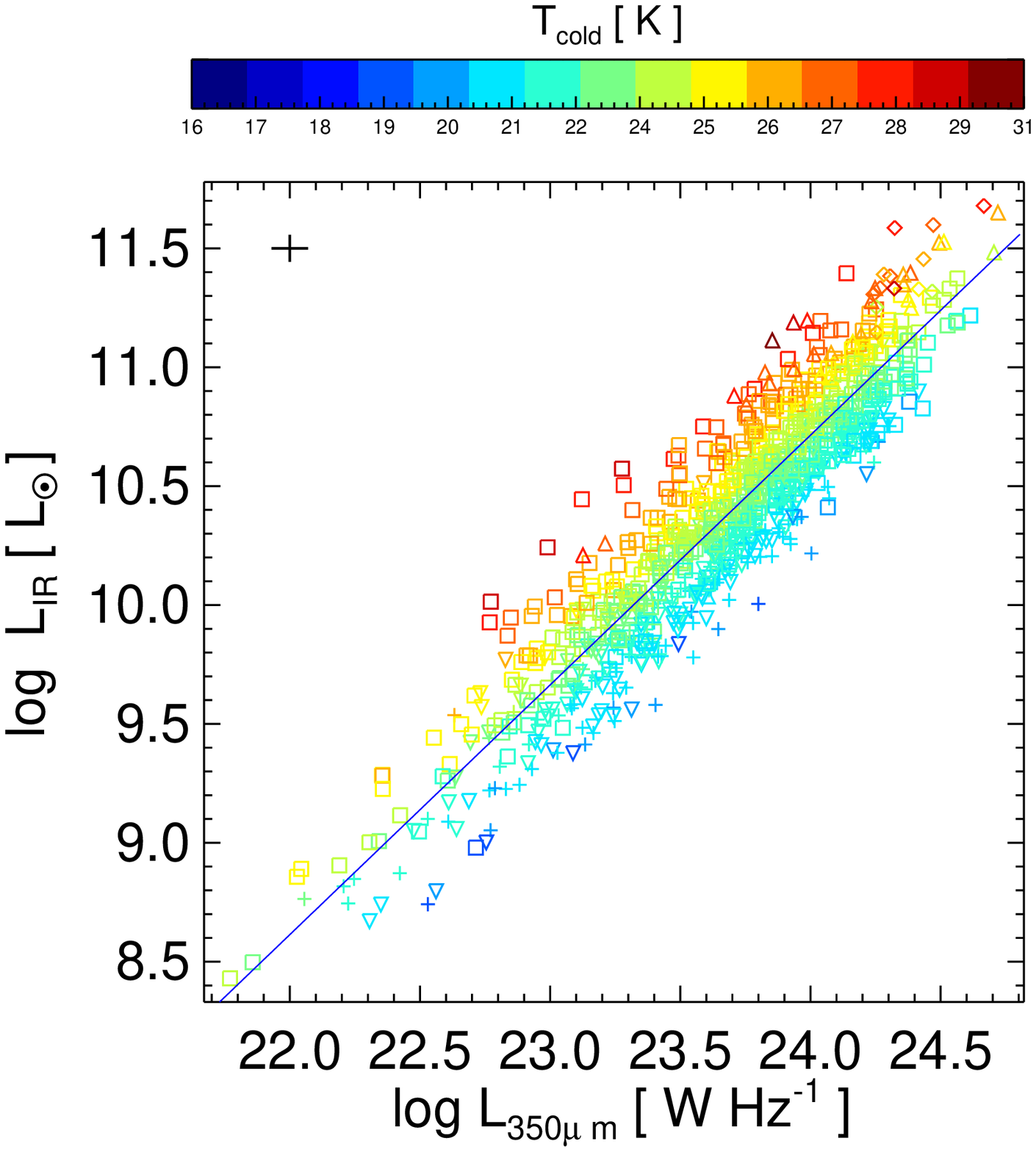} 
  \includegraphics[bb=9 121 475 654,width=0.33\textwidth,clip]{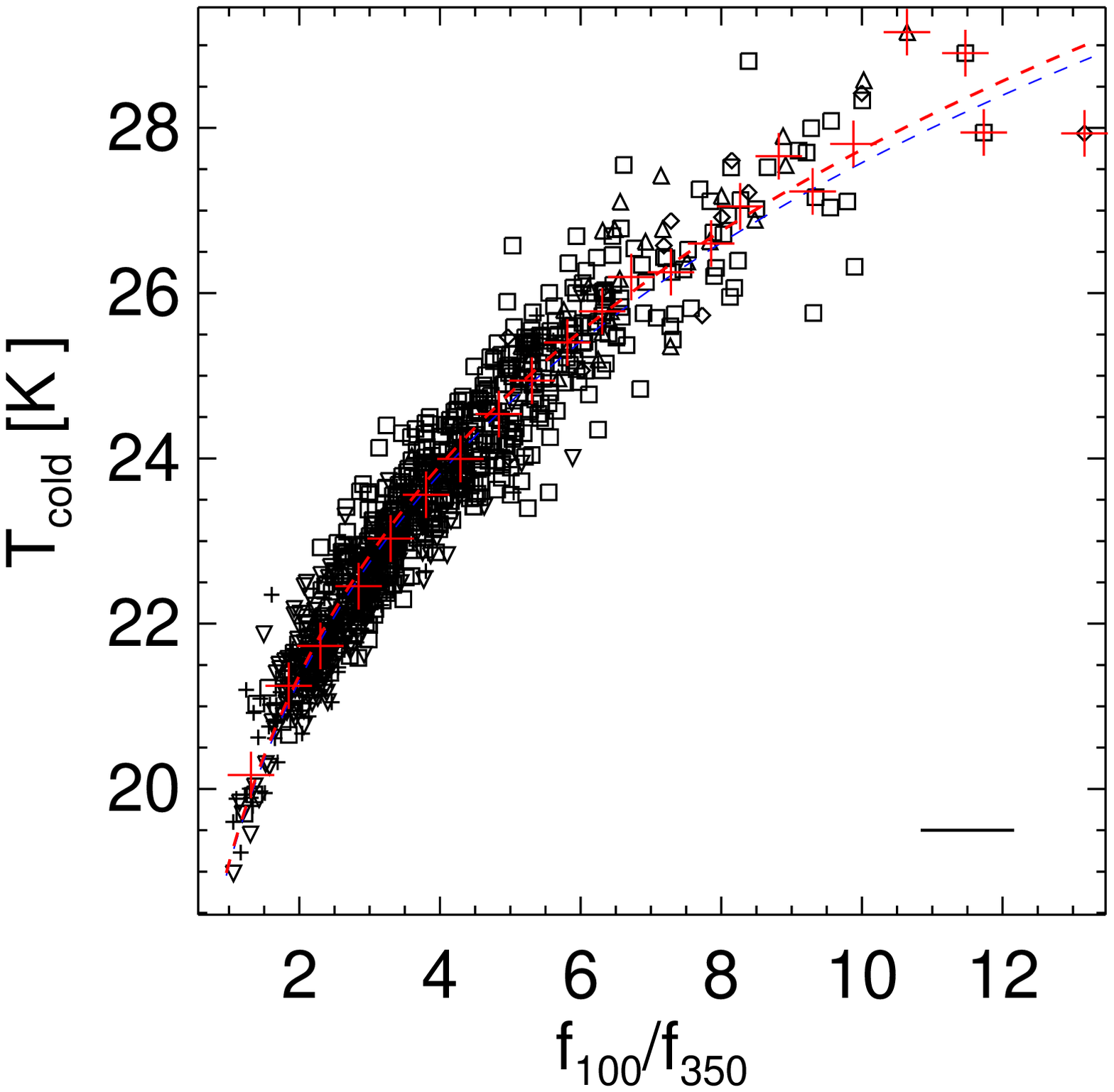}
  \caption{
  Left and middle panels: dependence of the dust mass and IR luminosity on the 350~\mum\ monochromatic luminosity. 
  Symbol colors show the temperature of the cold dust component,  and the  blue line shows the best fit.  
  The typical error is shown by the black cross.
  Equivalent figures for monochromatic luminosities at 550~\mum\ and 850~\mum\
  can be found in Appendix \ref{apn:lir-lfir} and \ref{apn:Mdust-RJtail}. 
  Right panel: the correlation between the temperature of the cold dust component and the
  100~\mum\ to 350~\mum\ flux ratio, with the best fit shown by the dashed blue line.
  Large, red crosses show the mean value of \tc\ in 0.5 mag color bins, and the red 
  dashed curve corresponds to  the best fit line to the red crosses.
  The typical error in the $x$-axis is shown by the black horizontal line.
  Similar plots relating \tc\ to IR and sub-mm colors can be found in Appendix \ref{apn:dust temp}.
  In all panels the symbols are the same as in Fig. \ref{fig:mass comp}. 
  }  
  \label{fig:l350 with lir,mdust and tdust}
  \end{center}
\end{figure*}

While the dust plane provides a powerful tool to relate the dust mass, total IR luminosity and 
dust temperature of the cold component, or derive any one parameter from the other two, 
the comprehensive dataset available for our large local
sample is difficult to obtain for other samples, especially those at high redshift. 
We thus provide several scaling relationships which can be used to estimate the location of a galaxy 
in the dust plane in the presence of limited data or, alternatively, to derive one or all parameters 
of the dust plane phase space. 

Since the dust plane is most sensitive to changes in \tc, we present several relations to estimate its value
using sub-mm to IR colors.  Previous works, e.g., \citet{soiet87,chanet07,hwaet10}, have typically
derived dust temperatures from IRAS colors: \iras\ 60/\iras\ 100. 
Here we present that based on the 100~\mum\ and 350~\mum\ color; see
Appendix \ref{apn:dust temp} for the equivalent results from other IR to sub-mm colors.
The relationship between the cold component dust temperature and the 100~\mum\ to 350~\mum\ flux ratio is
shown in the right panel of Figure \ref{fig:l350 with lir,mdust and tdust}.
The best fit to this relationship is: 
\begin{equation}
 \rm \frac{T_{cold}}{[K]}=10^{(1.280\pm0.001)}\left(\frac{\emph{f}_{100}}{\emph{f}_{350}}\right)^{(0.160\pm0.001)}
 \label{eq:tcold f350-f100}
\end{equation}

To obtain a cleaner relation, large crosses in Figure \ref{fig:l350 with lir,mdust and tdust} show the mean 
value of \tc\ in bins of 0.5 mag in the color $\rm f_{100}/f_{350}$.
Interestingly, the coefficients of the best fit (red dashed line) show that the values are consistent within the
errors with the coefficients obtained for the complete sample.

Both the dust mass and the total IR luminosity can be estimated 
from a single monochromatic luminosity in the RJ tail of the dust emission.
The dependence of \mdust\ and \lir\ on the 350~\mum\ luminosity is shown in 
the left and middle panels of Fig. \ref{fig:l350 with lir,mdust and tdust}, respectively.
At first glance, these relationships appear to have a large scatter. However, 
it is clear that this scatter can be fully explained (and removed)
by the use of the temperature of the cold dust component. Thus the estimation of \mdust\ and/or \lir\ can
be made very accurately in the presence of an estimate of the cold dust temperature (see previous paragraph)
or at least roughly in the absence of a cold dust temperature. We will address both scenarios below for
the case of using the 350~\mum\ luminosity for the RJ tail luminosity 
(see Appendix \ref{apn:lir-lfir} and \ref{apn:Mdust-RJtail}
for the results of using other sub-mm frequencies).

Using the monochromatic luminosity at 350~\mum\ in the presence of a value for \tc , we obtain two planes to
determinate \lir\ or \mdust:

\begin{equation}
\rm \log \left(\frac{L_{IR}}{[\lsun]}\right)  - 1.017~ \log \left(\frac{L_{350}}{[WHz^{-1}]}\right) 
\nonumber
\end{equation}
\begin{equation}
~~~~~ - 0.118~ \left(\frac{T_{\rm cold,~dust}}{[K]}\right) + 16.45 = 0
\nonumber
\end{equation}
\begin{equation}
 \rm \log \left(\frac{M_{dust}}{[\msun]}\right) - 0.940~ \log \left(\frac{L_{350}}{[WHz^{-1}]}\right)
\nonumber
\end{equation}
\begin{equation} 
~~~~~ + 0.0791~ \left(\frac{T_{\rm cold,~dust}}{[K]}\right) + 12.60 = 0 
\label{eq:lir-mdust with l350  and td}
\end{equation} 

In the absence of a \tc\ estimation, one has to accept the full scatter seen in 
the left and middle panels of Fig. \ref{fig:l350 with lir,mdust and tdust}, i.e.
a dispersion of 0.5 dex and 1 dex in the estimation of \mdust\  and \lir\ from
a single 350\mum\ luminosity.
The best fit to the data points (blue lines in Figure \ref{fig:l350 with lir,mdust and tdust}, 
right and middle panels) is: 
\begin{equation}
\rm \frac{L_{IR}}{[\lsun]}  =  10^{-14.388\pm0.002} \left(\frac{L_{350}}{[WHz^{-1}]}\right)^{1.046\pm0.005}
\nonumber
\end{equation}
\begin{equation}
\rm \frac{M_{dust}}{[\msun]} = 10^{-13.963\pm0.002} \left(\frac{L_{350}}{[WHz^{-1}]}\right)^{0.920\pm0.005}
\label{eq:lir-mdust with l350}
\end{equation}

Estimations of the dust mass and IR luminosity from other monochromatic 
IR or sub-mm fluxes, and estimates of the temperature of the cold dust 
component from other IR and sub-mm colors can be found in Appendix \ref{apn:dust temp},
\ref{apn:lir-lfir} and \ref{apn:Mdust-RJtail}. 

\subsection{Dust to Stellar Mass Ratio}
\label{subsec:anticorrelation}

\begin{figure}[ht]
  \begin{center}
   \includegraphics[bb=20 136 480 643,width=0.4\textwidth,clip]{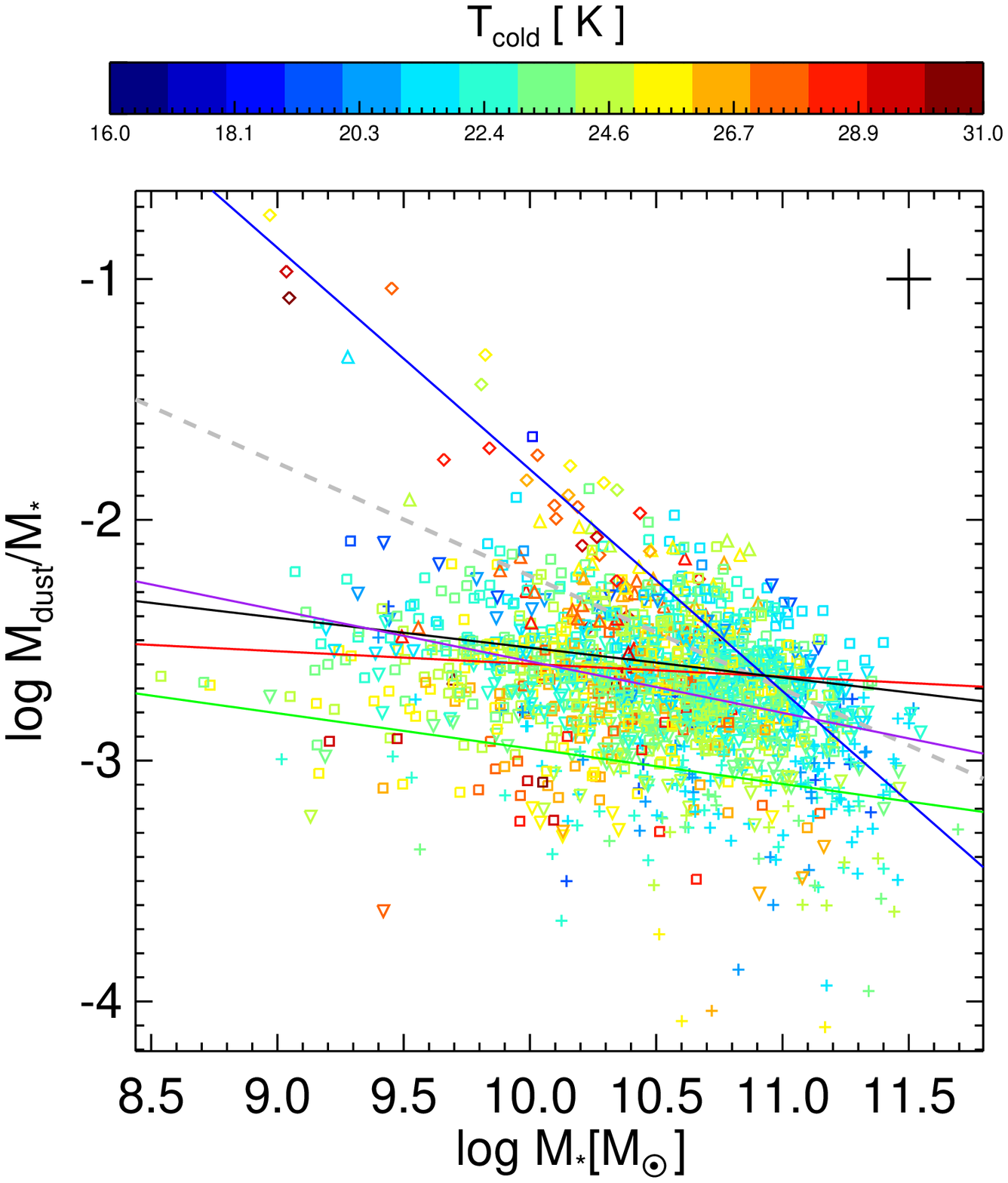}
   \includegraphics[bb=40 129 475 562,width=0.175\textwidth,clip]{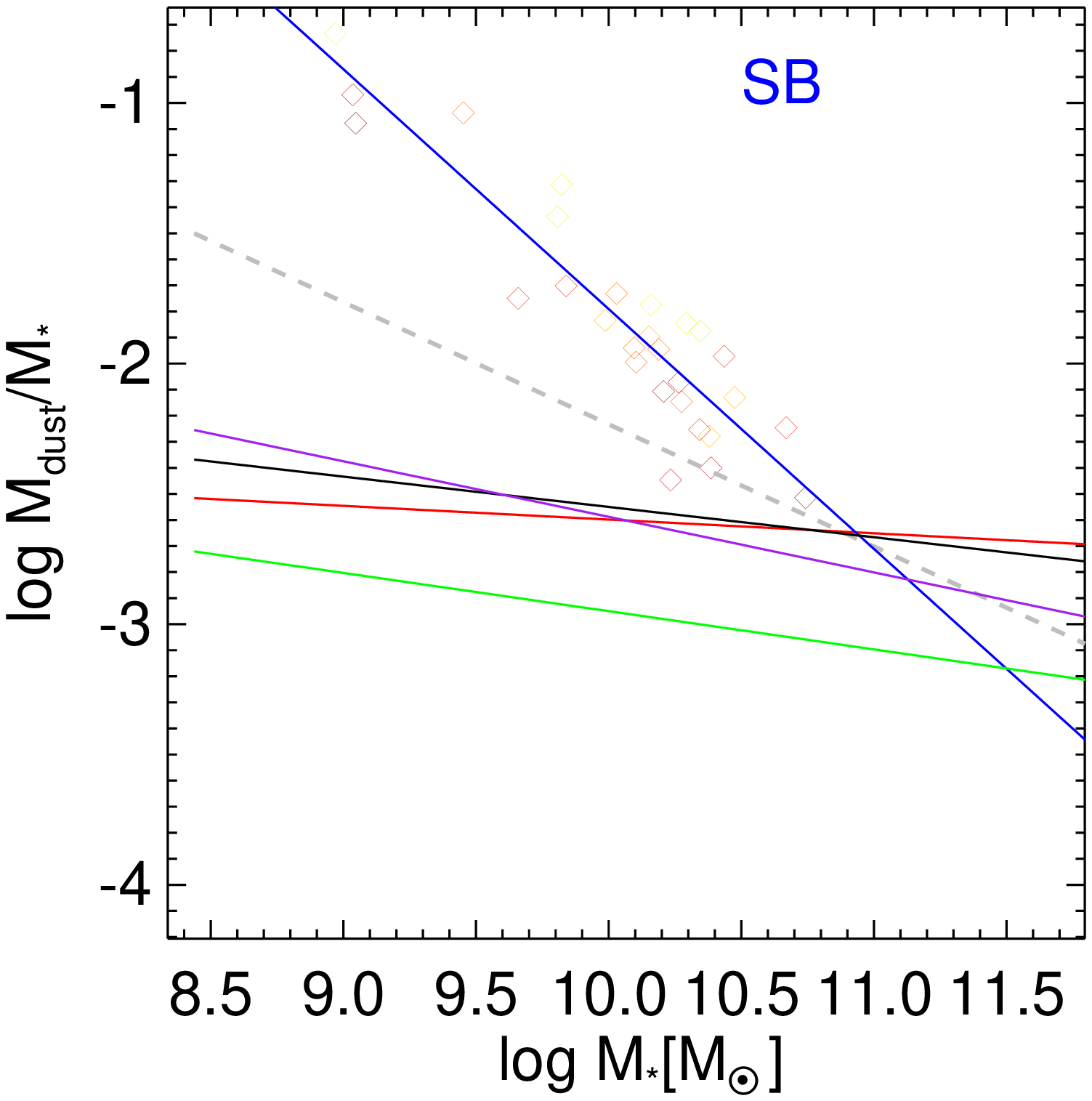}
   \includegraphics[bb=107 129 475 562,width=0.15\textwidth,clip]{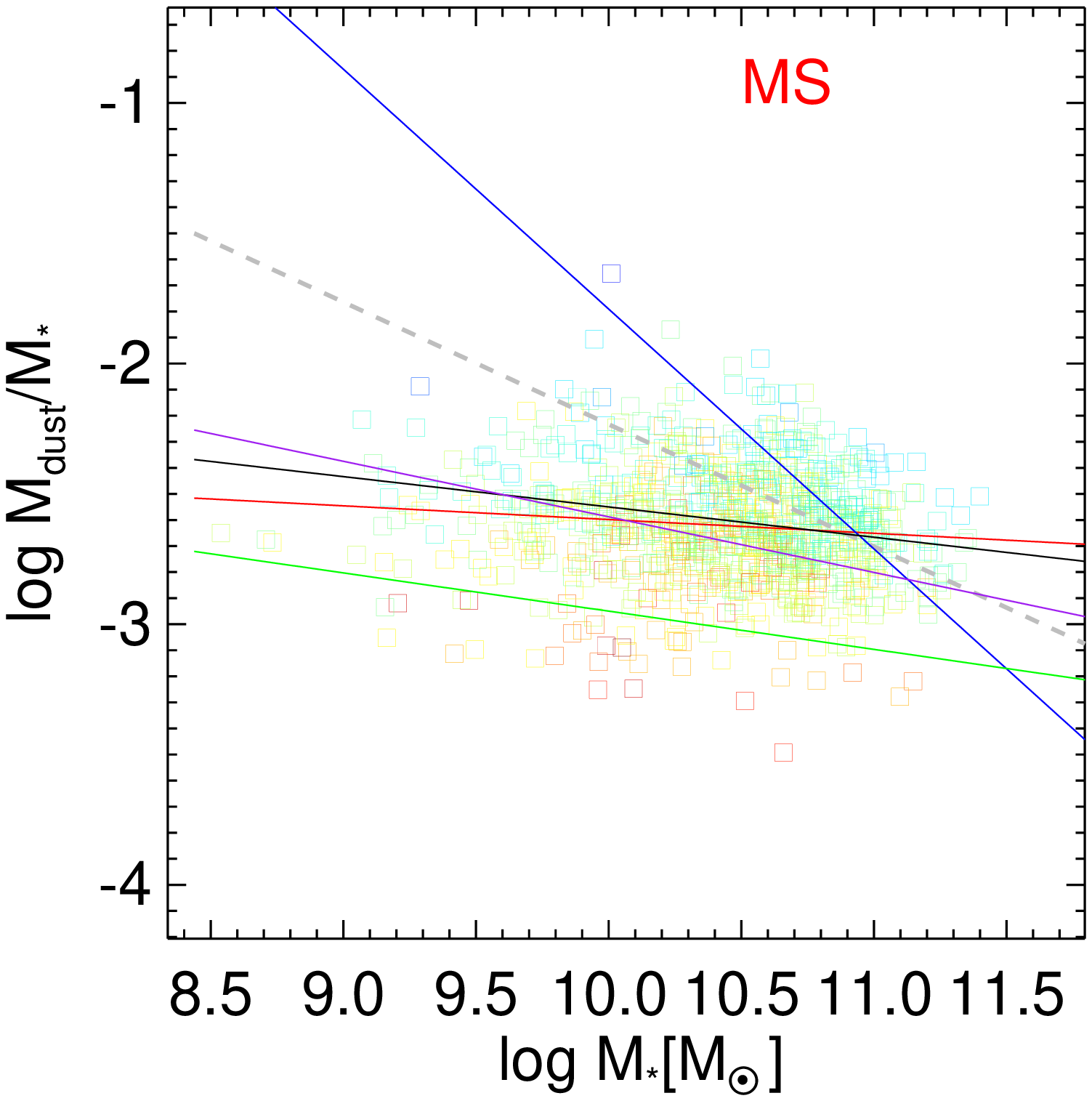}
   \includegraphics[bb=107 129 475 562,width=0.15\textwidth,clip]{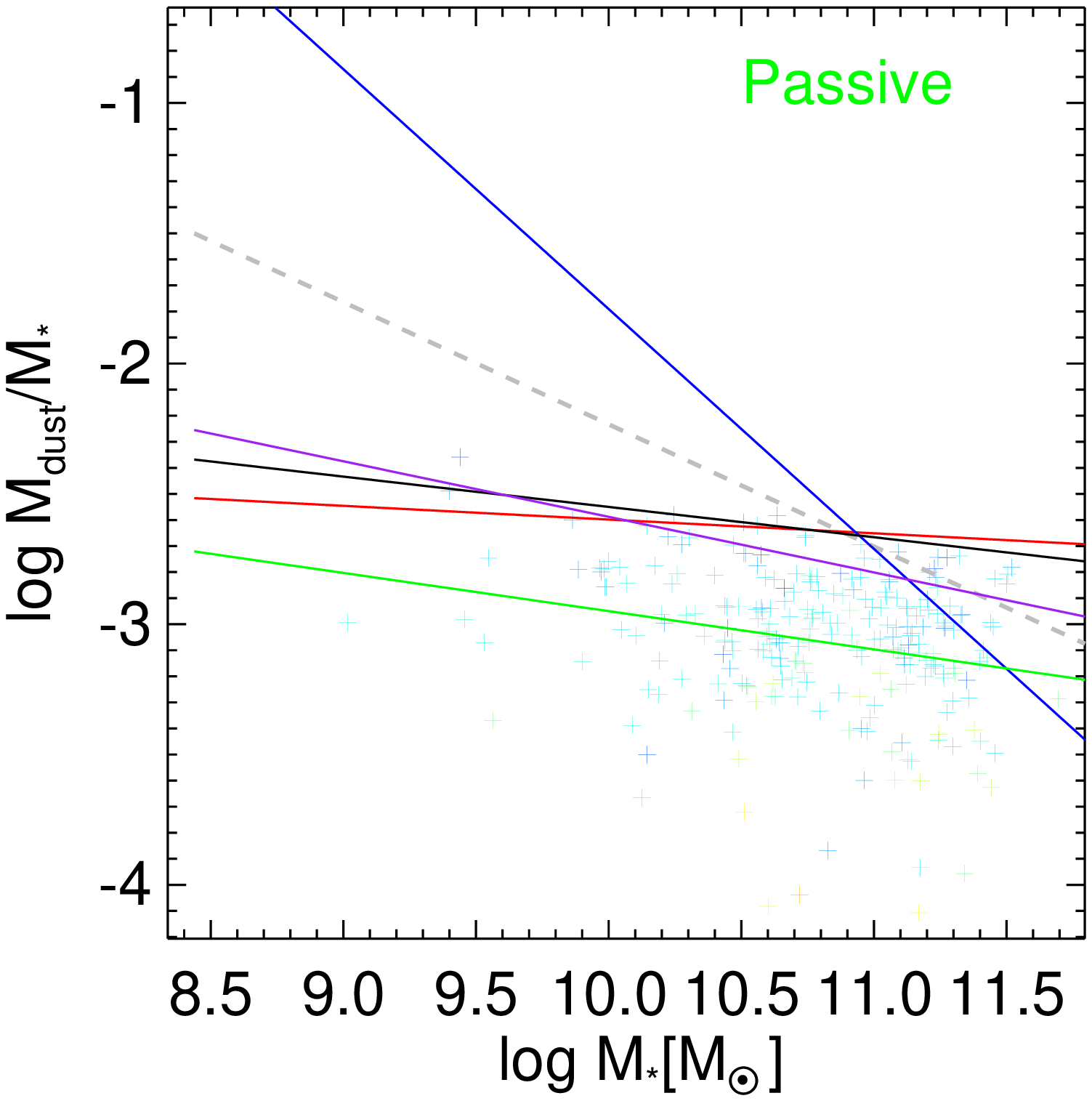}

         \caption{ Top panel: The dust to stellar mass ratio ($\rm\log \left(M_{dust}/M_{*}\right)$) 
         as a function of stellar mass. The dust mass is that obtained from the DL07 model fits.
         Symbols are the same as in Fig. \ref{fig:mass comp} and symbol colors denote the
         temperature of the cold dust component following the color bar.
         The dashed gray line shows the relation obtained by \citet{clemet10}.
         The linear fits to the data points of our sub-samples are shown in  
         blue for SB galaxies, red for MS galaxies, green for PAS galaxies, black for
         MS and SB galaxies combined, and purple for our entire sample.
         The typical error is shown by the black cross.
         Bottom panels: As in the top panel, but the panels separately plot SB (left), 
         MS (middle) and PAS galaxies (right). 
         }  
  \label{fig:mass mdust/m*}
  \end{center}
\end{figure}

 The typical $\mdust/\mstar$ ratios are 0.21\% and 0.25\% for our entire sample
and for all star-forming (non-passive) galaxies, respectively.
These ratios are smaller to those obtained by \citet{clemet13} 
(median $\mdust/\mstar= 0.46\%$ in a well defined sample of 234 nearby galaxies
detected by \planck) and \citet{claet15} (median $\mdust/\mstar= 0.44\%$ in the HAPLESS sample,
a blind sample of 42 nearby galaxies detected by \herschel).
Other three works, using the MAGPHYS code, show similar values for the dust to stellar mass ratio. 
\citet{dacet10} shows a value between 0.23\% to 0.14\% depending on the stellar mass bin used in a sample
of 1658 galaxies at z < 0.1; \citet{smiet12} obtain a dust to stellar mass of 0.22\% from a sample of 184 
galaxies at $z < 0.1$; while \citet{pappet16} show a value of 0.18\% in its main sample.

An anticorrelation between the \mdust/\mstar\ and \mstar\ has been
shown by \citet[][; in Virgo cluster galaxies]{coret12}, \citet{clemet13} 
and \citet{claet15}.

We see the same anticorrelation in our sample (Figure \ref{fig:mass mdust/m*}), 
with the points showing a large dispersion. The fits to our full sample (FS) 
(purple line) and to all our star-forming (SF) galaxies (non-passive galaxies; black line), are:

\begin{equation}
\rm\frac{M_{dust}}{M_*} = 10^{-0.4\pm0.2} \left(\frac{M_*}{[\msun]}\right)^{-0.28\pm0.02}~~ \rm FS
\nonumber
\end{equation}
\begin{equation}
\rm\frac{M_{dust}}{M_*} = 10^{-1.3\pm0.2} \left(\frac{M_*}{[\msun]}\right)^{-0.12\pm0.02}~~ \rm SF
\end{equation}

When we separate the galaxies by star-forming mode (bottom panels of
figure \ref{fig:mass mdust/m*}), we find two interesting results: 
(a) SB galaxies show a much steeper anticorrelation than the other sub-samples, and 
(b) the source of the scatter in the anticorrelation can be traced by the  temperature of the cold
dust component, \tc\ (or the weighted dust temperature $\rm T_{weight}$). 
The anticorrelations obtained individually for each subsample are: 
\begin{equation}
\rm\frac{M_{dust}}{M_*}= 10^{7.4\pm0.8} \left(\frac{M_*}{[\msun]}\right)^{-0.92\pm0.08} ~~ SB
\nonumber
\end{equation}
\begin{equation}
\rm\frac{M_{dust}}{M_*}=  10^{-2.1\pm0.2}\left(\frac{M_*}{[\msun]}\right)^{-0.09\pm0.01} ~~ MS
\nonumber
\end{equation}
\begin{equation}
\rm\frac{M_{dust}}{M_*}= 10^{-1.2\pm0.4}\left(\frac{M_*}{[\msun]}\right)^{-0.17\pm0.04} ~~ PAS
\end{equation}

Essentially, the dust to stellar mass ratio in SB galaxies (typically 1.3\%) 
is higher than that in MS galaxies (typically 0.24\%), though the
difference is smaller for galaxies at the highest stellar masses. 
Passive galaxies show a very low dust to stellar mass ratio, with a typical value of 
$\mdust/\mstar=0.1\%$.  Additionally, for all sSFR groups, at a given stellar mass, 
the galaxies with the highest \tc\ ($\rm T_{weight}$) have the
smaller dust to stellar mass ratios.

\subsection{Dust to Gas Mass Ratio}
\label{subsec:relation mdust mism}

\begin{figure*}[ht]
  \begin{center}
   \includegraphics[bb=32 136 475 566,width=0.33\textwidth,clip]{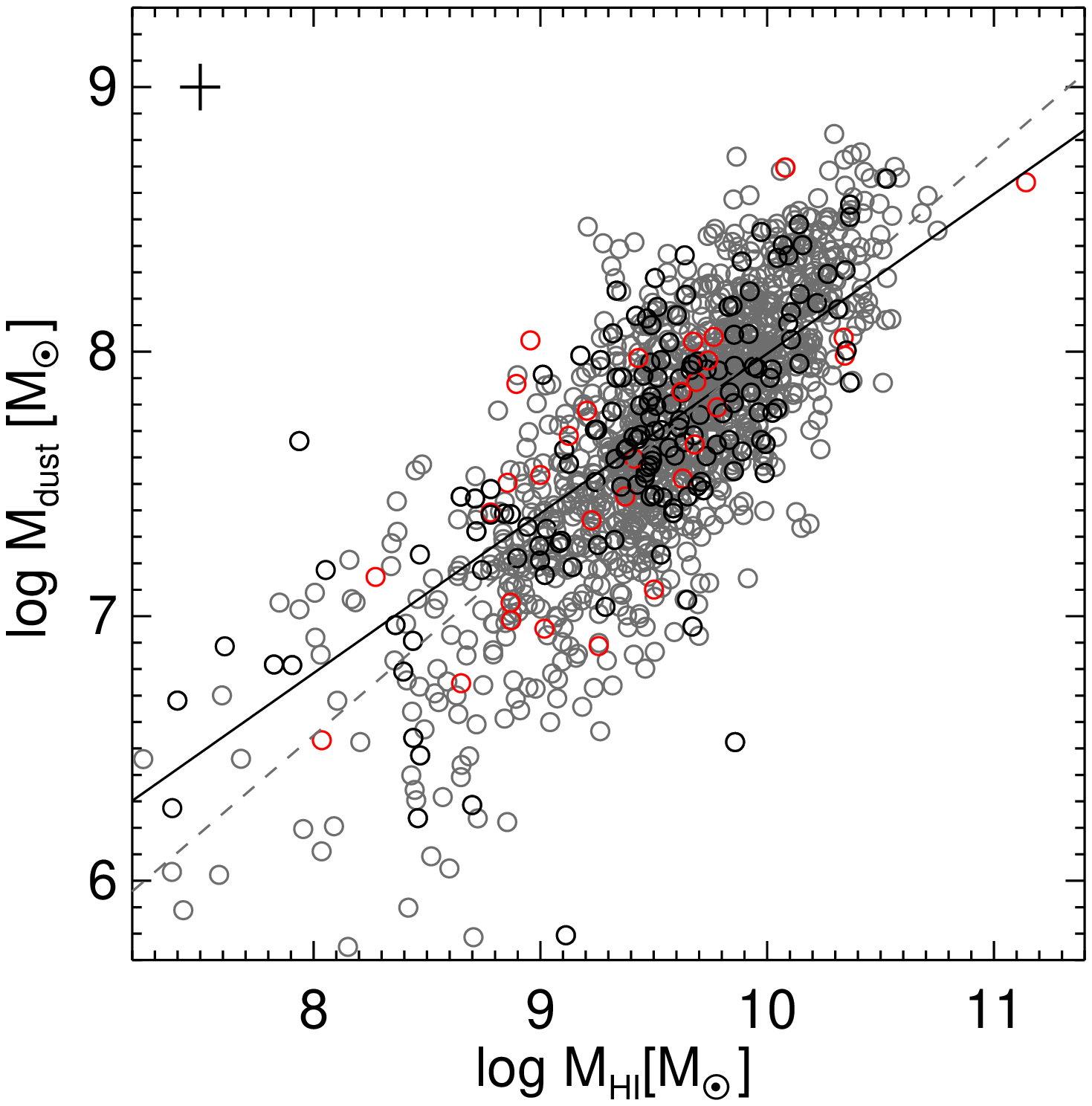}
   \includegraphics[bb=32 136 475 566,width=0.33\textwidth,clip]{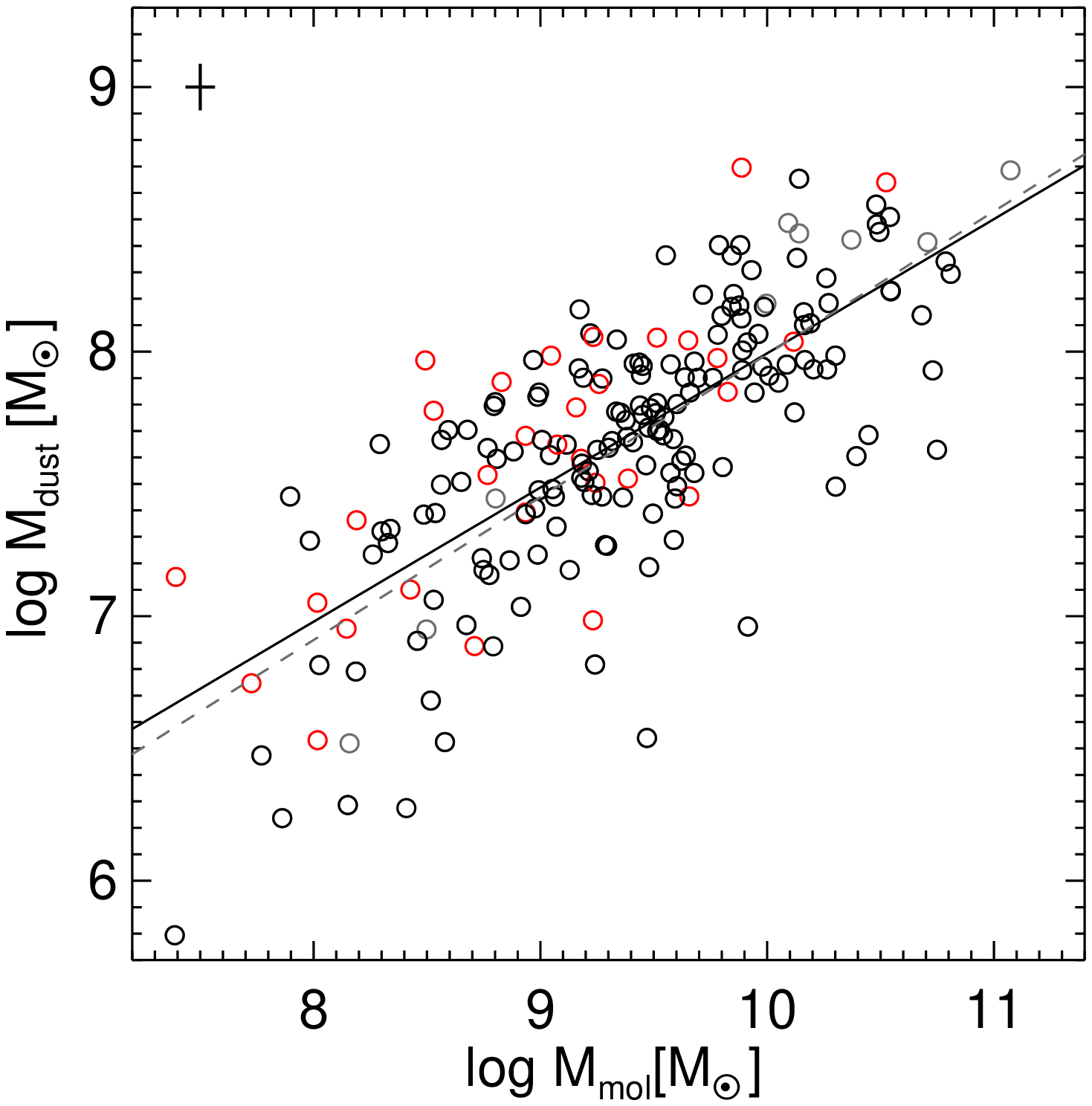}
   \includegraphics[bb=32 136 480 566,width=0.33\textwidth,clip]{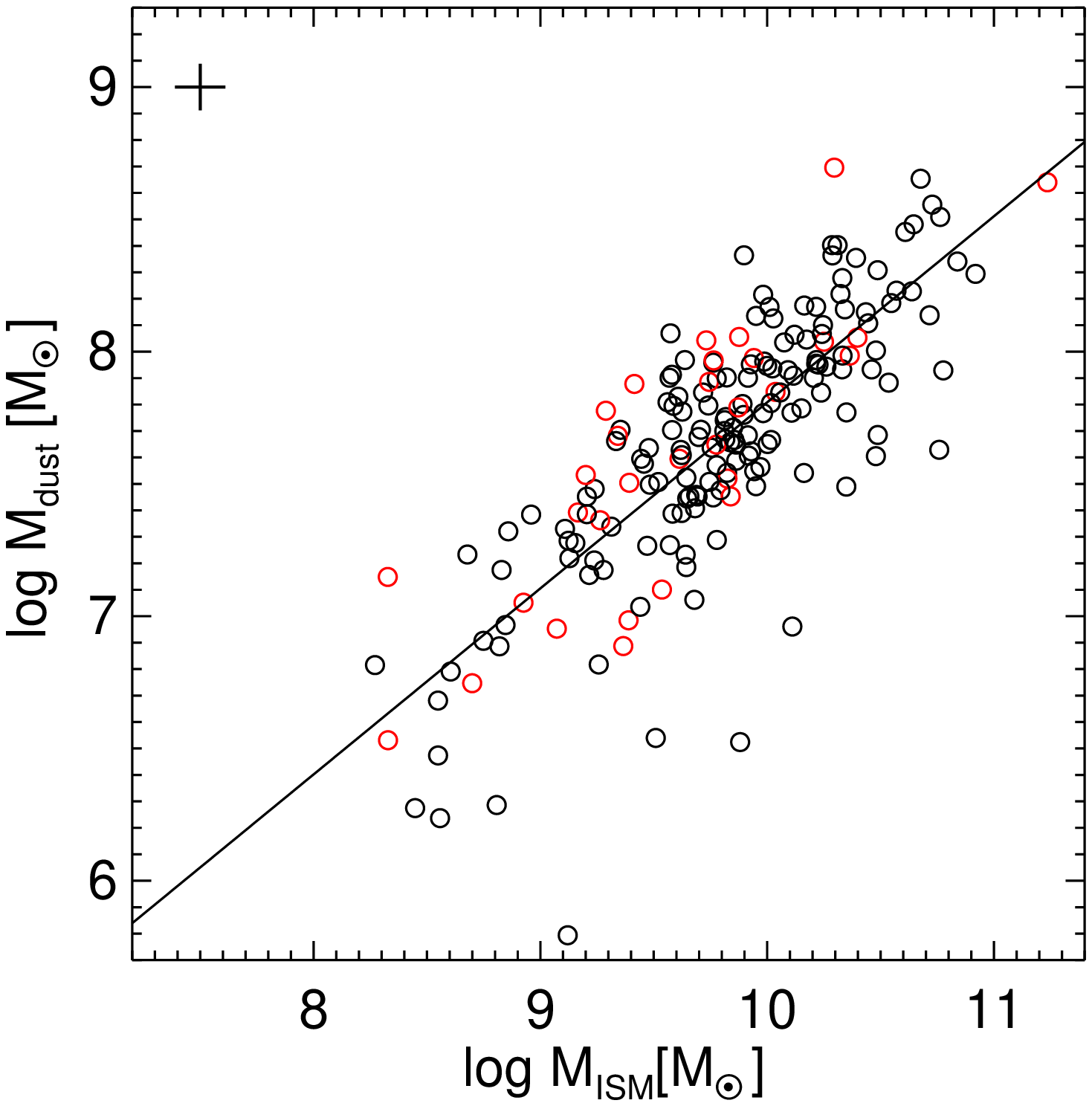}
         \caption{The relationships between dust mass and atomic gas mass (left), molecular gas mass 
         (middle) and total ISM mass (right). Black symbols are used for galaxies 
         with HI and CO observations and dust mass from DL07 models; 
         Red symbols are used for galaxies with HI and CO observations, but with 
         dust masses estimated from the luminosity at 350, 550 or 850~\mum\ (see Eq. 
         \ref{eq:lir-mdust with l350} and Appendix \ref{apn:Mdust-RJtail});
         Gray symbols denote galaxies with either HI or CO observations and dust mass from DL07 models. 
         The solid black lines in each panel show the best fit to all sources with 
          measurements of both $\rm M_{HI}$ and $\rm M_{mol}$,  and the gray dashed lines in the two 
          left panels show the fit to all points in the respective panel. 
          The typical error is shown by the black cross in each panel.
          }  
  \label{fig:mass mdust-mism}
  \end{center}
\end{figure*}

To more easily study the relation between dust and gas masses, we separate 
the total gas mass (hereafter referred to as interstellar medium mass or \mism) 
into two components: the molecular gas mass ($\rm M_{mol}$), and the atomic 
(Hydrogen) gas mass ($\rm M_{HI}$): $\rm M_{ISM}=M_{mol}+M_{HI}$.
Figure \ref{fig:mass mdust-mism} shows the relation between the dust mass 
and the atomic (left panel), molecular (middle panel) and total (right panel) gas masses.

The median gas fraction ($\rm M_{ISM}/(M_{*}+M_{ISM})$) for the full sample 
is $33\pm11$\%. The medians for MS, PAS, and SB galaxies are $32\pm8$\%, $5.0\pm0.8$\% and 
$68\pm25$\%, respectively.
For all sources with HI and CO measurements, the molecular to atomic 
gas mass ratios ($\rm M_{mol}/M_{H_I}$) show a median value of $0.8\pm0.2$.
 Even for MS and PAS galaxies is observed the same value, however SB galaxies show a huge value ($4\pm1$). 
(see Table \ref{tab:mas gas ratio Hi and H2}). 

\begin{table}[ht]
\centering
\caption{Sources with measurements in both $\rm M_{HI}$ and $\rm M_{mol}$}
\setlength{\tabcolsep}{4pt}
\begin{tabular}{c|ccccc}
              & \small{$\rm M_{mol}$/$\rm M_{HI}$} & \small{\mdust/$\rm M_{HI}$} & \small{\mdust/$\rm M_{mol}$} & \small{\mdust/$\rm M_{ISM}$} & \footnotesize{Galaxies}  \\
              &                            &  \%                 &  \%                  & \%                   & \#   \\                          
\hline
\small{sample}       & $0.8\pm0.2$         & $1.6\pm0.4$         &$2.1\pm0.3$          &$0.8\pm0.1$            & 189       \\
\small{MS}           & $0.8\pm0.2$         & $1.4\pm0.3$         &$1.8\pm0.3$          &$0.7\pm0.1$            &  82       \\ 
\small{SB}           & $4\pm1$             & $3.4\pm0.9$         &$0.9\pm0.2$          &$0.6\pm0.1$            & 8        \\
\small{PAS}          & $0.8\pm0.1$         & $2.1\pm0.1$         &$2.8\pm0.5$         &$1.4\pm0.2$            &  17 
\end{tabular}
\label{tab:mas gas ratio Hi and H2}
\end{table}

 \begin{table}[ht]
\centering
\caption{Sources with measurements in $\rm M_{HI}$ or $\rm M_{mol}$}
\begin{tabular}{c|cccc}
              & \mdust/$\rm M_{HI}$ & \# Sources         & \mdust/$\rm M_{mol}$ & \# Sources           \\
              &  \%                 & with $\rm M_{HI}$  &  \%                  &  with  $\rm M_{mol}$  \\  
\hline
sample        & $1.3\pm0.3$         &  1297              & $2.0\pm0.3$          &     189               \\
MS            & $1.2\pm0.3$         &   700              &$2.0\pm0.3$           &      84                \\
SB            & $2.0\pm0.7$         &   14               &$0.9\pm0.1$           &      10                \\
PAS           & $1.8\pm0.4$         &   207              &$2.7\pm0.1$           &      18         
\end{tabular}
\label{tab:mas gas ratio Hi or H2}
\end{table}

The median values of $\rm M_{dust}/M_{HI}$ are 1.6 \% and 1.3\% for the galaxies with 
both $\rm M_{HI}$ and $\rm M_{mol}$ and for those with only $\rm M_{HI}$ measurements, 
respectively (see tables  \ref{tab:mas gas ratio Hi and H2} and \ref{tab:mas gas ratio Hi or H2}). 
These values are smaller than those obtained by 
\citet[ 2.2\%]{clemet10} and \citet[ 3.9\%]{claet15}. 
PAS galaxies show the highest dust to atomic mass ratios, followed by SBs and 
finally MS galaxies. 
The median dust to molecular gas mass ratio is 2.1\% for  all galaxies with both $\rm M_{HI}$ and $\rm M_{mol}$,
and  2.0\% for all galaxies with only $\rm M_{mol}$. Passive galaxies show the largest dust to molecular gas
mass ratios followed by MS galaxies, with SB galaxies showing the lowest values. 
The median dust to ISM mass ratios are similar for the full sample, MS and SB
galaxies ($0.8\%$, $0.7\%$ and $0.6\%$, respectively), while PAS galaxies show larger values ($1.4\%$).

Figure \ref{fig:mass mdust-mism} shows a clear correlation between 
dust mass and molecular and atomic gas mass, individually. 
But the correlation between dust and \mism\ shows the smallest dispersion, with
the best fit: 
\begin{equation}
 \rm \frac{M_{ISM}}{[\rm M_{\sun}]}=6.03 \left(\frac{M_{dust}} {[\rm M_{\sun}]}\right)^{0.7}
 \label{eq:mdust-mism fit}
\end{equation}

This correlation between ISM and dust mass has been noted previously by \citet[in five Local Group galaxies]{leret11}
and \citet[in 35 metal-rich Virgo spirals]{corbet12}.

\begin{figure}[!pht]
  \begin{center}
    \includegraphics[bb=38 145 478 643,width=0.4\textwidth,clip]{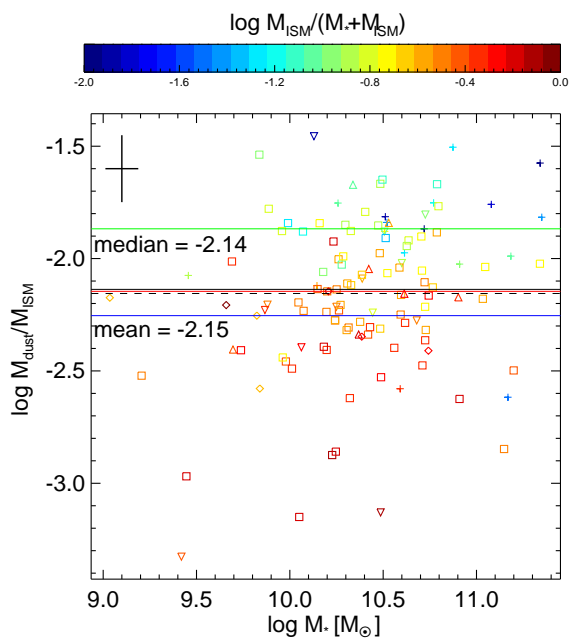}
         \caption{ The dust to ISM mass ratio as a function of stellar mass. Symbols
         are the same as in Fig. \ref{fig:mass comp} and  colors represent the gas fraction.
         The black solid and dashed lines show the median and mean dust to gas ratios, respectively, for the 
         entire sample.  Colored solid lines show median values 
         for SB (blue), MS (red) and PAS (green) galaxies.  The black cross shows the typical error.
          }  
  \label{fig:DGR vs Mstar}
  \end{center}
\end{figure}

The dust to gas ratio ($\rm \delta_{DGR}=M_{dust}/M_{ISM}$) does not systematically vary with the stellar mass 
for our sample galaxies (Fig. \ref{fig:DGR vs Mstar}), despite the expected metallicity range (between 8.4 and 9.1) 
for this stellar mass range ($10^9<$\mstar[\msun]$<10^{11}$) \citep{treet04}.
Our full sample shows a large scatter in $\rm \delta_{DGR}$ ($\pm$0.5 dex), but the scatter significantly
decreases when considering SB, MS, and PAS galaxies separately. 
MS and PAS galaxies show a dispersion in $\rm \delta_{DGR}$ of $\pm$0.3 dex, and SBs show a dispersion
of only $\pm$0.2 dex. The median values of $\rm \delta_{DGR}$ for MS and SB galaxies 
(red and blue lines in Fig. \ref{fig:DGR vs Mstar}, respectively)
are very similar to the median value of the full sample;
however, PAS galaxies show a higher median $\rm \delta_{DGR}$ (green line in Fig. \ref{fig:DGR vs Mstar}).
Additionally, the symbol colors show that the galaxies with higher gas fractions
have lower values of $\rm \delta_{DGR}$, especially MS and PAS galaxies.

\section{Discussion }
\label{sec:discussion}

We have modeled the dust emission using detailed DL07 dust models in an unprecedentedly large
sample of 1,630 nearby ($z<0.1$, $\rm < z >=0.015$) galaxies with uniform photometric data
from   \wise\ (3.4 to 22~\mum), \iras\ (12 to 100~\mum), \planck\ (350 to 850~\mum) 
and/or SCUBA (850~\mum).
This sample covers a significant parameter space in stellar mass and SFR, and thus sSFR, going from starburst 
(SB; LIRGs;  $\lir>10^{11}\lsun$) to passive (PAS; $\lir\sim10^{8.3}\lsun$) galaxies.

In comparison, previous studies that present detailed dust models to galaxies with sub-mm data
include the KINGFISH/SINGS sample \citep{dalet12}
of 65 nearby, \emph{normal} ($\lir<10^{11}\lsun$) spiral galaxies with photometry from
\iras, \spitzer, \herschel, and SCUBA, 
and the ERCSC sample \citet{neget13,clemet13} of 234 local  ($z<0.1$) galaxies with
photometry from \iras, \wise, \spitzer, \herschel\ and \planck.
The ERCSC sample was created by matching four \herschel\ samples,  
HRS, HeViCS, KINGFISH and H-ATLAS, to the \planck\ Early Release Compact Source Catalogue.

Note that the other large nearby sample of galaxies with dust modeling results, the
sample of 1,658 galaxies in \citet{dacet10}, used MAGPHYS modeling applied to UV, optical,
and limited IR data (2MRS, \iras): the lack of sub-mm data is expected to lead to less reliable
dust mass estimates \citep[e.g.,][]{draet07}.

The advantages of our sample over the ERCSC sample include:
a) a sample size $\sim7$ times larger with equivalent and consistent data observations; 
b) the use of the Second \planck\ catalog of Compact Sources, which contains only reliable detections
(in this release the \planck\ collaboration separated the less reliable detections into a
separate 'excluded' catalog which we do not use);
c) the fact that the catalogs of the \planck\ second release are deeper than those of the Early Release;
d) the use of flux corrections to the \planck\ catalog fluxes \citep{naget15} to ensure consistency
   with the SCUBA flux scale.  

We remind the reader that all galaxies fitted with dust models had between 
four and seven photometric points distributed between 12\mum\ and 850\mum.
In all cases these points populated both sides of the SED peak 
with $\geq$2 photometric points shortward of $\lambda\leq 100\mum\ $ and 
$\geq$ one photometric point at $\lambda\geq 350\mum\ $.
A further indication that the available number of photometric points sufficiently
constrained the fitted SED is that all templates matched to the photometry  
with $\chi_{r}^2< \hbox{min}\left(\chi^{2}_{r}\right)+1$ have similar spectral 
shapes and parameters (such as \mdust\ or \lir), 
despite the extremely large number of templates (24,200) in our work. 

The differences in the dust masses obtained by us and those obtained by 
\citet[][; who used similar modeling but more extensive photometric data at $\lambda\leq 100\mum\ $]{draet07} 
in 17 KINGFISH/SINGS galaxies, ranges between $-$0.3 dex and 0.4~dex, with a median of $-$0.07~dex. 
The equivalent differences in the IR luminosities obtained by us and those obtained
by \citet{draet07}  range between $-$0.017~dex and 0.5~dex with a median of 0.14~dex.
The largest discrepancies are seen in galaxies (e.g., NGC~6946) for which \citet{draet07} did
not use sub-mm data. 
More recent results on the dust modeling of the KINGFISH/SINGS sample, which includes
sub-mm fluxes for all 64 galaxies of the sample has been done in \citet{dalet12}, 
but the dust masses and IR luminosities obtained from the model fits were not explicitly listed. 
Additionally, the values of \lir\ and \lfir\ calculated from the dust model fits are 
in overall agreement with those estimated from \iras\ fluxes using the \citet{sanet91} equations.
In fact we use our SED-derived \lir\ and \lfir\ values to recalibrate the equations to estimate
these values from only \iras\ photometry (Appendix \ref{apn:lir-lfir}).

 In general, the advantage of our samples, compared to each sample which makes it up 
(listed in table \ref{tab:papers Neil}), is that now we have more than 
one sub-mm data (usually this samples contain only one SCUBA data at 850 \mum, e.g. the SLUGS sample), 
for sources with high sub-mm emission (flux at 850\mum\ $>$ 1 Jy). Even, many of these samples 
have sub-mm information (e.g., 2MRS, ATLAS-3D, FCRAO, \citet{sanet91}, etc.), for first time. 
This means that for the first time, we have a large amount of nearby galaxies (more than ten times larger 
than previous works) with sub-mm information. Additionally, as a consequence of the all-sky observation 
by \planck, our sample reduced its selection effects only to the cutoffs of \planck\  (its reliable zones), 
without selection effects for a specific kind of galaxy, e.g., the KINGFISH/SINGS sample only 
select star forming galaxies.

\subsection{Interpreting the dust temperature}
\label{sectdis-a}

We can interpret the dust temperature of the cold component (\tc)
as the equilibrium temperature of the diffuse ISM. This
interpretation is supported by the strong correlation between \tc\ and the 
starlight intensity parameter of the diffuse ISM ($\rm U_{min}$) in DL07. 
This correlation has been demonstrated by \citet[in a sample of two resolved galaxies]{aniet12} 
and used by \citet[in the HRS sample]{cieet14} to compare their SED fitting results with 
the results of other works. 
The problem with this interpretation is that the 
temperature of the diffuse ISM (thus \tc) is not expected to correlate with the
SFR (produced in PDRs rather than the diffuse ISM). 
However, the dust plane shown here implies a close 
connection between \lir, \mdust\ and \tc\ (see Section \ref{subsec:fundamentalplane}). 
Furthermore, the SFR - \mstar\ relation also shows 
a dependence on \tc: \tc\ increases as one moves from sources with higher \mstar\ and lower SFR 
to sources with lower \mstar\ and higher SFR. To solve this contradiction, 
\citet{clemet13} propose that the lower dust masses in star-forming regions allow
the escape of UV photons thus increasing the starlight intensity impinging on the diffuse
ISM, which in turn would increase its temperature. 

In agreement with the \citet{clemet13} results, we find that \lir, \mdust\ and other parameters 
do not correlate with the warm dust component temperature (\tw). 
The parameters $\rm U_{min}$ and $\gamma$ (the percentage of dust mass PDRs) 
show only a weak correlation with \tw. 
Nevertheless, the importance of using a warm component is that
when it is used the cold component temperature (\tc) (rather than the temperature obtained from
a single component fit) shows much cleaner correlations with \lir, \mdust, and the dust plane, and
also better explains the systematics in, e.g., the \mdust/\mstar\ vs. \mstar\ anticorrelation. 
Further, \tc, rather than, e.g., T$_{\rm weight}$, is more cleanly correlated with the IR to sub-mm colors.

\begin{figure}[!pht]
  \begin{center}
   \includegraphics[bb=35 129 481 647,width=0.35\textwidth,clip]{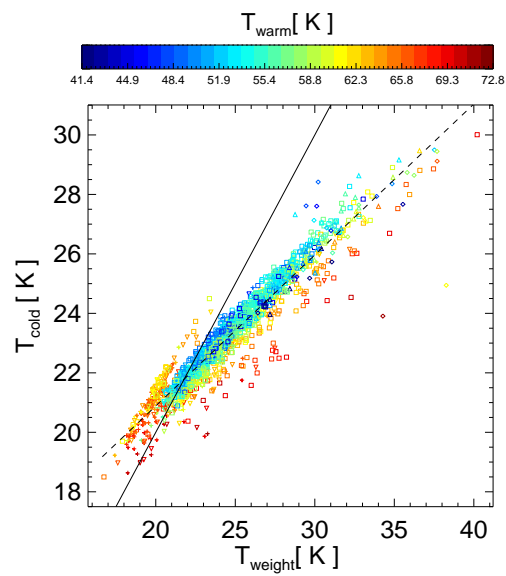}
         \caption{ A comparison of the temperature of the cold dust component (\tc; from the two component fit)
         and the luminosity weighted temperature (T$_{\rm weight}$) for our sample.
         The different symbols show SBs (diamonds), 
         MS (squares), PAS (crosses), intermediate SB (upward triangles) 
         and intermediate PAS (downward triangles) galaxies. 
         Symbols are color coded by the temperature of the warm  dust component following the color bar.
         The dashed and solid lines show the best fit and the line of equality, respectively. 
         } 
  \label{fig:tweight}
  \end{center}
\end{figure}

 It is useful to compare the luminosity weighted dust temperature ($\rm T_{weight}$) 
 with the temperatures obtained from the two component fits, since $\rm T_{weight}$ 
 is independent of our assumptions made in the two component fits. 
 Figure \ref{fig:tweight} shows the relationship between $\rm T_{weight}$ 
 and the \tc: the scatter in the data points is primarily correlated with \tw. 
 The tightest correlation is found for MS and PAS galaxies with \tw\ $\leq$ 58~K. 
 SBs and Intermediate SB galaxies (typically at \tw\ $\geq$ 58~K) show a larger scatter. 
 A cleaner correlation is observed if we compare $\rm T_{weight}$ with a temperature 
 obtained from a weighted addition of \tc\ and \tw, where the weights are the respective 
 flux fractions (e.g. $\rm S_{c}/(S_w+S_c)$). 
 That is, the increase in the scatter seen in Figure \ref{fig:tweight} for \tw\ $\geq$ 58~K 
 is a consequence of the increment in the fractional flux of the warm component. 
 The use of $\rm T_{weight}$ instead of \tc\ in all the relevant relations presented 
 in Sect.~\ref{sec:results} produces similar results but with a larger scatter. 
 Furthermore, IR and sub-mm colors (see appendix \ref{apn:dust temp}) can be used to estimate \tc\
 with a smaller scatter than $\rm T_{weight}$.

We obtain a better estimation of the dust temperature of the cold component using a color. 
This color is based on a IR flux ($\lambda<100~\mum$) and a sub-mm ($\lambda>350~\mum$) flux.
These estimates show a stronger relation than compared to using only two IR fluxes or only two sub-mm fluxes 
(e.g. the color 60 to 100~\mum\ ). 
The lowest end of our wavelength range (12 and 22~\mum) shows the least reliable estimates of 
\tc\ using an IR and sub-mm color. 
This is a consequence of the warm component influence within these fluxes. 
However, the $\rm T_{weight}$ shows a tight correlation with the color obtained using 
the flux at 22~\mum\ and one sub-mm flux, but not with the color at 12~\mum. This is a 
consequence to its proximity to the PAHs emission.
 
\subsection{Dust mass as a tracer of total gas mass}

\begin{figure*}[!pht]
  \begin{center}
   \includegraphics[bb=28 129 481 658,width=0.30\textwidth,clip]{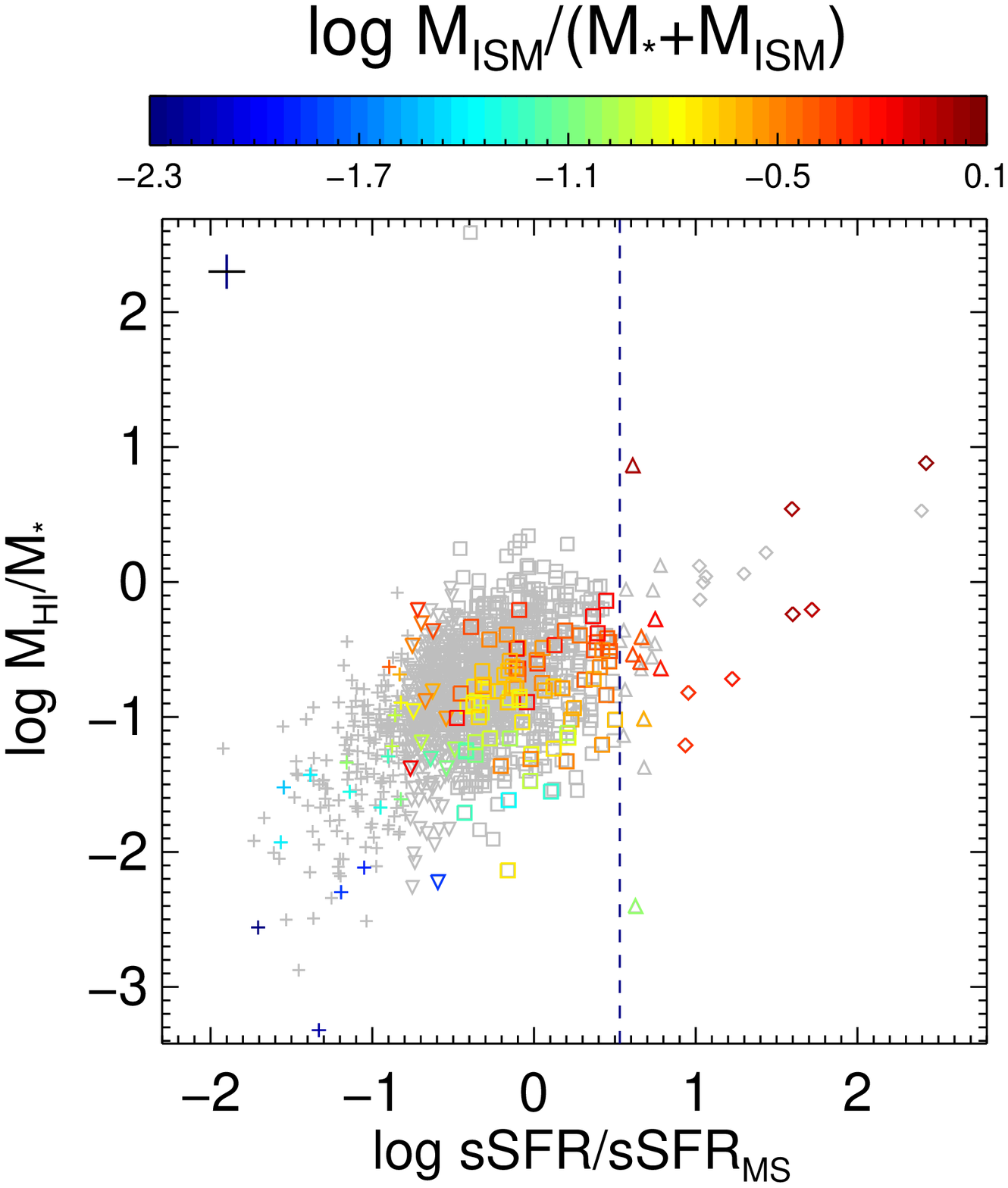}
   \includegraphics[bb=28 129 481 658,width=0.30\textwidth,clip]{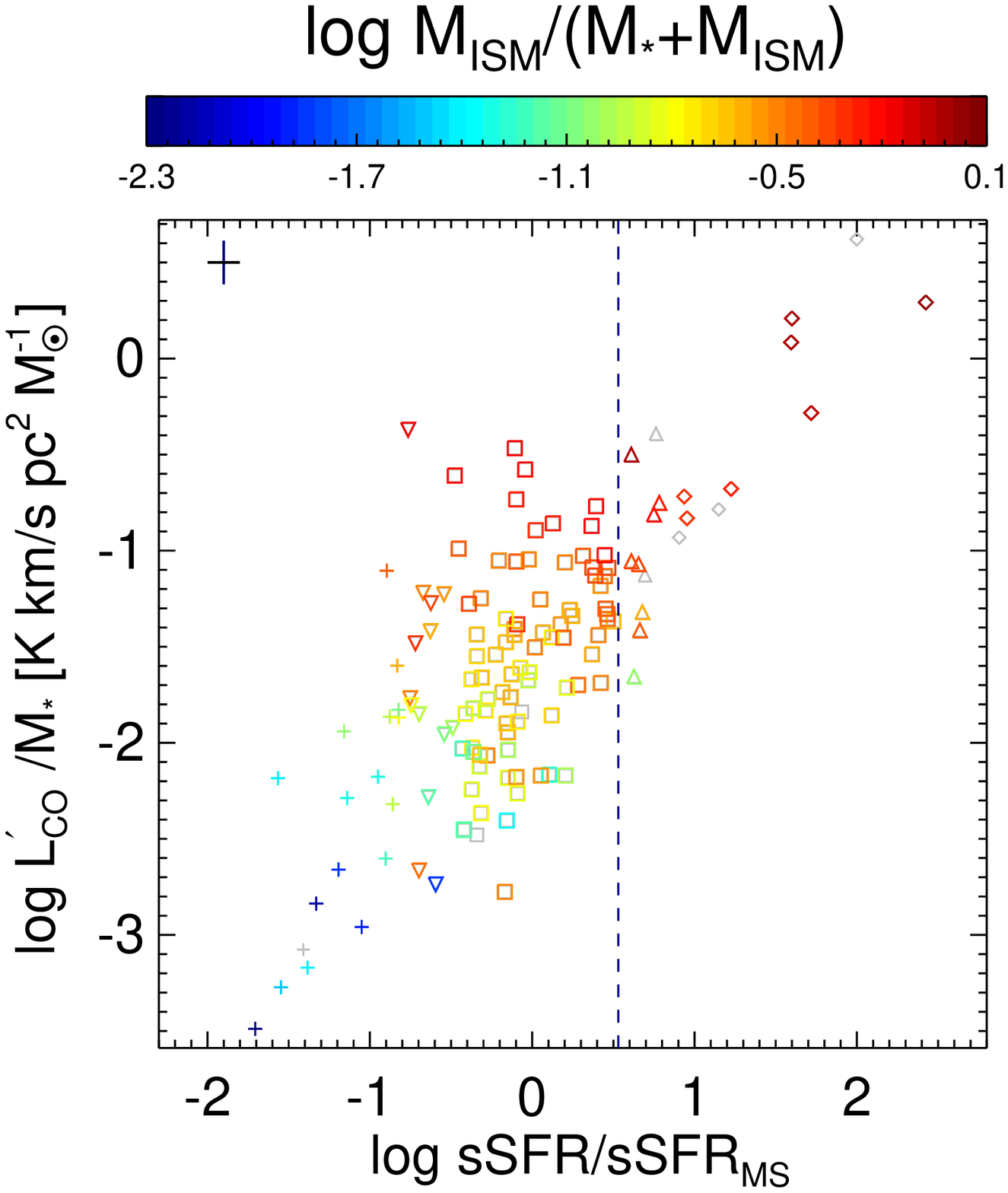}
   \includegraphics[bb=28 129 481 658,width=0.30\textwidth,clip]{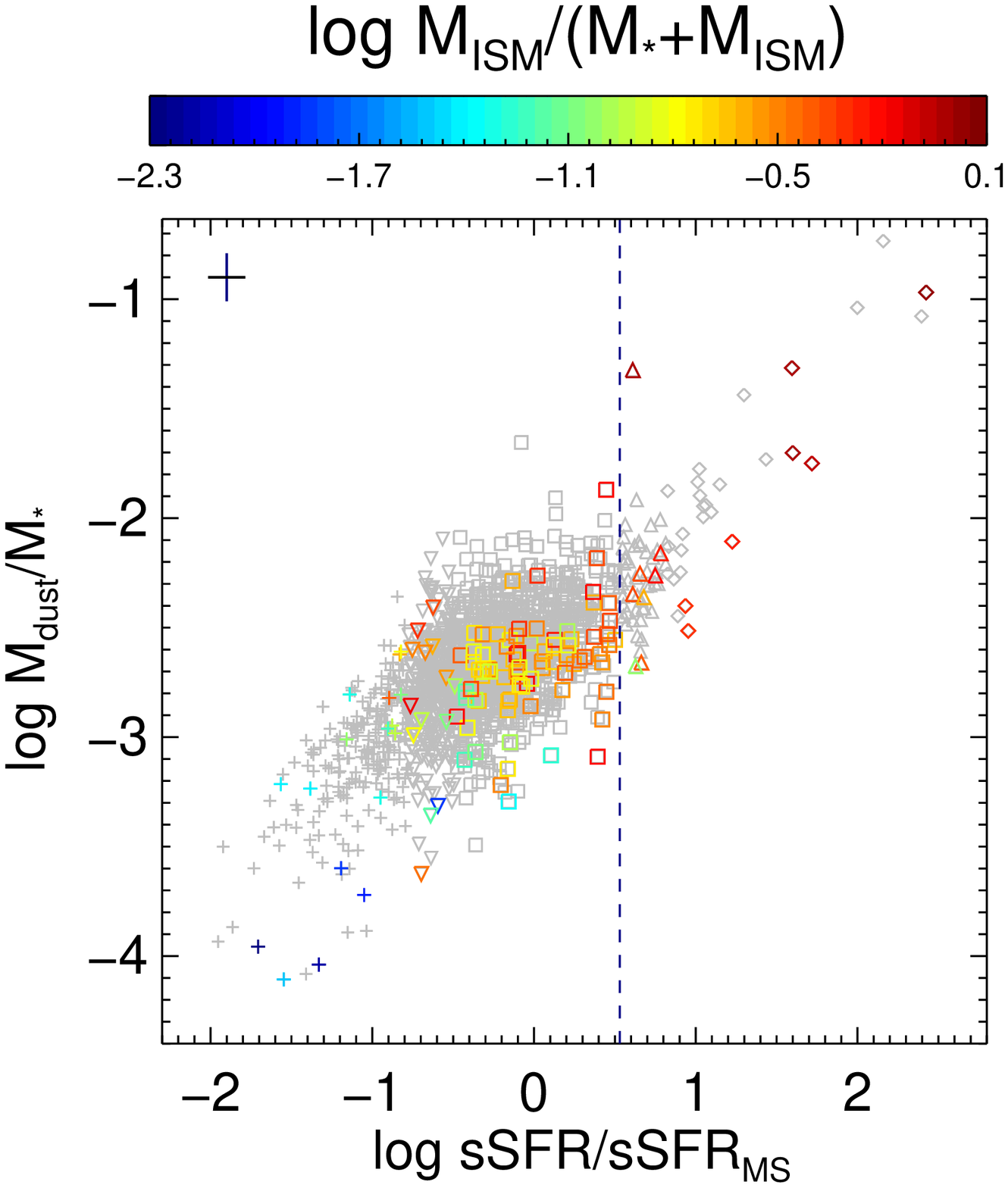}

         \caption{ Atomic hydrogen mass (left), \lpco\ (middle) and dust mass (right) to stellar mass ratio as
         a function of the distance to the MS (sSFR/$\rm sSFR_{MS}$). Symbols are the same as figure \ref{fig:tweight}.
         Symbols are colored by gas fraction, with gray symbols used for sources with unknown gas fraction. 
         The vertical line divides the galaxies with low and normal sSFR (PAS, InPAS and MS to the left), 
         and galaxies with high sSFR (InSB and SB to the right). The black cross shows the typical error.
                  }  
  \label{fig:md-mhi-lco m*ratio vs ssfr}
  \end{center}
\end{figure*}

Figure \ref{fig:md-mhi-lco m*ratio vs ssfr} shows the relationships between the atomic gas mass 
to stellar mass ratio, the \lpco\ to stellar mass ratio and the dust to stellar mass ratio, 
with the distance of the galaxy from the MS (log (sSFR/$\rm sSFR_{MS}$)). 
These plots reveal that log (sSFR/$\rm sSFR_{MS}$) is weakly
correlated with \mhi\ and \lpco, but tightly correlated with the dust to stellar mass ratio.

The evolution of the dust to stellar mass ratio as tracer of star formation mode (sSFR classification) 
was previously studied in galaxies within the redshift range $\rm 0 < z < 4$ by \citet{tanet14}, 
who use different dust to stellar mass ratios for  normal star-forming galaxies 
and extreme SB galaxies (ULIRGs) over the redshift range 0 to 2. The largest difference was found
at z$\sim$0 where for normal galaxies they used the averaged value of the normal galaxies in
the \citet{dacet10} sample.
In comparison, the normal galaxies in our sample show a large scatter ($\sim$ 1.3 dex) 
in their dust to stellar mass ratios (see right panel of Figure \ref{fig:md-mhi-lco m*ratio vs ssfr}),
so that they overlap with the median dust to stellar mass ratio of extreme starbursts. 
This scatter is related to the gas fraction of these galaxies, where gas-rich galaxies show 
higher dust to stellar mass ratios. 
Our sample does show that SB and InSB galaxies have higher dust to stellar mass ratios than
MS and PAS galaxies; in fact their dust to stellar mass ratios are higher than the data point
of extreme SBs at z$\sim$0 in \citet{tanet14} even though our SBs and InSB samples are made up of
LIRGs instead of ULIRGs. Thus overall, we observe median dust to stellar mass ratios higher 
than those used by \citet{tanet14}, and an overlap in the dust to stellar mass distributions 
of normal and SB galaxies.

\citet{grabeset10} have showed that galaxies with higher dust masses have higher stellar masses.
However, in our work we reveal a more complex picture, in which the dust-to-stellar-mass ratio  anticorrelations 
with the stellar mass, and the strength  of the anticorrelation changes in sub-samples of galaxies separated by their 
star-forming mode (i.e. by sSFR; see Sect.~\ref{subsec:anticorrelation}). 
Thus, low stellar mass SB galaxies are dustier than massive PAS galaxies. 
This result is in agreement with the discussion in \citet{clemet13}, where they conclude that 
despite ($\delta_{DGR}$) and \mstar\ being directly proportional to 
metallicity (Z), galaxies with smaller \mstar\ show higher $\rm M_{ISM}$/\mstar\ 
and this effect prevails over the tendency of galaxies with smaller \mstar\ to have smaller Z.

Passive galaxies of course show the lowest sSFR, dust to ISM mass ratios, and 
ISM to stellar mass ratios (Fig.~\ref{fig:md-mhi-lco m*ratio vs ssfr}). 
But Fig.~\ref{fig:DGR vs Mstar} shows that even though PAS galaxies have smaller gas
fractions, they have larger $\delta_{DGR}$ in comparison with MS and SB galaxies. 
That is, at low SFRs, decreasing SFR results in the ISM mass decreasing faster than
the dust mass.  

\subsection{Estimating ISM mass from a single sub-mm measurement}

\begin{figure}[!pht]
 \begin{center}
 \includegraphics[bb=34 142 481 642,width=0.4\textwidth,clip]{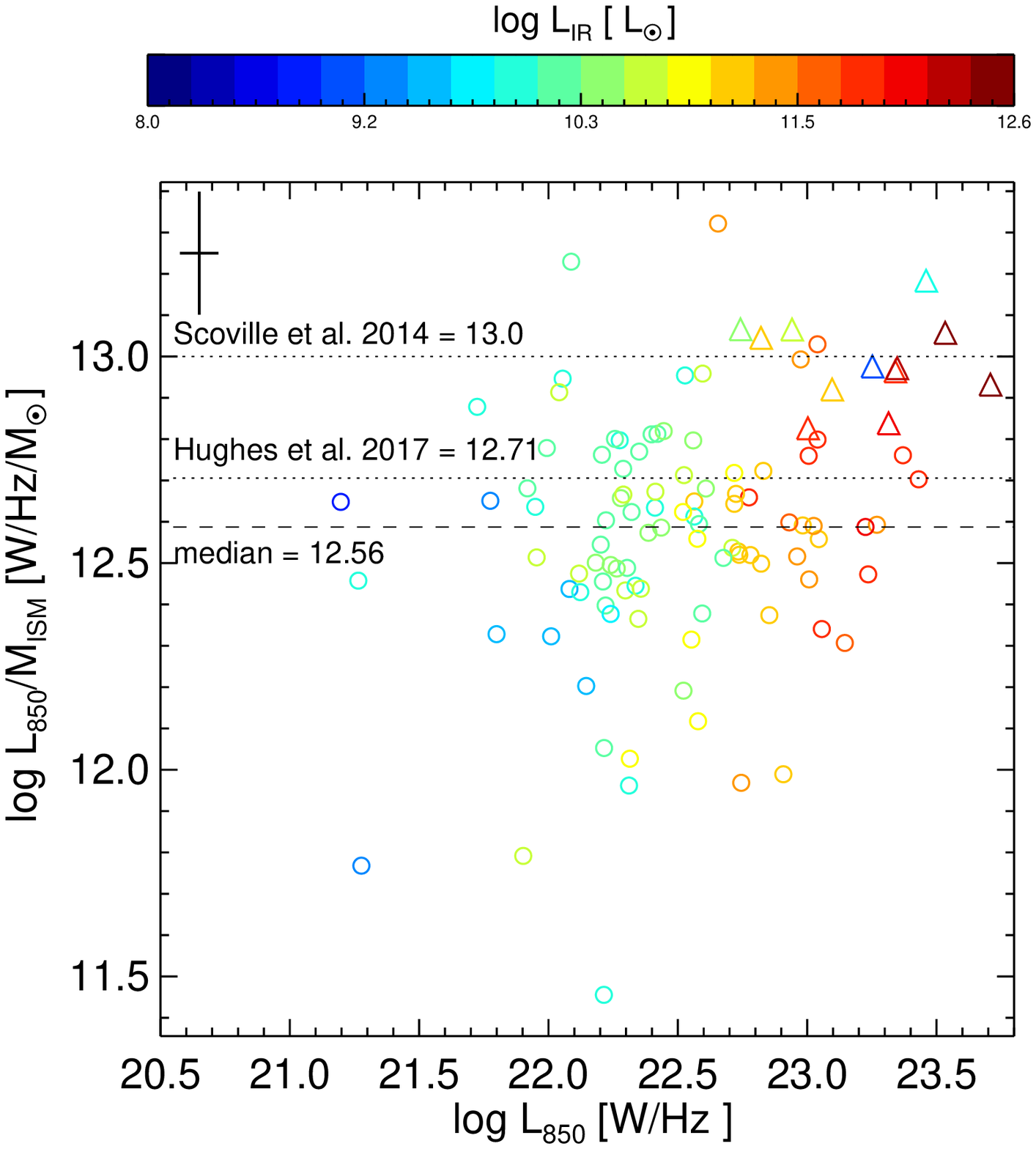}
         \caption{The ratio of the monochromatic luminosity at 850~\mum\ to the ISM mass as a function of 
         monochromatic luminosity at 850~\mum. 
         Circles show galaxies of our sample and large triangles show galaxies from \citet{scoet14}.
         Symbols are colored by the \lir\ following the color bar. 
         The median value for our sample is shown by the dashed line, while dotted 
         lines show the median values obtained by \citet{scoet14} and by \citet{huget17}. 
         The black cross shows the typical error.}
  \label{fig:alpha850}
  \end{center}
\end{figure}

The technique of using a single rest-frame monochromatic luminosity at 850~\mum\ to obtain
a reasonably accurate ISM mass,
developed by \citet{sco12,scoet14,scoet16}, is based on two primary assumptions: 
(1) a constant $\delta_{DGR}$ 
for galaxies with \mstar\ between $10^{9}$ and $10^{11}$ $\rm \msun$, and 
(2) the assumption of a uniform dust temperature for all galaxies, justified because
    a change in factor $\sim2$ in the dust temperature will introduce an error of only
    factor $\sim2$ (0.3 dex) in the ISM mass.
With these assumptions, $\rm \alpha_{850} $ (= $\rm L_{850}$/\mism) can be used to
convert a sub-mm luminosity to an ISM mass. 
In our sample the median value of $\rm \alpha_{850}$ is 12.56, with a rms of
$\sim0.3$ dex (Figure \ref{fig:alpha850}). This value is 
0.44 dex (factor 2.8) lower than that obtained by \citet{scoet14} 
and 0.15 dex (factor 1.4) lower than that obtained by \citet{huget17} using the VALES 
sample that contains 67 main sequence galaxies at 0.02 < z < 0.35. 
The scatter, and its systematics, could be explained by the two steps 
required to calculate $\rm \alpha_{850} $. First, the conversion
of \lsubmm\ to \mdust\ (e.g., Figs.~\ref{fig:l350 with lir,mdust and tdust} and \ref{fig:L850-L550-mdust}) 
is relatively straightforward: the two quantities are highly proportional and almost all the scatter 
(factor 1.33 or 0.12~dex; take rms from the figure \ref{fig:l350 with lir,mdust and tdust}) seen in 
this relationship is a consequence of the variation of cold dust temperature.
The conversion of dust mass to ISM mass, however, is subject to a larger (and dominant)
scatter (rms=0.25 dex; Fig.~\ref{fig:mass mdust-mism}). 
We note that we found no dependence of $\rm \alpha_{850}$ on either the IR luminosity or 
the galaxy stellar mass.
 The \lir\ values (colored bar in Figure \ref{fig:alpha850}) 
reveal that the \citet{sco12,scoet14,scoet16} sample (large triangles) contains LIRGs galaxies, 
ULIRGs and galaxies with $log \lir < 11$ and do not show any correlation with the constant 
$\rm \alpha_{850}$ and the luminosity at 850 \mum. However, our galaxies (circles) show weak correlation 
between the $\alpha_{850}$ and the \lir, where galaxies with smaller \lir\ show smaller values of 
$\alpha_{850}$. 
 A brief analysis on the images and SEDs on the \citet{sco12,scoet14,scoet16} sample, with data taken from NED, 
we observe that the SED in all its galaxies of that sample show prominent synchrotron emission and an infrared
peak emission at wavelength smaller than 100 \mum\ which is consistent with a cold dust component
with high temperature. 
Additionally the images of the sample show that all the galaxies have irregular  and/or disturbing morphologies 
consistent with ongoing galaxy interactions. 
This suggests that \citet{sco12,scoet14,scoet16} sample could have a possible bias, caused by selection effects, 
which allows the tight and constant value for $\alpha_{850}$. In the other hand, the VALES sample, 
exclude the interacting galaxies, and its median $\alpha_{850}$ is closer to our value.

\section{Conclusions}
\label{sec:conclusion}

We have presented reliable and accurate dust mass properties obtained by
fitting DL07 dust models to a large (1,629) sample of nearby galaxies.
The principal results from our analysis are the follows:

\begin{itemize}
\renewcommand{\labelitemi}{$\blacksquare$} 
\item The dust temperature of the cold dust component (\tc) and the weighted
 dust temperature of the fitted SED model are closely correlated 
 to both \lir\ and \mdust, forming a plane which we refer to as the \emph{dust plane}.\\ 

\item The dust mass (and infrared luminosity) can be estimated from a single monochromatic luminosity 
   within the Rayleigh Jeans tail of the dust emission. The error in this estimation is
   0.12 (0.20) dex. This error can be significantly reduced by using an estimate of the dust 
   temperature of the cold component: errors in the estimation now reduce to 0.05 (0.10) dex.\\
 
\item The dust mass is better correlated with the total ISM mass 
   (\mism\ $\propto$ \mdust$^{0.7}$) than with the molecular or
   atomic gas mass separately. \\

\item The conversion factor between the single monochromatic luminosity at $850~\mum$ and the total 
   ISM mass ($\alpha_{850~\mum}$) shows a large scatter (rms = 0.29 dex) for our sample galaxies
   and a weak correlation with the \lir. \\ 
   
\item The star formation mode of a galaxy (based on its sSFR) is correlated
   with both the gas and dust masses: the dustiest (high \mdust/\mstar) galaxies are 
   gas rich and have high SFRs.\\

\item the detailed DL07 model fits, and their resulting parameters, have been used to 
   determine accurate estimators of \lir, \lfir, \mdust, dust temperature from limited
   photometric data, as detailed in the paper and the appendices.\\

\item The results of fitting a large number of passive, MS, and SB galaxies with DL07 templates
   have allowed us to determine typical ranges of value for the DL07 input parameters for
   galaxies with different star forming modes, 
   potentially easing model fitting (i.e. reducing the number of input templates)
   to galaxies with limited photometry.\\

\end{itemize}

Determinations of the true dust temperatures and gas masses of galaxies require detailed studies of
the star formation regions and the conversion factor $\rm \alpha_{CO}$ in a statistically 
relevant sample of galaxies with different star forming modes. 
Such studies are extremely expensive and almost impossible to perform at the present. 
Thus, the empirical scaling relations presented in this work are very useful
to study the global gas and dust properties of galaxies, and constrain their evolutionary stage.

\begin{acknowledgements}

 We would like to thank the referee all for his/her very useful comments and suggestions.
This work was supported by the European Commission through the FP7 SPACE project ASTRODEEP (Ref.No: 312725)
and Programa Financiamiento Basal PFB-06 ETAPA II.
G.O. acknowledges the support provided by CONICYT(Chile) through Programa Nacional de Becas de Doctorado 2012 (21120496),
and FONDECYT postdoctoral research grant no 3170942. 
P.C.-C. acknowledges the support provided by CONICYT(Chile) through Programa Nacional de Becas de Doctorado 2014 (21140882)
P.C. acknowledges the support provided by FONDECYT postdoctoral research grant no 3160375.
R.L. acknowledges support from Comité Mixto ESO-GOBIERNO DE CHILE, GEMINI-CONICYT FUND 32130024, and FONDECYT Grant 3130558.
We acknowledge the usage of the following databases and codes: HyperLeda (http://leda.univ-lyon1.fr), 
NED (https://ned.ipac.caltech.edu), IRSA (http://irsa.ipac.caltech.edu), EDD (http://edd.ifa.hawaii.edu/), 
SDSS (www.sdss3.org), NASA-Sloan Atlas (http://www.nsatlas.org/) and
MPFIT (http://www.physics.wisc.edu/~craigm/idl/idl.html)
\end{acknowledgements}

\appendix 
\addcontentsline{toc}{chapter}{APPENDICES}

\section{Distribution of DL07 model parameters }
\label{apn:parameters}

\begin{figure}[ht]
  \begin{center}
  \includegraphics[bb=79 72 287 139,width=0.5\textwidth,clip]{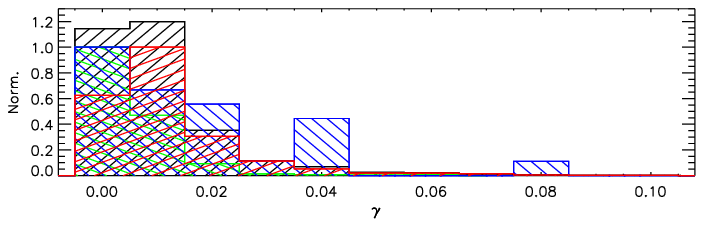}
  \includegraphics[bb=79 72 287 139,width=0.5\textwidth,clip]{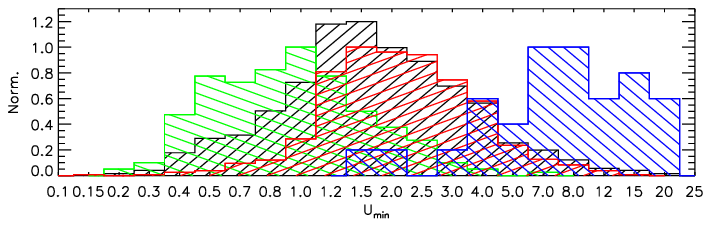}
  \includegraphics[bb=79 72 287 139,width=0.5\textwidth,clip]{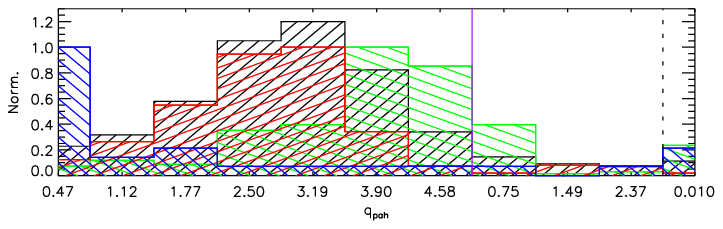}
         \caption{ Normalized distributions of 
         the dust mass fraction heated by PDRs ($\gamma$; top panel), 
         the constant intensity of the radiation field ($\rm U_{min}$; middle) 
         and the polycyclic aromatic hydrocarbon index ($\rm q_{PAH}$; bottom) 
         of the DL07 templates used in our sample.
         Histograms show the distribution of the full sample (black), 
         SB (blue), MS (red), and PAS (green) galaxies . 
         In bottom panel the $\rm q_{pah}$ histogram is divided in three groups of values,
         showed by the purple solid and black dashed lines: Milky Way models (left side of the solid line), 
         the Large Magellanic Cloud (between purple solid and black dashed lines), and the Small Magellanic Cloud
         (right side of the black dashed line).
         }
  \label{fig:parameters}
  \end{center}
\end{figure}

The Draine \& Li models (DL07) need a large number of parameters (6 parameters in total) 
to characterize the emission from the IR to the sub-mm wavelengths. Despite some restrictions 
in DL07 libraries and \citet{draet07}, the amount of avoided SED templates 
is large (in our case 24,200 templates).

In this work, we summarize the use of three DL07 parameters ($\rm U_{min}$, $\gamma$ and $\rm q_{PAH}$). 
Additionally, the main sequence (MS) classification allows a reduction of the number of 
SED templates for samples with less available data.

Figure \ref{fig:parameters} shows the distribution of the parameter $\gamma$ (top panel), 
$\rm U_{min}$ (middle) and $\rm q_{pah}$ (bottom panel), using the values restricted 
by the available libraries for our complete sample (black histogram). The range of values 
for $\gamma$ is between 0.0 and 0.24. Moreover, 91\% of the selected values are between 0.0 and 0.02, 
and just 2 sources show $\gamma > 0.1$.
For $\rm U_{min}$, the range expands between 0.2 and 15, but 75\% of the subsample show values between 1.0 and 3.0. 
The $\rm q_{pah}$ parameter shows a total range between 0.047 and 0.01, 
where values between 0.47 to 4.58 come from models based on the Milky Way (MW), 
between 0.75 to 2.37 are based on the Large Magellanic Cloud (LMC) and the value 0.01 
is a model based on the Small Magellanic Cloud (SMC). 
 Figure \ref{fig:parameters} show this groups of values separated by the purple solid and black dashed lines:
values based on MW models are on left side of the solid line, based on LMC are between purple solid and black 
dashed lines and based on SMC are in the right side of the black dashed line \footnote{The values showed in the x-axis 
are not order consecutively}.
The $\rm q_{PAH}$ parameter ranges between 1.77 and 3.90 including 74\% of the sample, 
all of them belong to models based on the MW. The $\rm q_{PAH}$s based on MW model are
the most relevant values since they cover the 91\% of the complete sample.

\begin{table}[ht]
\centering
\caption{DL07 parameters for MS galaxies}
\begin{tabular}{c|ccc}
              & Total        & Typical    &  \% of sources in \\
Parameter     & Range        & Range      & the typical range  \\
\hline
$\gamma$      & 0.0 - 0.09   & 0.0 - 0.02 & 90\%               \\
$\rm U_{min}$ & 0.2 - 15     & 1.0 - 5.0  & 89\%               \\
$\rm q_{pah}$ & 0.47 - 0.010 & 1.77 - 3.9 & 81\% 
\end{tabular}
\label{tab:params MS}
\end{table}

\begin{table}[ht]
\centering
\caption{DL07 parameters for SB galaxies}
\begin{tabular}{c|ccc}
              & Total      & Typical    & \% of sources in  \\
Parameter     & Range      & Range      & the typical range  \\
\hline
$\gamma$      & 0.0 - 0.23 & 0.0 - 0.04 & 89\%            \\
$\rm U_{min}$ & 2.0 - 25   & 4.0 - 20   & 82\%            \\
$\rm q_{pah}$ & 4.47 - 0.010 & 0.47 - 1.77 & 68\% 
\end{tabular}
\label{tab:params SB}
\end{table}

\begin{table}[ht]
\centering
\caption{DL07 parameters for Passive galaxies}
\begin{tabular}{c|ccc}
              & Total      & Typical    & \% of sources in  \\
Parameter     & Range      & Range      & the typical range  \\
\hline
$\gamma$      & 0.0 - 0.09 & 0.0 - 0.02 & 92\%             \\
$\rm U_{min}$ & 0.3 - 8.0  & 0.4 - 2.0  & 86\%             \\
$\rm q_{pah}$ & 0.47 - 0.010 & 2.5 - 0.75 & 84\% 
\end{tabular}
\label{tab:params Passive}
\end{table}

Following the sSFR classification, it's possible to select 
more specific values for these parameters. See tables \ref{tab:params MS}, 
\ref{tab:params SB} and \ref{tab:params Passive} for more details.

\section{Robustness test}
\label{apn:robutsness test}

\begin{table*}[ht]
\centering
\setlength{\tabcolsep}{4pt}
\begin{tabular}{lc|ccccccc}
  Rejected                   & \#  points &  \#         & $\chi^2/\chi^{2(*)}$  & $\rm \chi_{r}^{2}/\chi_{r}^{2(*)}$ &  \mdust/\mdust$^{(*)}$\  &  \lir/\lir$^{(*)}$\  & \tc/\tc$^{(*)}$\ & \tw/\tw$^{(*)}$\   \\
    point(s)                 &  SED fit   &  templates$^{(*)}$  &              &                    &                &                &               &              \\                          
                             &            & (1)         & (2)          &   (3)              & (4)            &  (5)           & (6)           & (7)           \\  
\hline
f$_{350}$                    & 6          & 0.9$\pm$0.2 & 0.5$\pm$0.2  & 0.6$\pm$0.3        & 0.90$\pm$0.08  & 0.99$\pm$0.05  & 1.01$\pm$0.01 & 1.00$\pm$0.01 \\
f$_{12}$,f$_{22}$           & 5          & 13 $\pm$7   & 0.6$\pm$0.2  & 1.0$\pm$0.4        & 1.02$\pm$0.04  & 0.96$\pm$0.09  & 0.99$\pm$0.01 & 1.00$\pm$0.06 \\
f$_{350}$,f$_{550}$         & 5          & 0.9$\pm$0.4 & 0.3$\pm$0.2  & 0.5$\pm$0.3        & 0.8 $\pm$0.1   & 1.01$\pm$0.07  & 1.02$\pm$0.02 & 0.99$\pm$0.05 \\
f$_{550}$,f$_{850}$         & 5          & 0.7$\pm$0.3 & 0.3$\pm$0.3  & 0.6$\pm$0.5        & 1.1 $\pm$0.1   & 1.02$\pm$0.07  & 0.99$\pm$0.01 & 1.02$\pm$0.03 \\
f$_{12}$,f$_{22}$,f$_{350}$ & 4          & 14$\pm$7    & 0.2$\pm$0.1  & 0.4$\pm$0.3        & 0.9 $\pm$0.1   & 0.9$\pm$0.1    & 1.00$\pm$0.01 & 1.01$\pm$0.07 \\
f$_{12}$,f$_{22}$,f$_{850}$ & 4          & 12$\pm$7    & 0.2$\pm$0.1  & 0.4$\pm$0.3        & 1.1 $\pm$0.1   & 0.9$\pm$0.1    & 0.98$\pm$0.02 & 1.01$\pm$0.06
\end{tabular}
\caption{Robustness test statistics.
The values represent the mean value of the distribution of parameters, normalized to the value of the parameter using 7 points.
The symbol (*) shows that the parameters are normalized. 
The error quoted in each column corresponds to the standard deviation of the distribution.
Columns: (1) number of template; (2)-(3) $\chi^{2}$ and reduced $\chi^{2}$ for the new fitting, (4) dust mass, (5) IR luminosity, (6)-(7) 
the dust temperature of the cold and warm components, respectively}
\label{tab:robustness}
\end{table*} 

As we mention in Section \ref{subsec:robustness}, we prepare a robustness test using 24 galaxies with 
7 photometric observations and good SED fits. 
In this test, we explore the robustness of our SED fitting considering our SED fitting criteria: 
a) observations at 60 and 100~\mum\ ;
b) observations at 12 and 22~\mum\ or 22~\mum\ ;
c) minimum of 4 points with two possible distributions, 3 at $\lambda\le100~\mum$ and 1 at $\lambda\ge350~\mum$
or 2 at $\lambda\le100~\mum$ and 2 at $\lambda\ge350~\mum$.
This criterion plus the technique of weighted geometric mean to obtain the final value of the parameters 
(described in section \ref{sec:model dust emission}), shows to be very robust. 
In our test, we systematically discard different amounts of points at different wavelength, this technique evaluates 
the reduction of the measurements and the influence of the position of this measurement in the SED fitting.
 Table \ref{tab:robustness} shows the rejected points (e.g., f$_{350}$ is the flux at 350~\mum)  and the number 
of points in each SED fitting test.
To characterize the change of the parameter values for each restriction, we study
the distribution of the value for the parameter normalized with the value of them, using the fitting
with 7 points.
Columns 1 to 7 show the mean value of each normalized distribution; the quoted error is the standard deviation.
The normalized number of templates (column 1) shows a great increment ($\sim$10 times) of accepted templates 
in the SED fitting, in absence of the wavelengths at 12 and 22~\mum\ . 
However, in absence of the sub-mm data, the amount of accepted templates is only reduced by a factor of 0.1 to 0.3. 
The value of the $\chi^{2}$ and the reduced $\chi^{2}$ show a direct dependence on the amount of points used and 
a weak dependence on the wavelength of the rejected points. 
On the other hand, the dust mass, IR luminosity and the temperature of the cold and warm dust components show, in general, 
factors of difference smaller than 0.1 with a standard deviation smaller than 0.1 
(the dust mass shows greater factors of difference and greater standard deviation when the sub-mm data are rejected, 
as is expected \citep{drali07}).

\section{Stellar mass comparison}
\label{apn:stellar mass}

\begin{figure}[!pht]
  \begin{center}
  \includegraphics[bb=32 188 585 641,width=0.35\textwidth,clip]{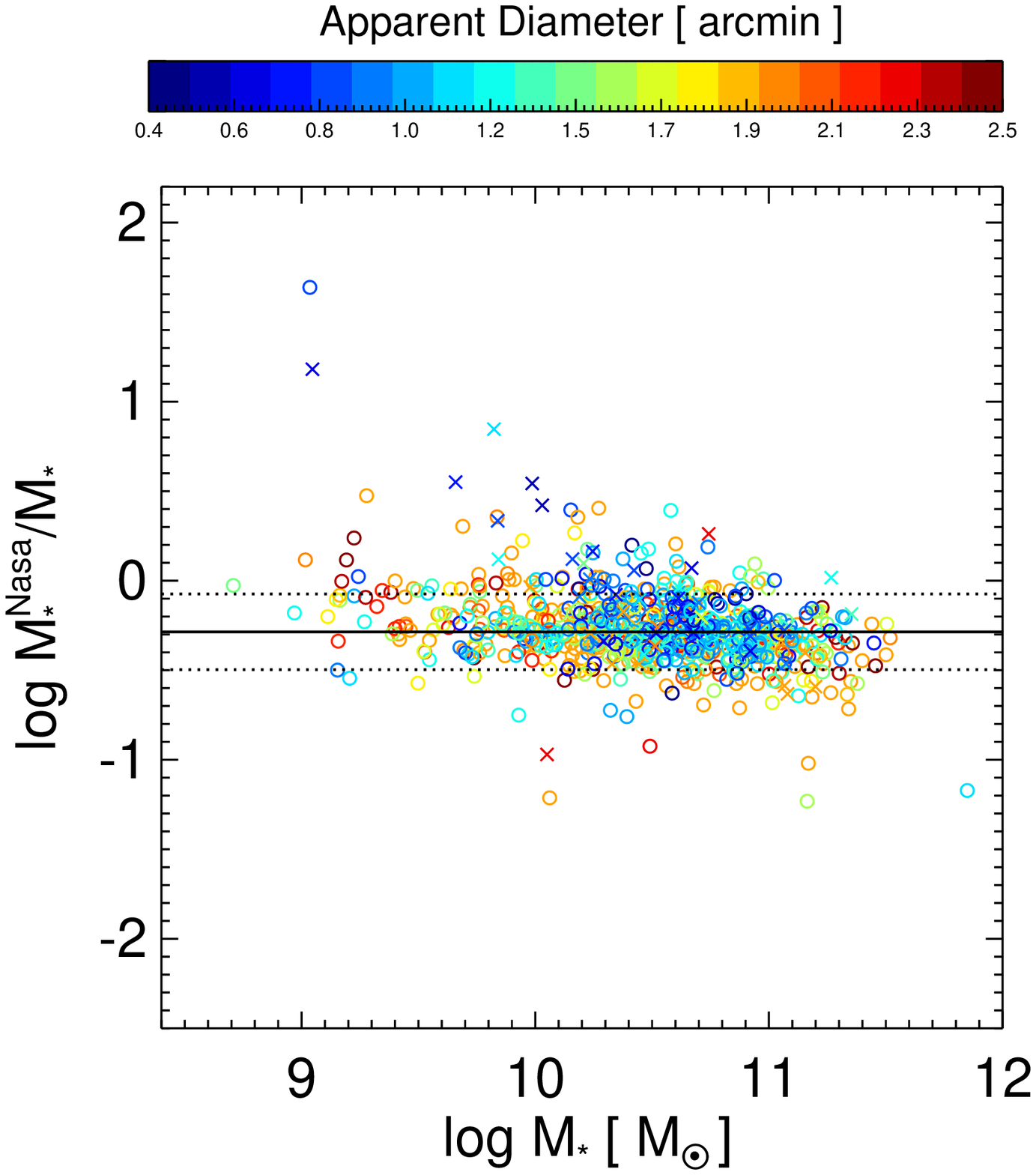}
  \includegraphics[bb=32 188 585 561,width=0.35\textwidth,clip]{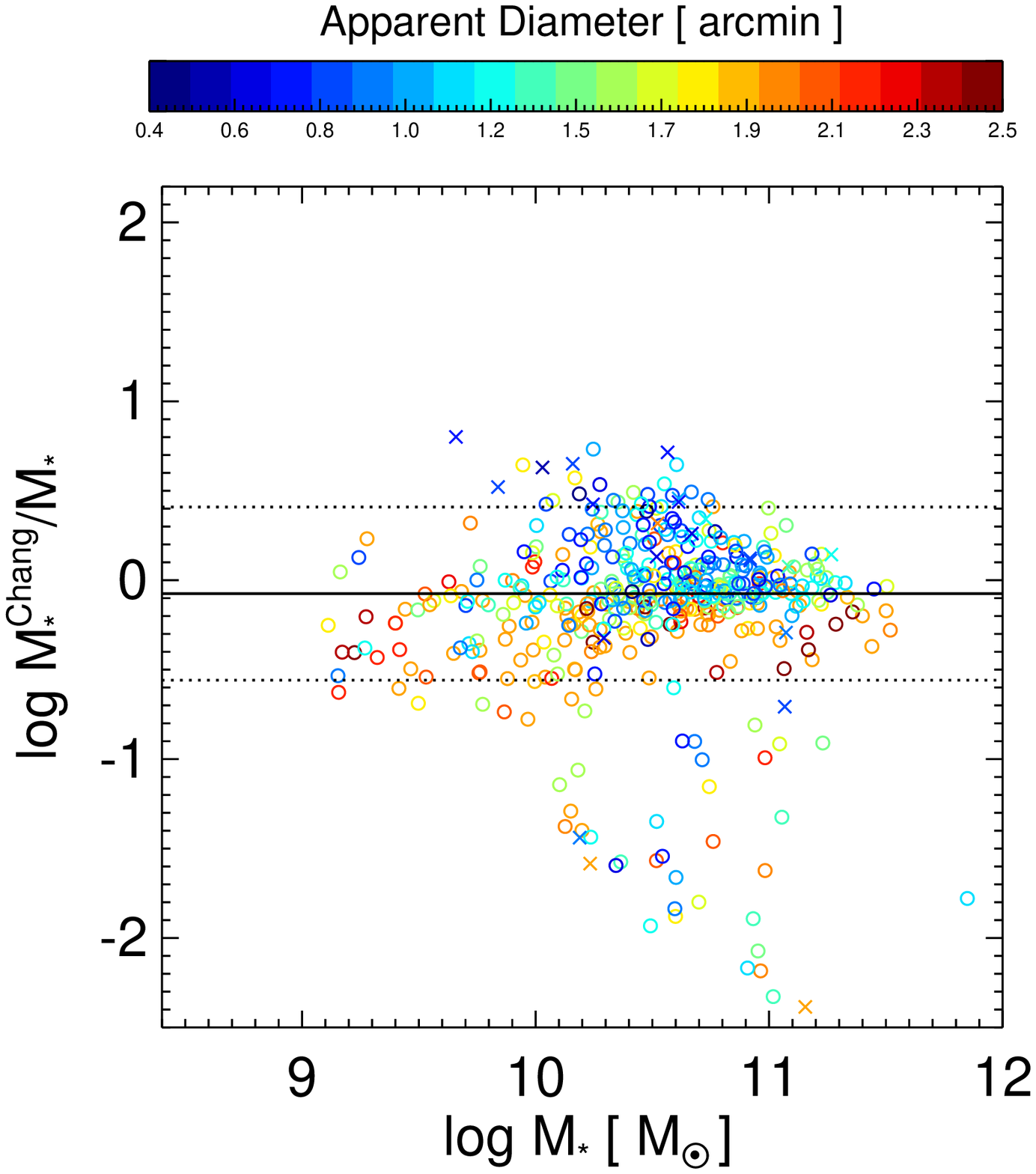}
  \includegraphics[bb=32 130 585 561,width=0.35\textwidth,clip]{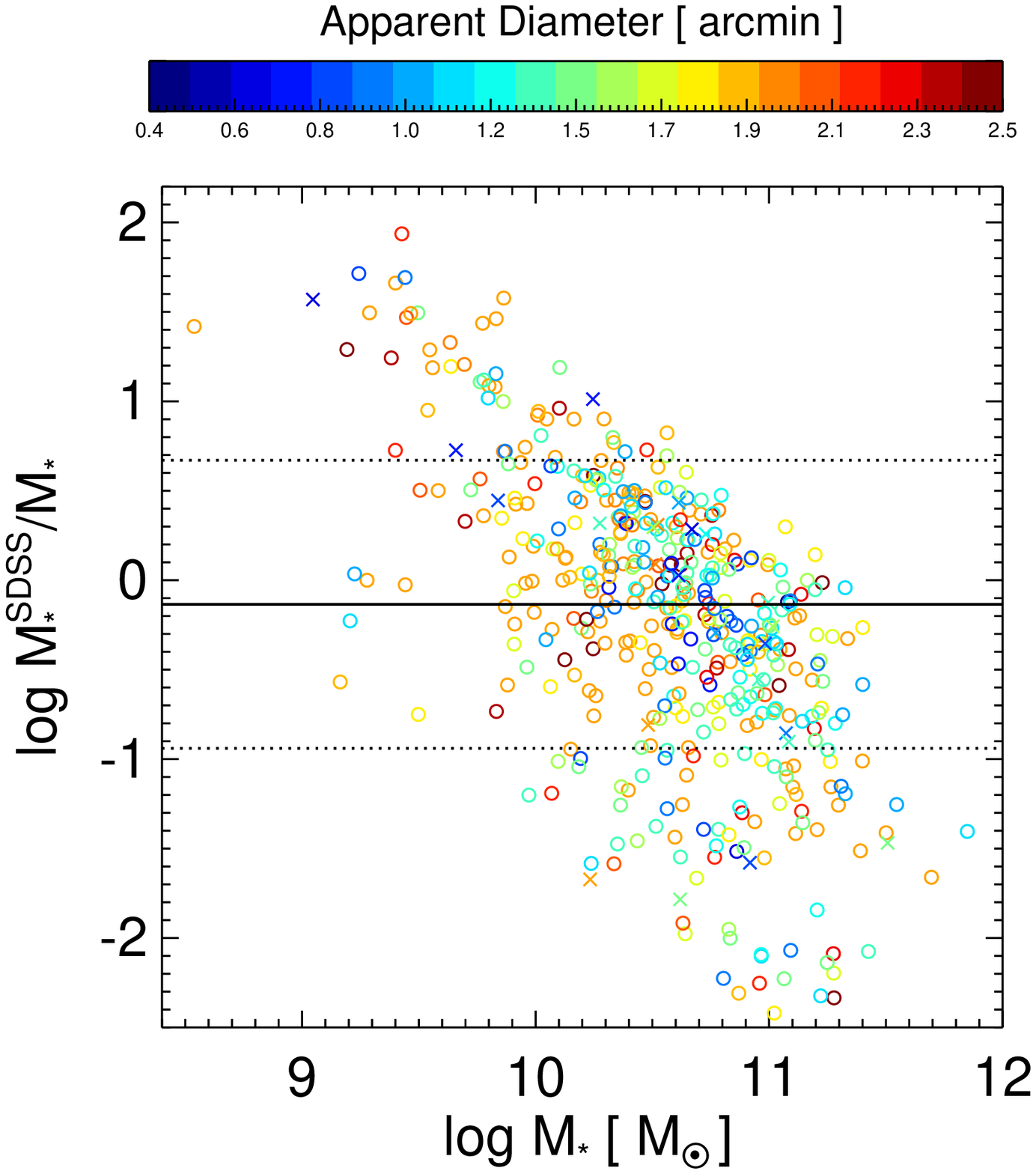}
         \caption{ Comparison between stellar mass estimates and the 
         ratio between the NASA Sloan Atlas
         estimation (top), \citep{chanet15} estimation (middle) and MPA-JHU 
         estimation (bottom) with our stellar mass. 
         The solid line shows the median value and dashed lines show the $\pm 1\sigma$ dispersion.
         Colors represent the apparent diameter of the galaxies in arcminutes. 
         Crosses show sources classified as interacting galaxies and circles show 
         non-interacting galaxies.
         }
  \label{fig:stellar mass}
  \end{center}
\end{figure}

A frequently used parameter to characterize galaxies and their evolution 
is the stellar mass (\mstar). Different methods are used in order to obtain 
the \mstar\ from different data and models. For our study, we use the relations 
obtained by \citet{cluet14}, who uses W1 and W2 (at 3.4 and 4.6~\mum\ , respectively) 
from \wise\ observations. The advantage is that no correction by opacity is needed,
in contrast to other methods using optical photometry.
To compare our results, we use three other samples: 
(1) The Nasa Sloan Atlas (NSA)\footnote{nsatlas.org} based on all sources from the Sloan Digital 
Sky Survey (SDSS) data release 8 (DR8) at redshifts z$<$ 0.005 with observations from the 
Galaxy Evolution Explorer (GALEX). 
(2) the Chang catalog \citep{chanet15}, which is a compilation of catalogs based on 
SDSS and \wise\ photometry for sources with z$<0.6$ . 
(3) the MPA-JHU database with redshift z$<$ 0.1, which is based on the SDSS data release 8 
(DR8). 

The Nasa Sloan Atlas derives its \mstar from  SDSS ugriz photometry, 
assigning a mass to light ratio according to the galaxy broad-band 
colors and a Salpeter IMF \citep{belet03}. 

\citet{chanet15} obtain its \mstar from SDSS ugriz photometry and \wise\ W1-W2 photometry using 
the MAGPHYS code \citep{dacet08}, which contains a large library of SED templates 
covering the UV to IR range.

The MPA-JHU's \mstar\ is estimated using a Bayesian methodology from SDSS spectroscopy 
and ugriz galaxy photometry with a correction for nebular emission \citep{kauet03}.

 The stellar mass from NSA (Fig. \ref{fig:stellar mass} top panel) 
shows a tight correlation with our stellar masses, the mass ratio ($\rm\mstar^{NSA}/\mstar^{our}$)
show 1 $\sigma$ dispersion of 0.25 dex (factor 1.8) and at 3 $\sigma$ 0.7 (factor 5.0). 
However, NSA stellar masses are systematically larger than our stellar masses 
for interacting galaxies (crosses), while some other non-interacting galaxies with
apparent diameter $> 1.2\arcmin$, show smalls stellar masses than our estimation.

The stellar mass from \citet{chanet15} sample (Fig. \ref{fig:stellar mass} middle panel) 
shows a good correlation with our stellar masses, where $1\sigma$ dispersion between 
the stellar masses is 0.5 dex (factor of 3.2). Interacting galaxies 
show greater stellar masses for \citet{chanet15} estimation, but many non-interacting galaxies 
with apparent diameter $> 1.0\arcmin$ show greater values in our stellar mass estimation.

The stellar mass from SDSS MPA-JHU sample (Fig. \ref{fig:stellar mass} bottom panel) 
shows no consistency with our stellar masses, since the $1\sigma$ dispersion is 0.8 dex (factor 6.3). 
The MPA-JHU estimation seems to overestimate the stellar mass of the majority of our sources with 
\mstar$<10^{9.5}$\msun\ and to underestimate the stellar mass of half of the sample by \mstar$>10^{10}$\msun.

Comparing the SDSS photometry, the NSA has the strongest data because its standard pipeline 
to calibrate SDSS images oversubstract the sky in large sources. To do this, 
the NSA SDSS photometry uses the technique developed by \citet{blaet11} and the pipeline
developed by \citet{lupet01}. Additionally, its flux measurements 
use the SDSS petrosian magnitude and azimuthally-averaged profiles optimizing extended sources. 
On the other hand, the SDSS MPA-JHU estimates 
stellar masses from SDSS spectroscopy measurements with a fiber aperture of 3\arcsec. These
technique is optimized for sources with sizes less than 0.5\arcmin. 

\citet{chanet15} sample uses \wise\ 'pro' magnitude, which is optimized for point-sources and 
underestimates the magnitude from extended sources.

\section{Estimating Dust temperature from IR and sub-mm data}
\label{apn:dust temp}

The dust temperature of the cold component is a important parameter 
as dust mass \mdust\ and the total IR luminosity \lir correlate with it (see Section \ref{subsec:fundamentalplane}).

\begin{figure*}[!pht]
  \begin{center}
  \includegraphics[bb=22 121 500 564,width=0.245\textwidth,clip]{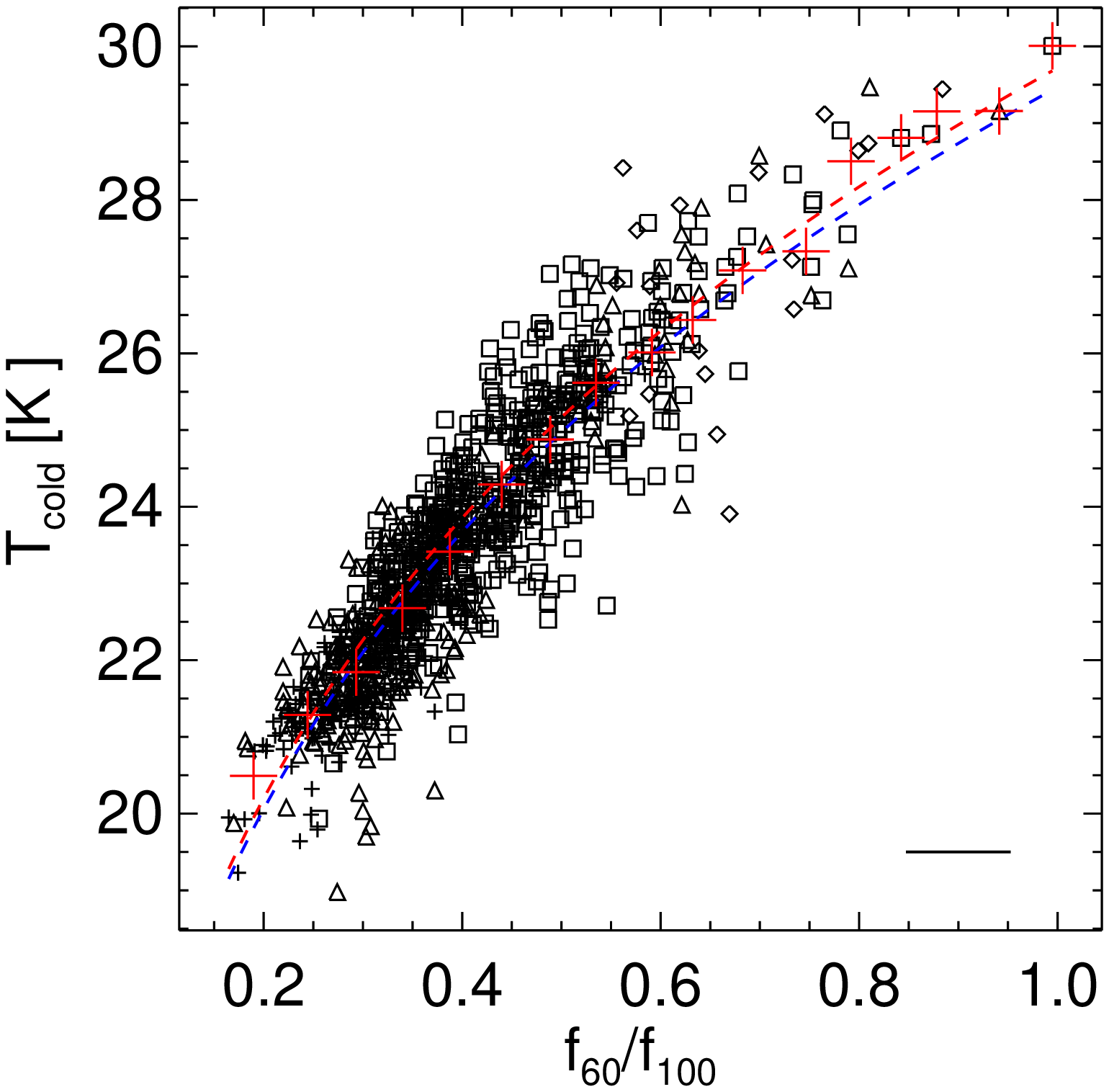}
  \includegraphics[bb=22 121 500 564,width=0.245\textwidth,clip]{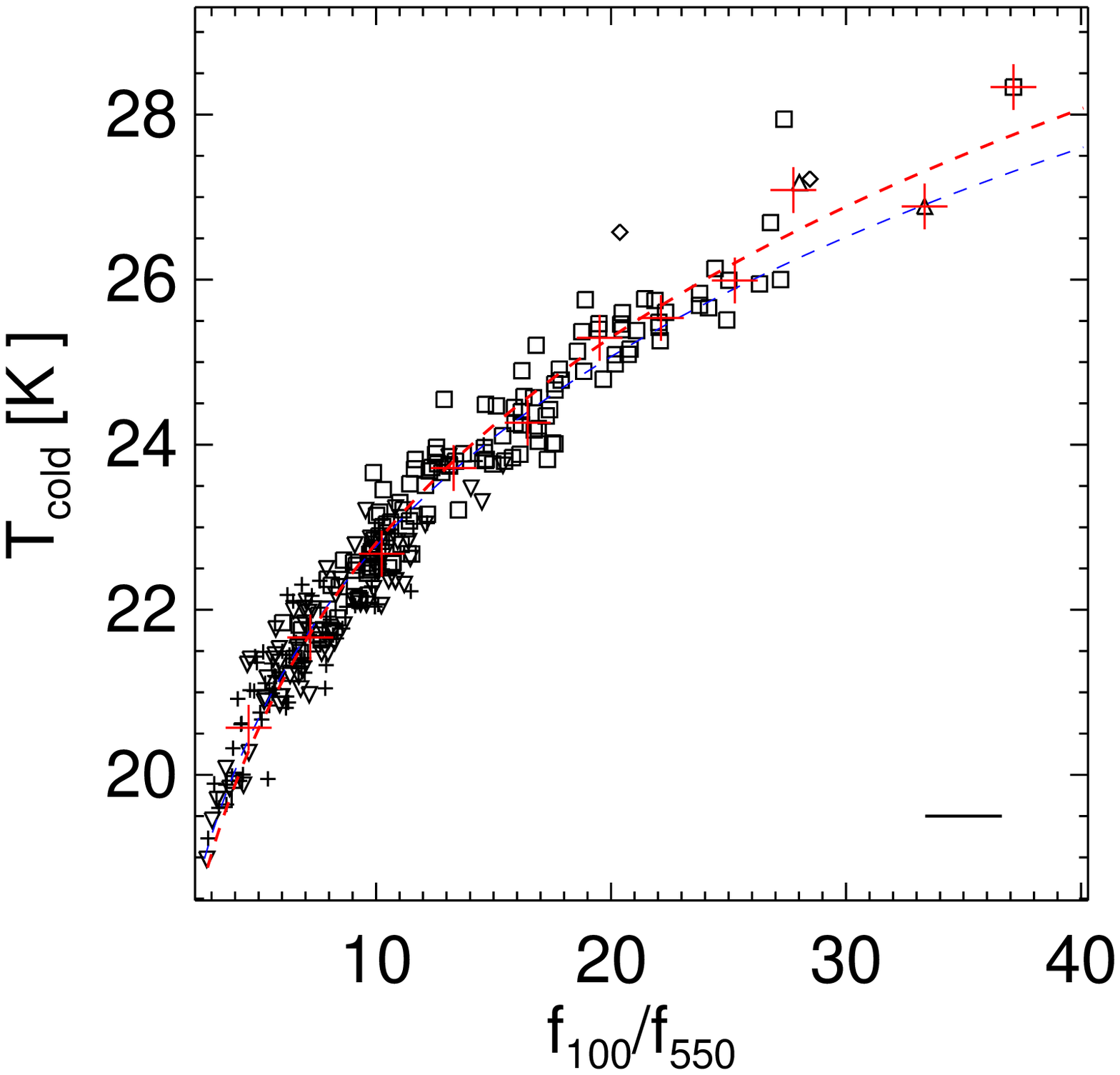}
  \includegraphics[bb=22 121 500 564,width=0.245\textwidth,clip]{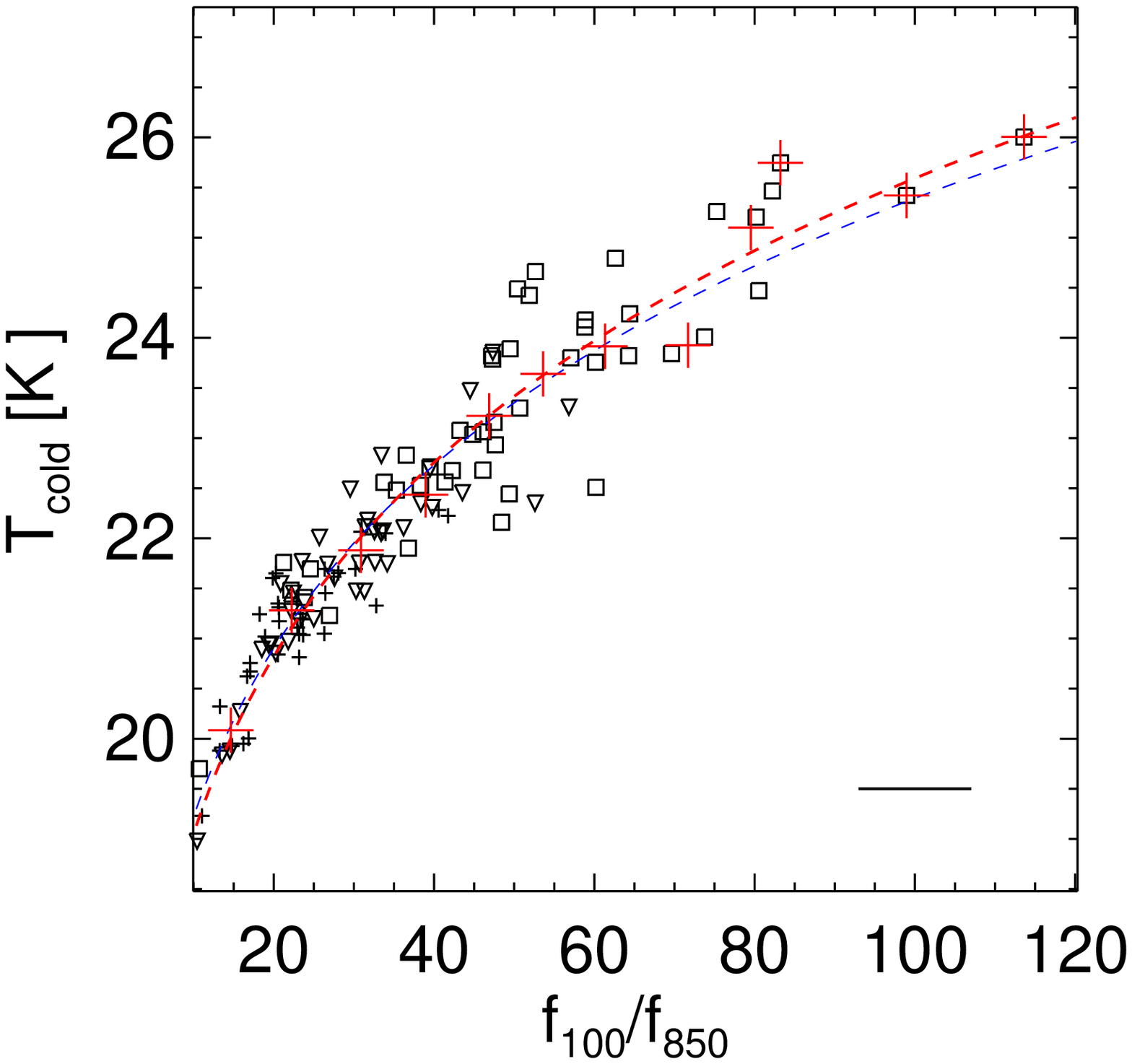}
  \includegraphics[bb=22 121 500 564,width=0.245\textwidth,clip]{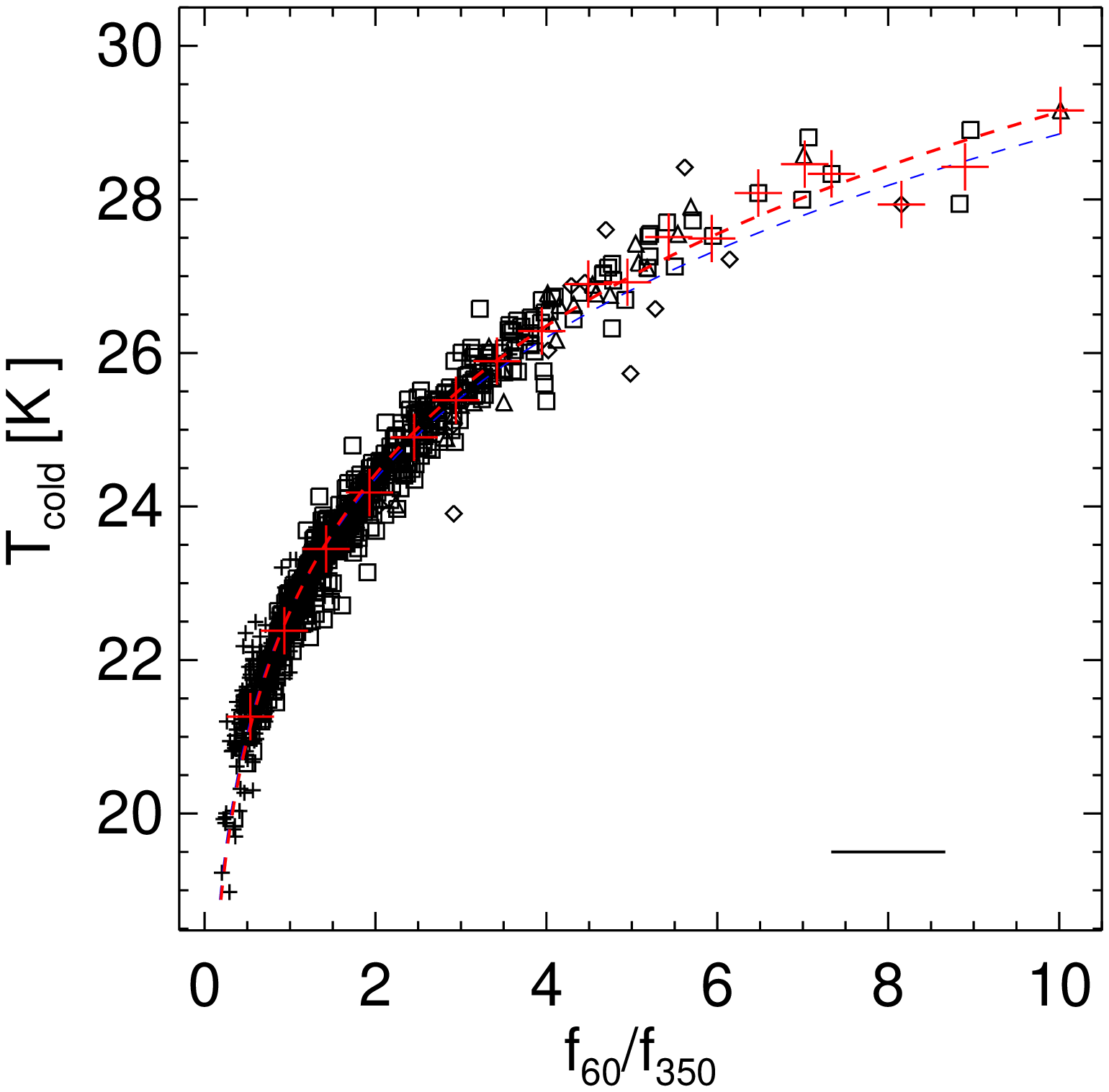}
  \includegraphics[bb=22 121 500 564,width=0.245\textwidth,clip]{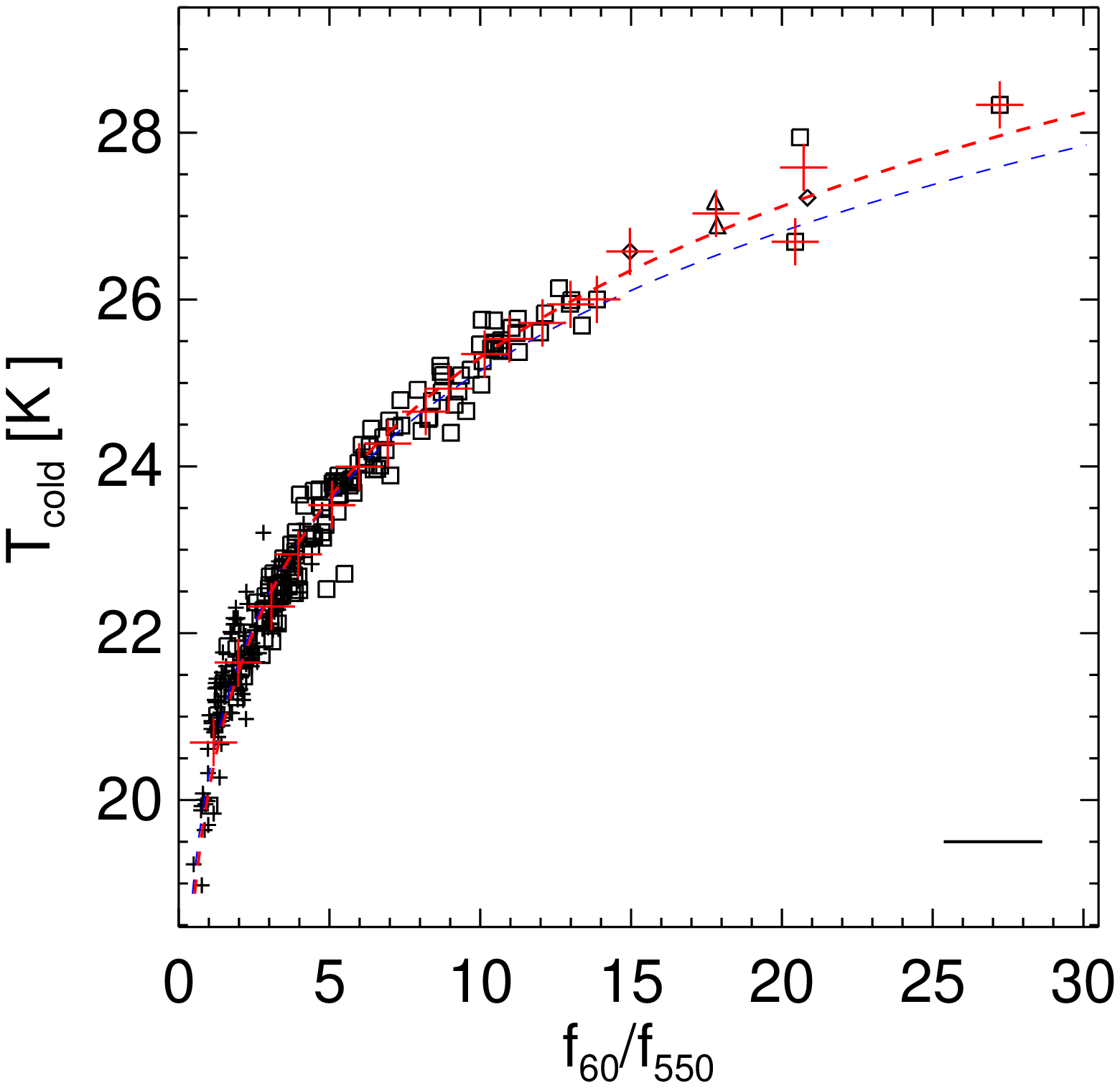}
  \includegraphics[bb=22 121 500 564,width=0.245\textwidth,clip]{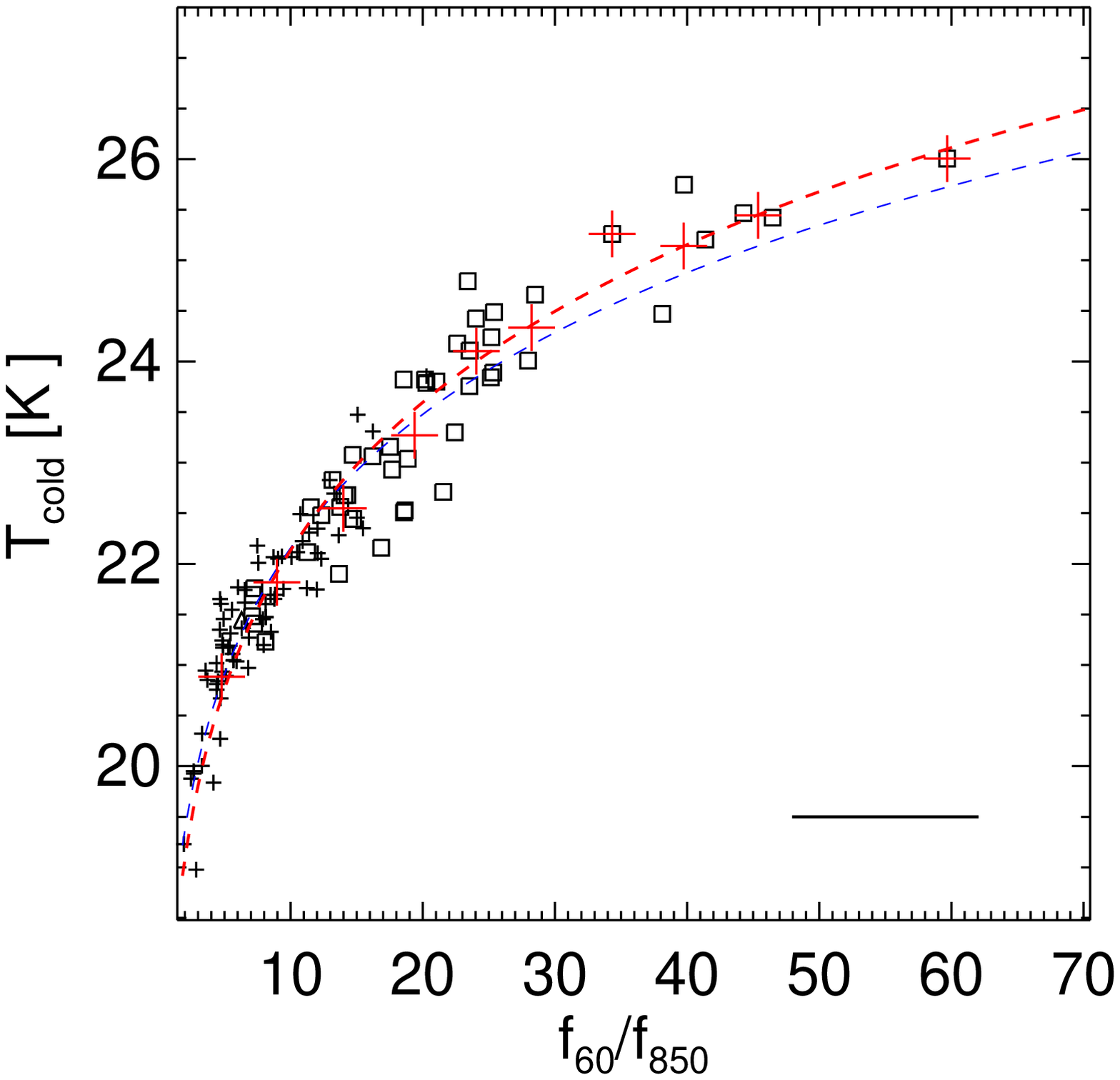} 
  \includegraphics[bb=22 121 500 564,width=0.245\textwidth,clip]{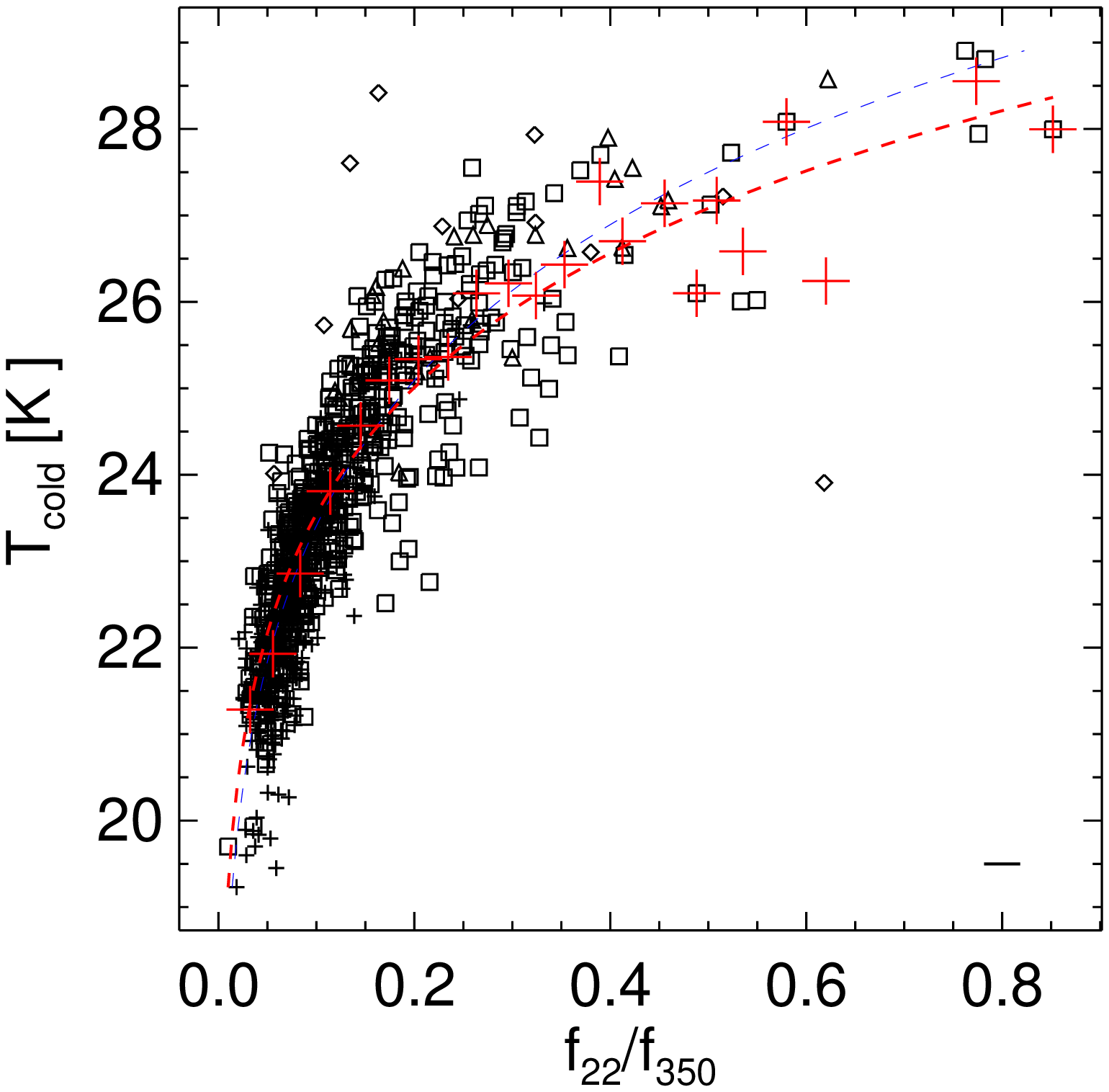}
  \includegraphics[bb=22 121 500 564,width=0.245\textwidth,clip]{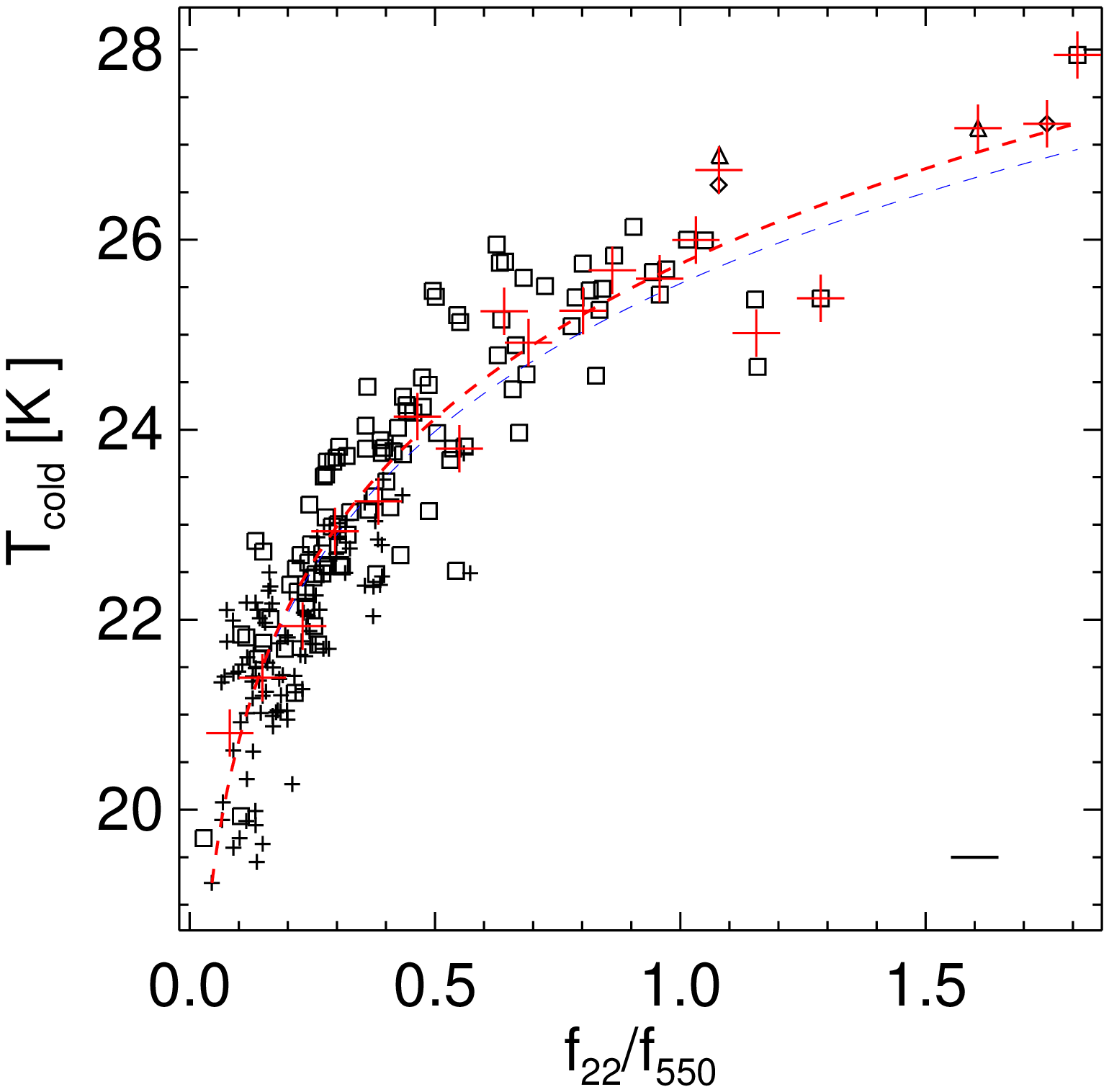}
  \includegraphics[bb=22 121 500 564,width=0.245\textwidth,clip]{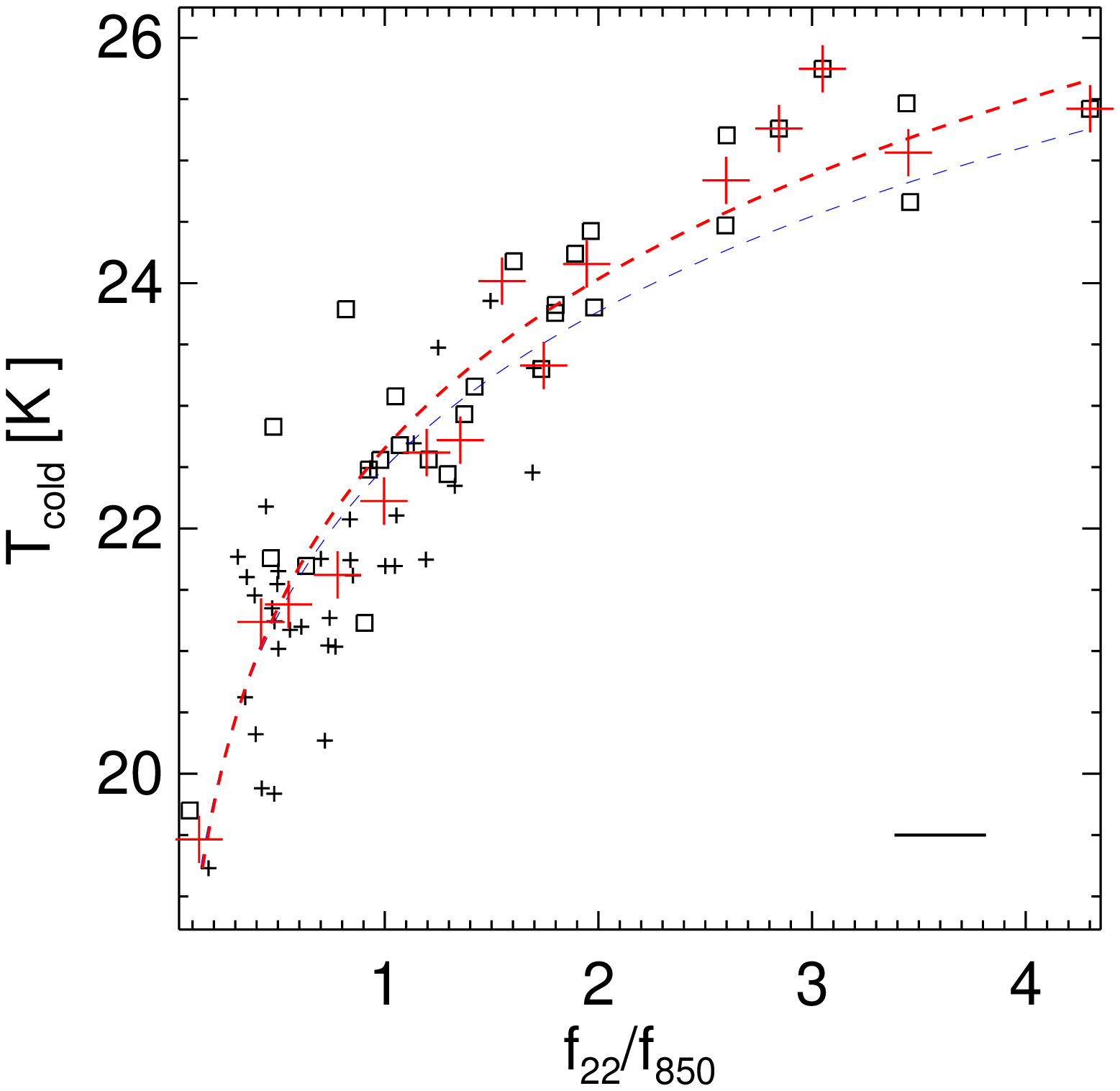} 
  \includegraphics[bb=22 121 500 564,width=0.245\textwidth,clip]{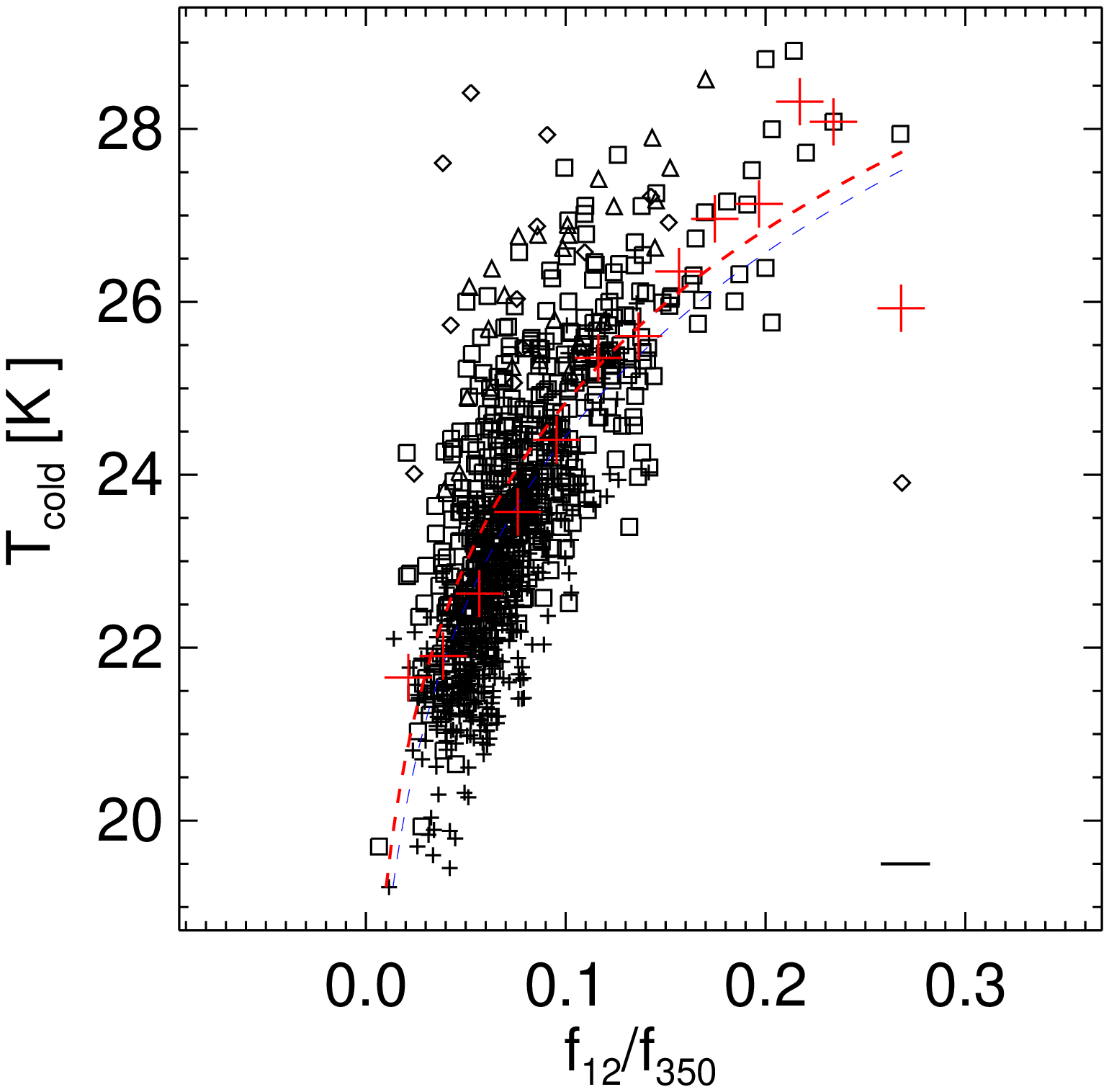}
  \includegraphics[bb=22 121 500 564,width=0.245\textwidth,clip]{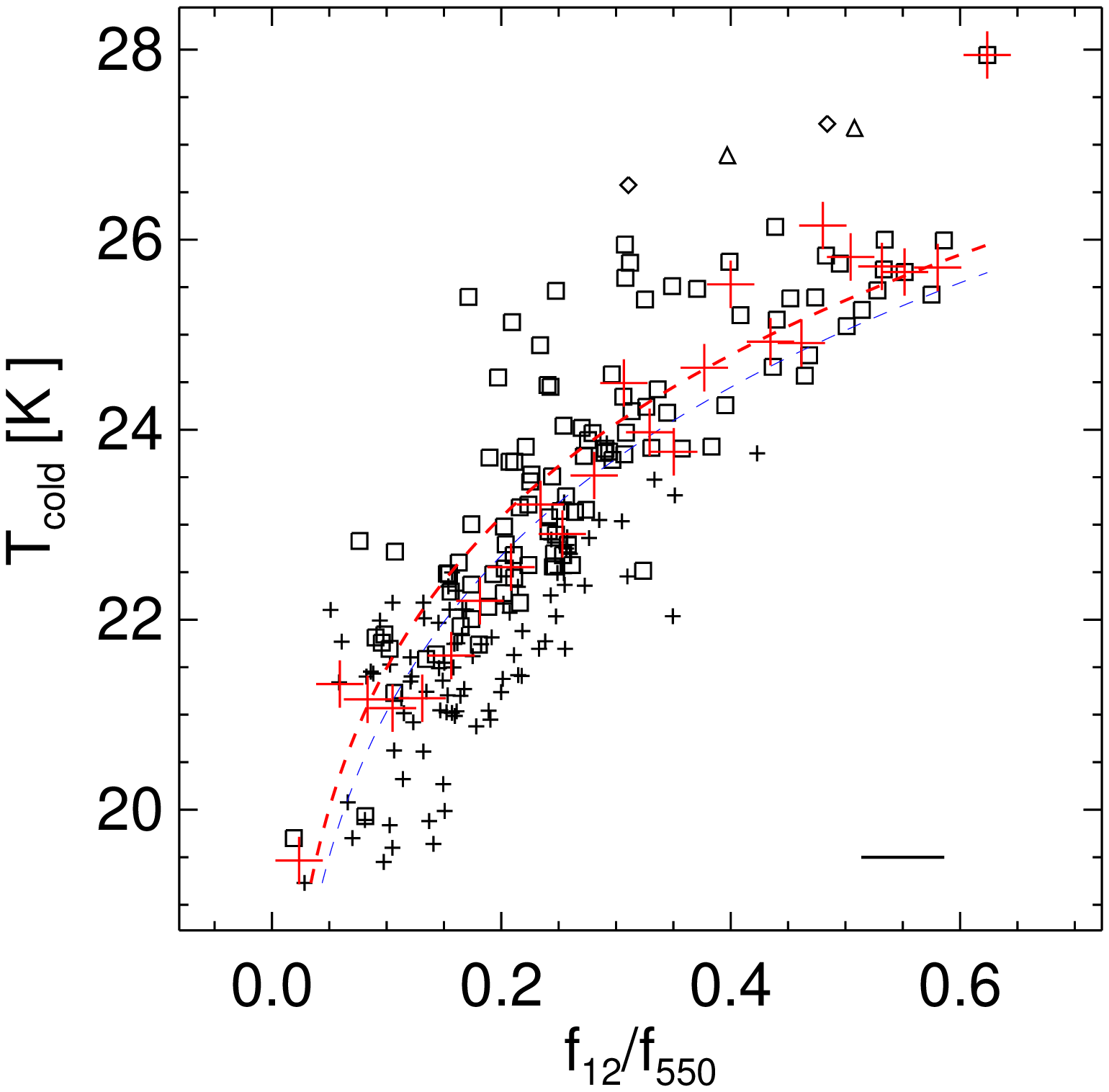}
  \includegraphics[bb=22 121 500 564,width=0.245\textwidth,clip]{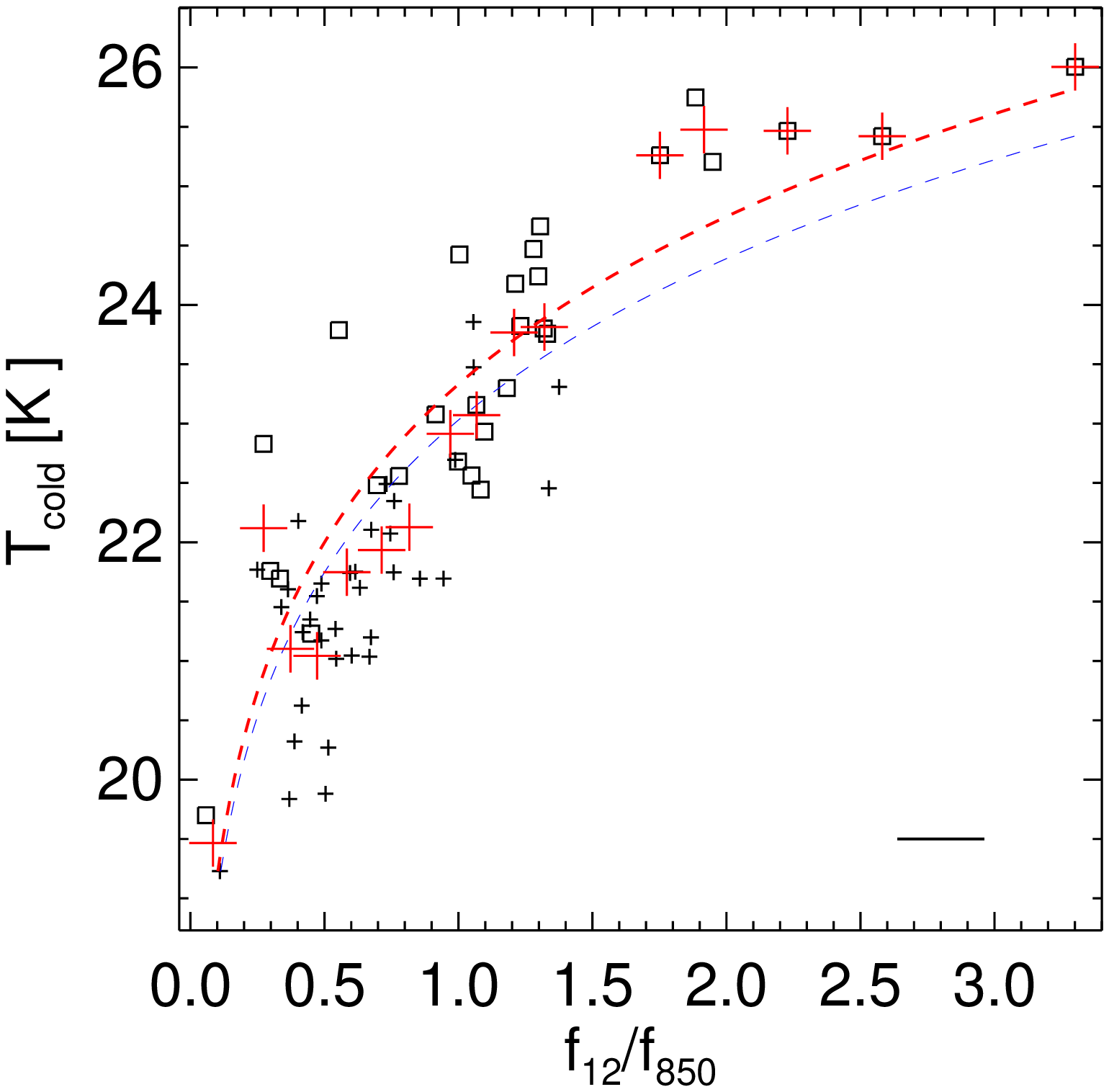} 
  \caption{  Relation between IR and sub-mm colors (with exception of the first panel, which is the color $\rm f_{60}/f_{100}$) 
      and the dust temperature of the cold  component ($f_{\lambda}$, $\lambda$ in \mum). 
     This relation is shown only for galaxies with 5 or more photometric measurements and $\rm\chi_r^2<1.0$. 
     The large, red crosses show the mean values of \tc\ in different color bins.
      The blue dashed curves are the best fit lines for all points, while the red dashed 
     curves are the best fit lines for the large red crosses.
    Symbols represent MS galaxies (squares), SB galaxies (diamonds), PAS galaxies (crosses) and 
    intermediate SB galaxies (triangles). The black horizontal line shows the typical error in the 
    flux ratio.}
  \label{fig:color tdust}
  \end{center}
\end{figure*}

Figure \ref{fig:color tdust} shows the correlation between the dust temperature of the cold
component with different colors at IR and sub-mm wavelengths. 
The best fit of each panel is given by the equation
\begin{equation}
 \rm \frac{T_{cold}}{[K]}=10^{a\pm\Delta a}\left(\frac{f_{\lambda_1}}{f_{\lambda_2}}\right)^{b\pm\Delta b}
 \label{eq:tccolor}
\end{equation}
where the values of the parameters ($\rm a,~\Delta a,~b,\Delta b$) are shown in Table 
\ref{tab:params tc-colors}.

\begin{table}[!pht]
\centering
\caption{Factors for best fit, described by eq. \ref{eq:tccolor} }
\begin{tabular}{cc|cc|cc}
 $\lambda_1$  & $\lambda_2$  & a       & $\rm \Delta a$ & b     & $\rm \Delta b$   \\
  $[\mum]$    & $[\mum]$     &         & $\pm$          &       &    $\pm$         \\
\hline
 100          & 350          &  1.280  &  0.001         & 0.160 & 0.001            \\
 100          & 550          & 1.218   &  0.002         & 0.140 & 0.002            \\
 100          & 850          & 1.156   &  0.005         & 0.156 & 0.004            \\
\hline
  60          & 100          & 1.469   &  0.001         & 0.239 & 0.003            \\
  60          & 350          & 1.355   &  0.001         & 0.105 & 0.001            \\
  60          & 550          & 1.307   &  0.001         & 0.093 & 0.001            \\
  60          & 850          & 1.259   &  0.002         & 0.086 & 0.002            \\
\hline
  22          & 350          & 1.469   &  0.002         & 0.100 & 0.002            \\
  22          & 550          & 1.407   &  0.002         & 0.091 & 0.003            \\
  22          & 850          & 1.352   &  0.002         & 0.077 & 0.005            \\
\hline
  12          & 350          & 1.508   &  0.005         & 0.120 & 0.004            \\
  12          & 550          & 1.434   &  0.004         & 0.109 & 0.006            \\
  12          & 850          & 1.362   &  0.002         & 0.083 & 0.007            
\end{tabular}
\label{tab:params tc-colors}
\end{table}

\begin{figure*}[!pht]
 \begin{center}
 \includegraphics[bb=31 128 480 562,width=0.25\textwidth,clip]{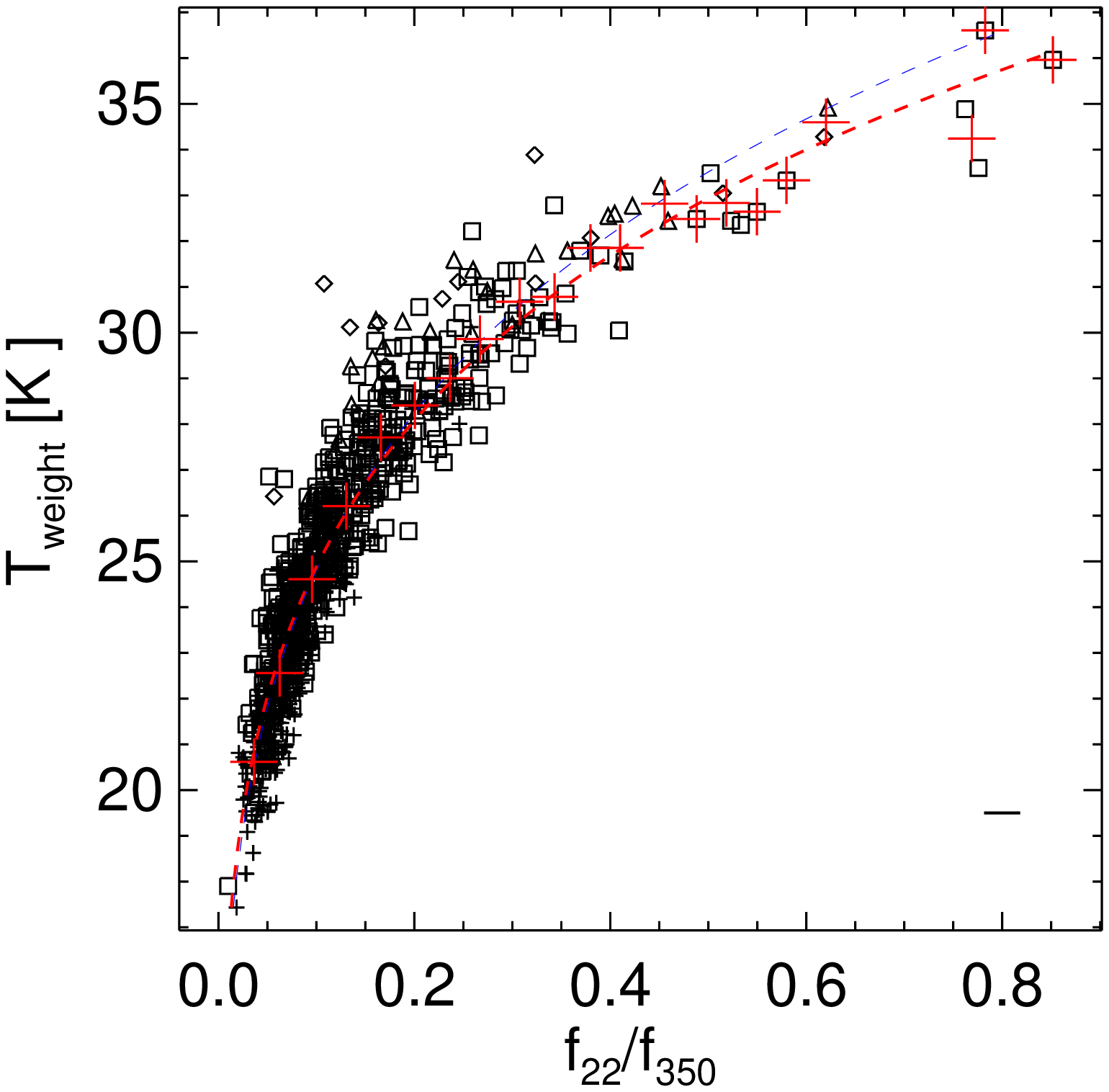}
 \includegraphics[bb=31 128 480 562,width=0.25\textwidth,clip]{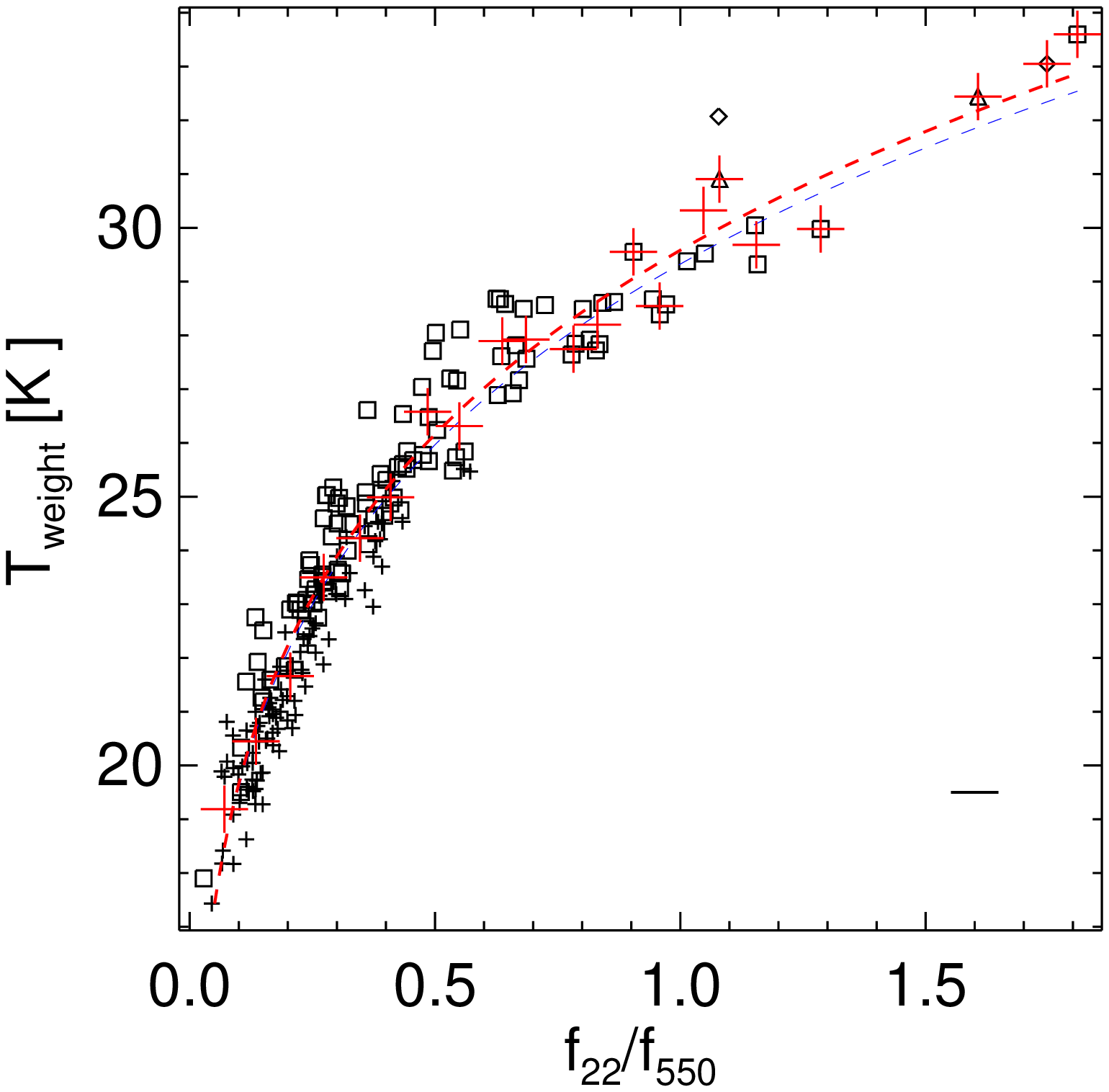}
 \includegraphics[bb=31 128 480 562,width=0.25\textwidth,clip]{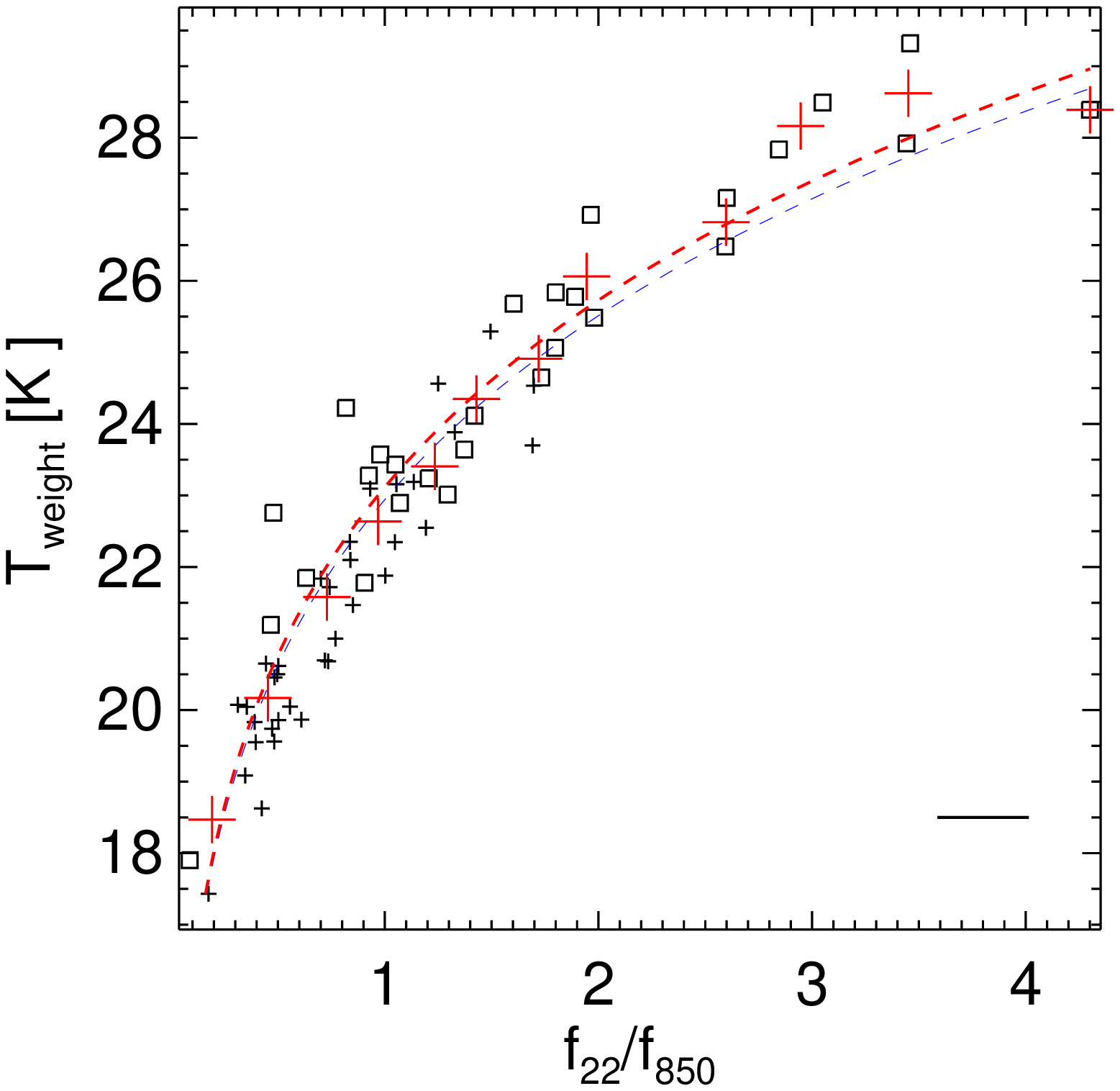}
  \includegraphics[bb=31 128 480 562,width=0.25\textwidth,clip]{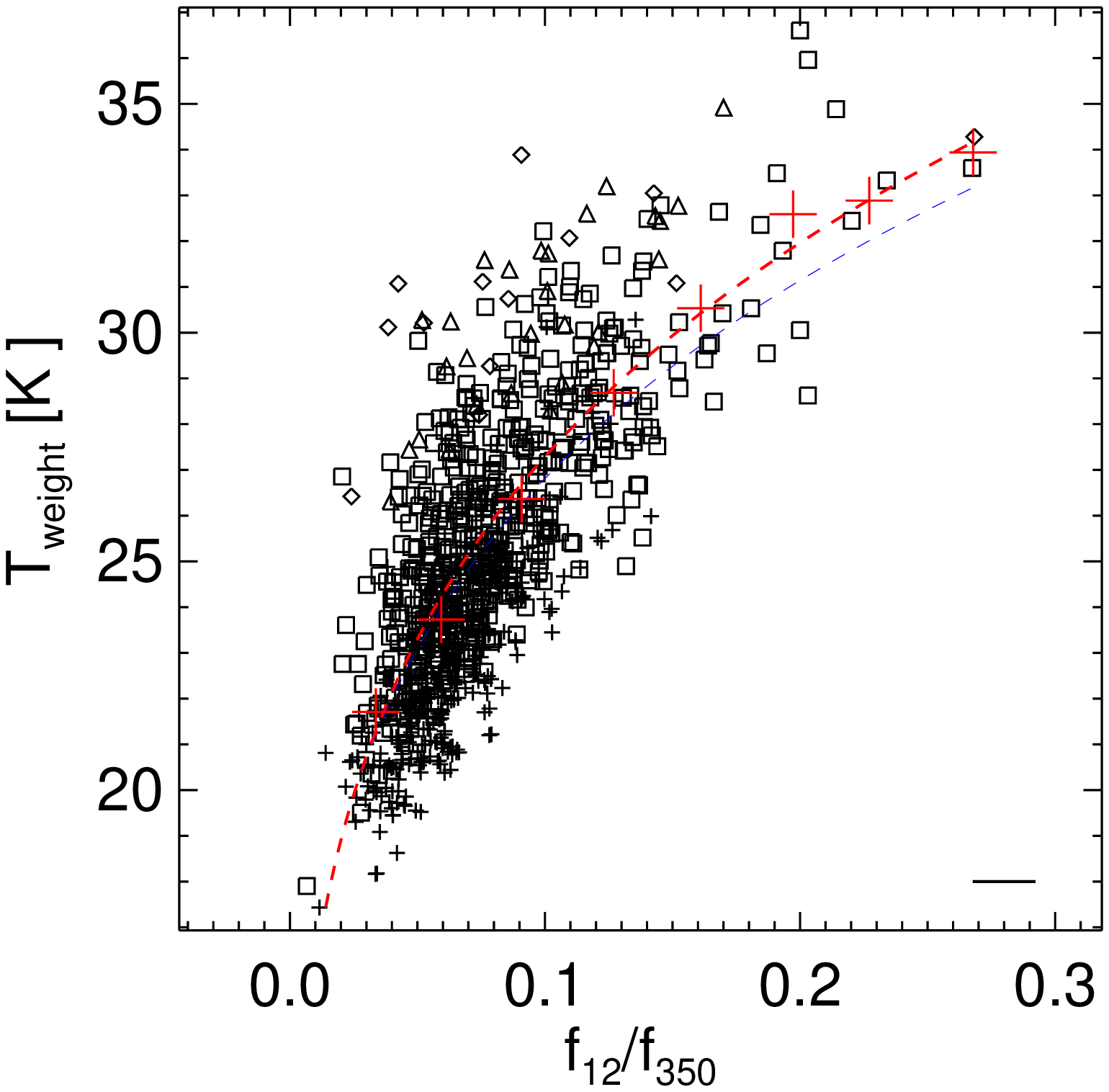}
 \includegraphics[bb=31 128 480 562,width=0.25\textwidth,clip]{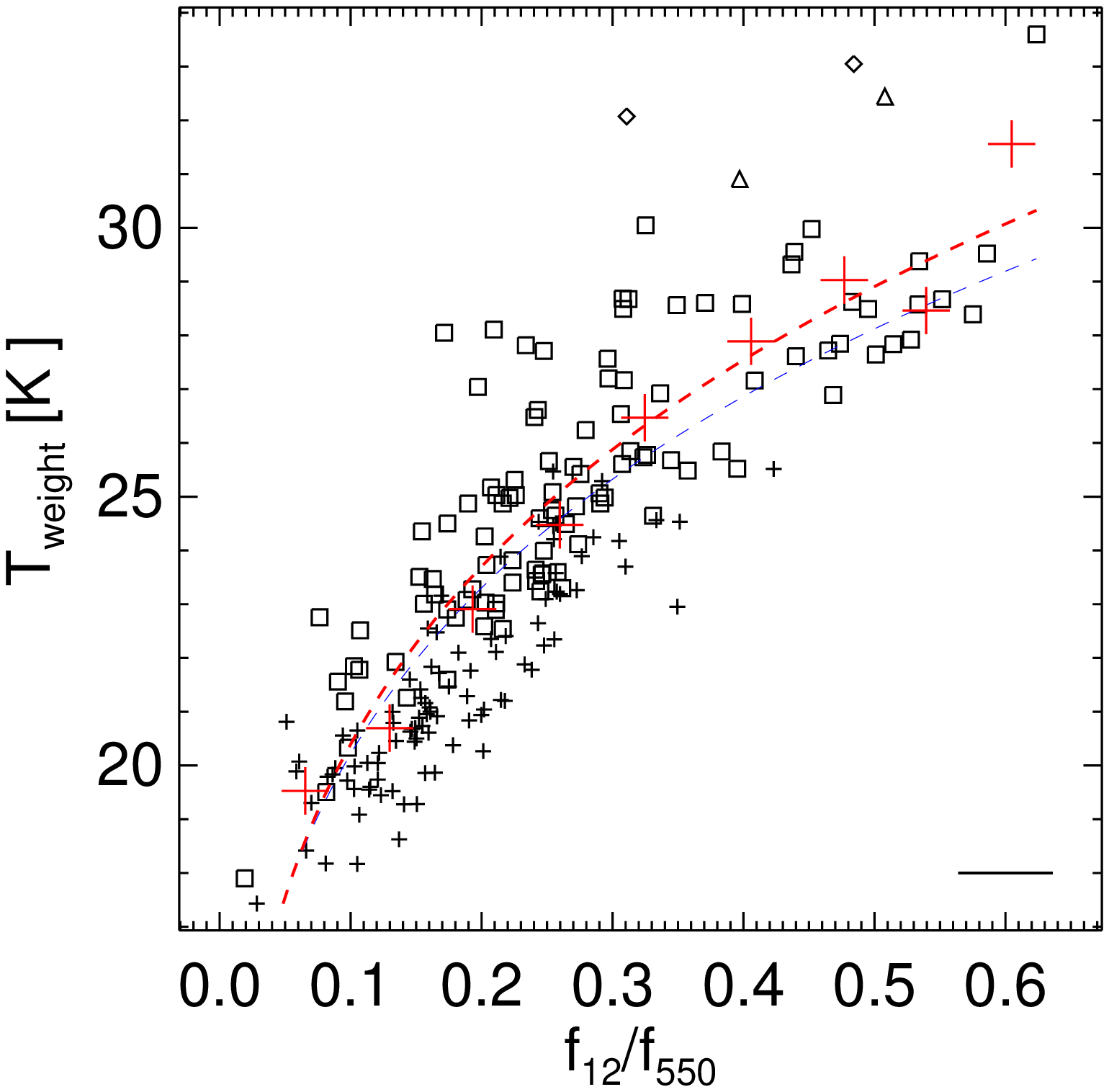}
 \includegraphics[bb=31 128 480 562,width=0.25\textwidth,clip]{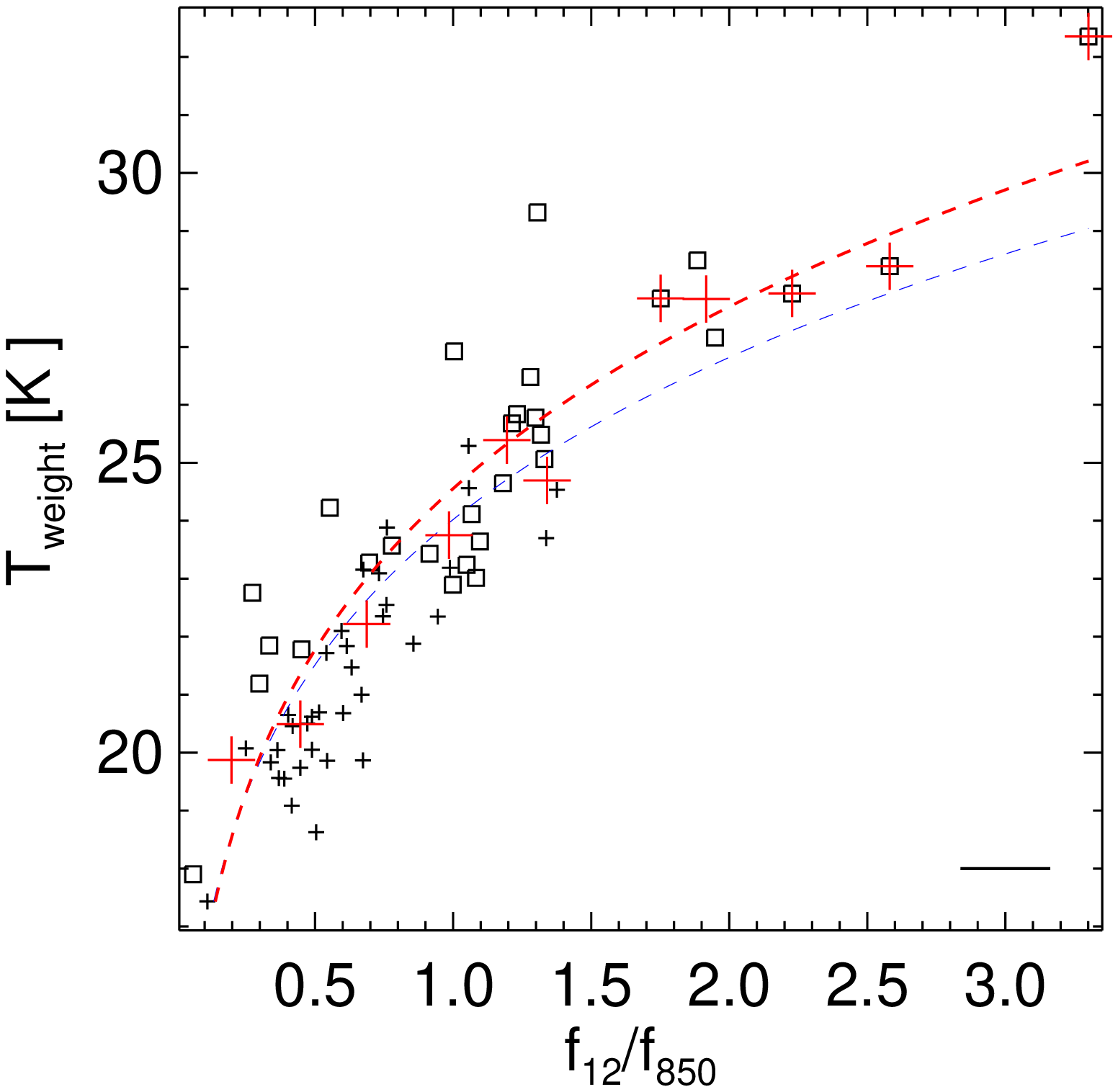}

  \caption{ As in Fig. \ref{fig:color tdust}, but for the IR flux at 22~\mum\ (top) and at 12~\mum\ (bottom) 
  and the light weighted dust temperature. }
 \label{fig:color tdust2}
 \end{center}
\end{figure*}

\begin{table}[ht]
\centering
\caption{Factors for best fit, described by eq. \ref{eq:twhcolor} }
\begin{tabular}{cc|cc|cc}
 $\lambda_1$  & $\lambda_2$ & c       & $\rm \Delta c$ & d     & $\rm \Delta d$   \\
 $[\mum]$     &   $[\mum]$  &         & $\pm$          &       &    $\pm$         \\
\hline
 350          &   22        & 1.581   &  0.003         & 0.187 & 0.002            \\
 550          &   22        & 1.467   &  0.002         & 0.176 & 0.004            \\
 850          &   22        & 1.361   &  0.002         & 0.153 & 0.007            \\
 \hline
 350          &   12        & 1.644   &  0.008         & 0.216 & 0.007           \\
 550          &   12        & 1.510   &  0.007         & 0.205 & 0.009            \\
 850          &   12        & 1.380   &  0.004         & 0.16  & 0.01  

 \end{tabular}
\label{tab:params tweight-colors}
\end{table}

The dust temperature of the cold component shows greater point dispersions for colors based on fluxes at
22 and 12~\mum. The light weighted dust temperature shows tight correlations with the colors based on the flux at 22~\mum\
and at 12~\mum, due the relevance of the warm component at this wavelength.  
Figure \ref{fig:color tdust2} shows a tight correlation between the color based at 22~\mum\ (12\mum) 
and the light weighted dust temperature, where the best fits are:
\begin{equation}
 \rm \frac{T_{weight}}{[K]}=10^{c\pm\Delta c}\left(\frac{f_{\lambda_1}}{f_{\lambda_2}}\right)^{d\pm\Delta d}
 \label{eq:twhcolor}
\end{equation}
where the factors ($\rm c,~\Delta c,~d,~\Delta d$) are shown in Table \ref{tab:params tweight-colors}.
Other colors (based on 60 or 100~\mum) correlate with the light weighted dust temperature in a similar 
way to that in which they correlate with \tc, although with larger point dispersions.

The red large crosses in Figures \ref{fig:color tdust} and \ref{fig:color tdust2} show the mean values of \tc\ different bins, 
to obtain a cleaner correlation. 
The best fit coefficients of the crosses (red dashed line) have values that are consistent within the errors with the best fit 
coefficients for all points.

\section{Estimating \lir\ and \lfir\ from IR and sub-mm data}
\label{apn:lir-lfir}

To study the SFR of a galaxy, either the estimation of the total infrared luminosity (\lir) 
or the far infrared luminosity (\lfir) is fundamental. The absence of photometric measurements at IR and sub-mm 
wavelengths obstructs the use of good SED fittings to estimate any of these parameters. 
Thus it is crucial to establish a scaling relation between a few measurements and the IR luminosity.

\begin{figure}[ht]
\begin{center}
  \includegraphics[bb=43 144 472 565,width=0.35\textwidth,clip]{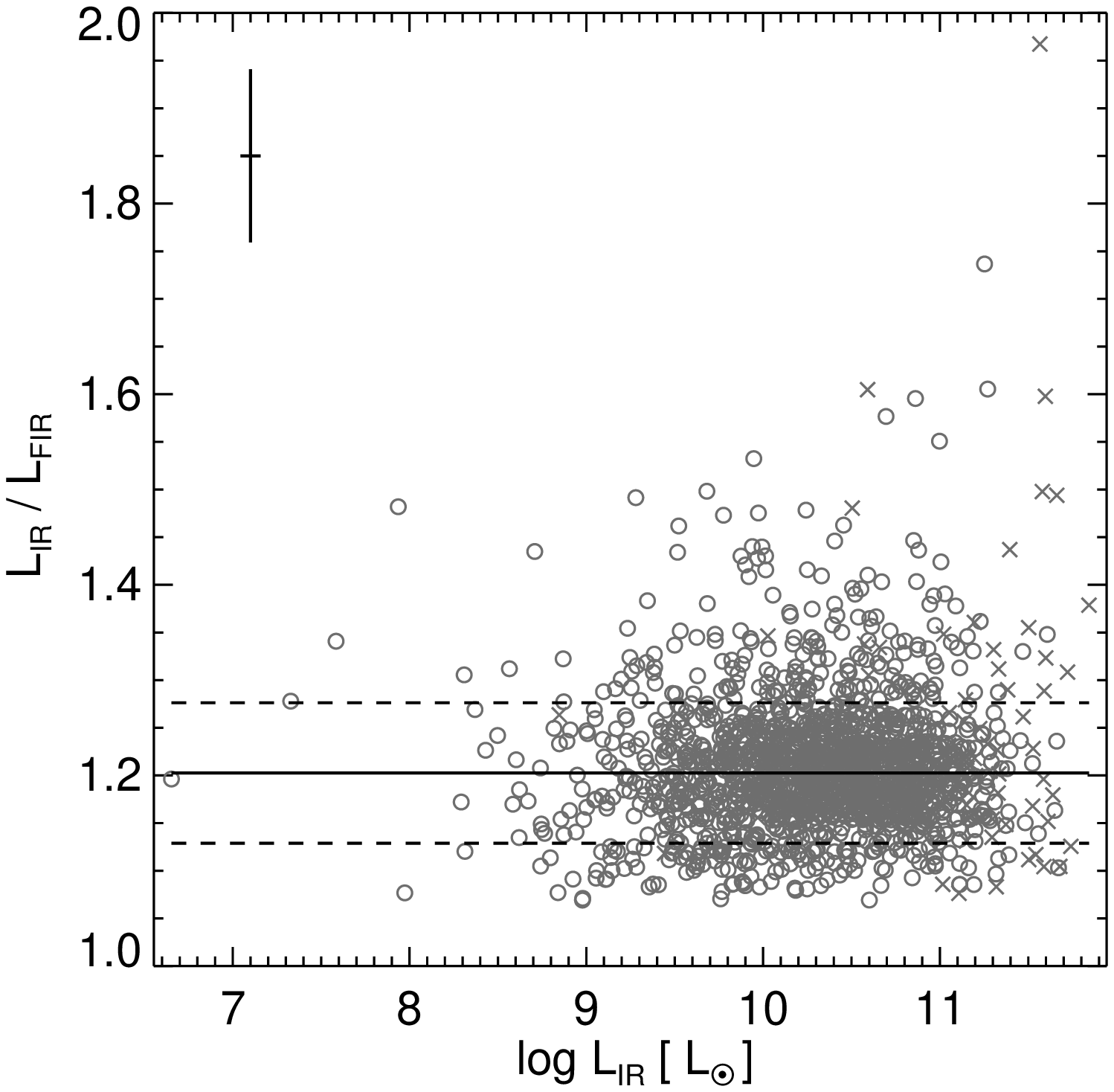}
         \caption{Ratio of \lir/\lfir as a function of \lir.
         Circles show non-interacting galaxies and crosses show interacting galaxies. 
         The median ratio for the sample is 1.20 (solid black line).
         The dashed lines show the $1 \sigma$ dispersion. 
         }
\label{fig:lirvslfir}
\end{center}
\end{figure}
  
The total \lir\ is defined between $\lambda=$ 8 and 1000~\mum, and the \lfir, between 
40 and 500~\mum. In our sample the direct integration of the SED fitting shows a 
\lir\ ranging from $3.9\times10^{6}\lsun$ to $ 7.0\times10^{11}\lsun$, 
with a typical error of 9\%, and median value of $2.2\times10^{10}\lsun$. \lfir\ ranges 
between $3.3\times10^{6}\lsun$ and $ 5.1\times10^{11}\lsun$, 
with a typical error of 12\%, and median value of $ 1.8\times10^{10}\lsun$.

Galaxies show a linear relation between the SED 
integrated quantities \lir\ and \lfir\ (Figure \ref{fig:lirvslfir}), 
with a median value of 1.20 and rms of 0.08 for \lir/\lfir.
These values can be contrasted to those obtained by 
\citet[median \lir/\lfir $\sim$1.3]{garet08} in a sample of 
17 local (z$\leq$ 0.7) LIRGs and ULIRGs, and \citet[median \lir/\lfir = 1.35;]{dacet10} 
in a sample of $\sim$3,000 nearby (z$\leq$ 0.1) SDSS galaxies.

\begin{figure}[ht]
\begin{center}
  \includegraphics[bb=49 153 480 422,width=0.35\textwidth,clip]{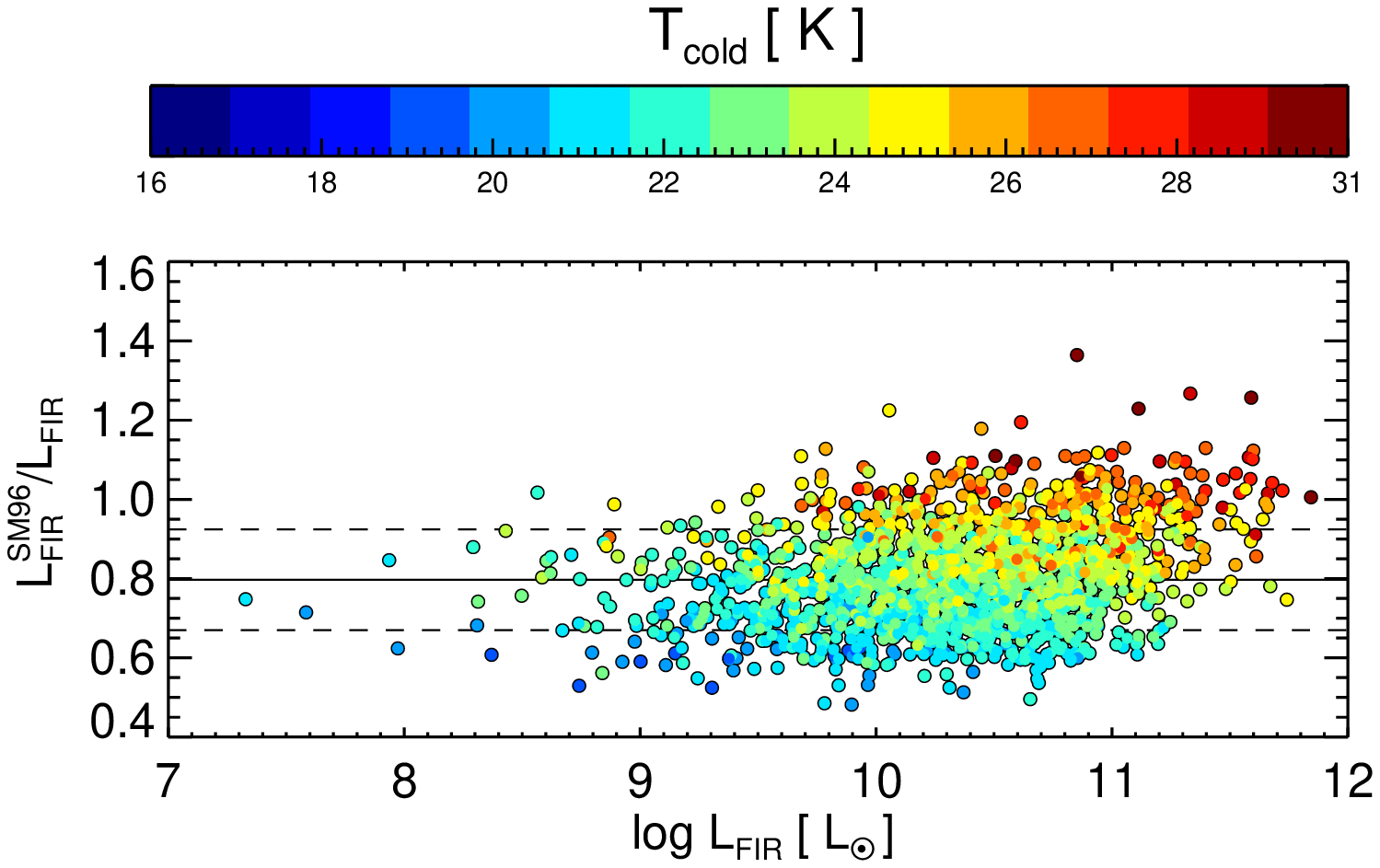}
  \includegraphics[bb=49 153 480 350,width=0.35\textwidth,clip]{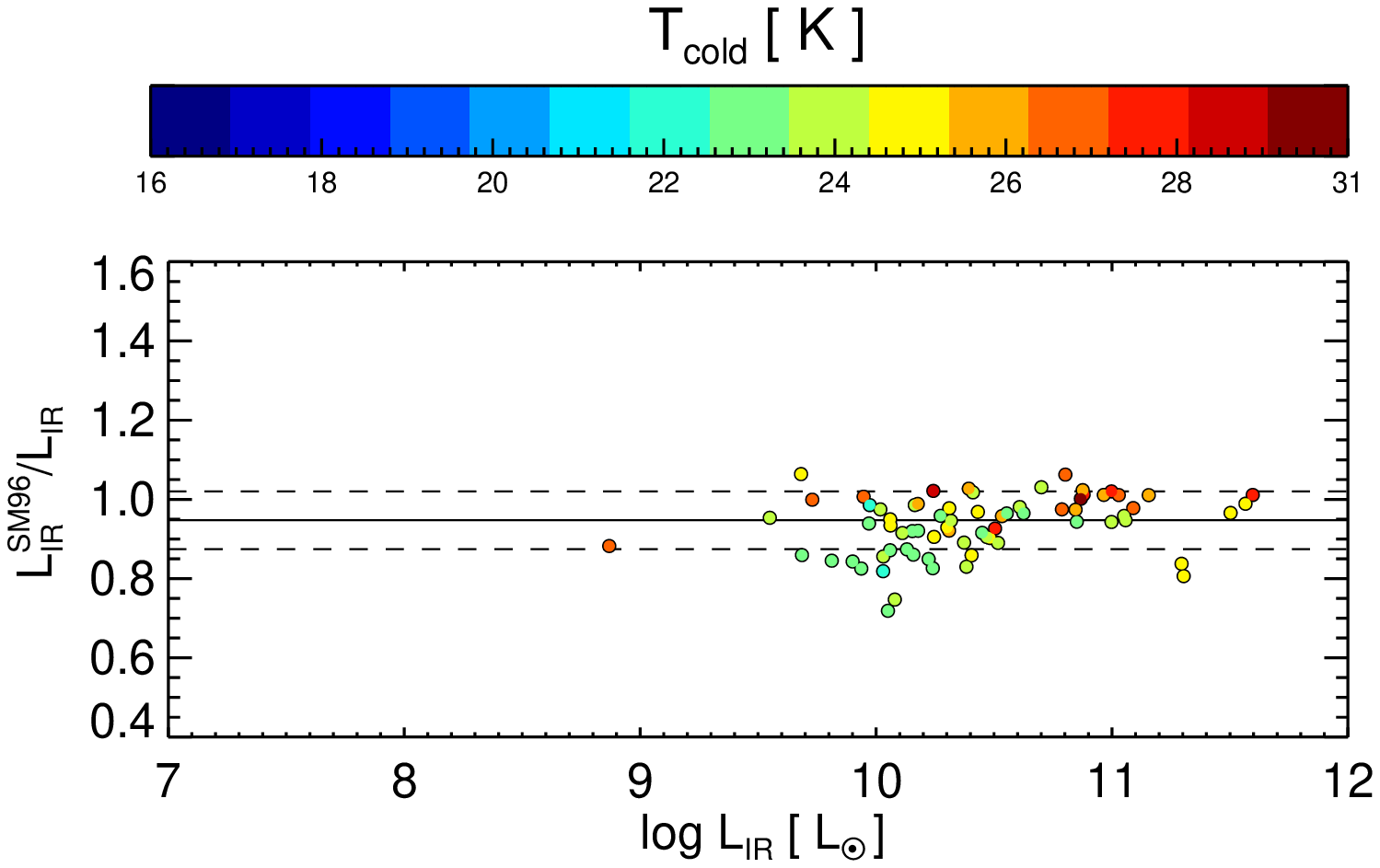}
  \caption{ The top panel (Bottom-panel) shows the comparison between the ratio of \lfir\ ( \lir ) 
        obtained by \citet{sanmir96} equations and with the direct integration of our final templates. 
        The solid line shows the median value and colors represent the dust temperature (in K) of the cold component. 
        Dashed lines represent the $1 \sigma$ dispersion. }  
\label{fig:sanders}
\end{center}
\end{figure} 

\citet{sanmir96} (hereafter SM96) show the most often used relation to obtain \lfir\ 
and \lir\ as a linear combination of fluxes at 12, 25, 60 and 100~\mum 
(observed frame).

They have used the following equations:
\begin{equation}
\rm F_{IR} = 1.8\times10^{-14}\left(13.48f_{12}+5.16f_{25}+2.58f_{60}+f_{100}\right)\left[Wm^{-2}\right] 
\nonumber
\end{equation}
\begin{equation}
 \rm L_{IR}(8-1000~\mu m)= 4\pi D_L^2~F_{IR}\left[\lsun\right] 
\nonumber
\end{equation}
\begin{equation}
\rm F_{FIR} = 1.26\times10^{-14}\left(2.58f_{60}+f_{100}\right)\left[Wm^{-2}\right] 
\nonumber
\end{equation}
\begin{equation}
 \rm L_{FIR}(40-500~\mu m)= 4\pi D_L^2CF_{FIR}\left[\lsun\right]
\label{eq:sanders and Mirabel}
\end{equation}
where $\rm f_x$ is the respective \iras\ flux expressed in Jy, $\rm D_L$ 
is the source luminosity distance in Mpc, and C is the color 
correction constant normally related to the $\rm f_{100}/f_{60}$ 
color, which is related to the IR peak emission. 

Fig. \ref{fig:sanders} top panel shows the relation between the ratio of \lfir\
from SM96 ($\rm L_{FIR}^{SM96}$) and our \lfir\  values obtained from direct integration of the final SED fits. 
Considering a constant value of C=1.4  \citep{sanmir96} in SM96's equation for \lfir, we observe an underestimation 
of the \lfir\ compared with our results. The median value of the $\rm L_{FIR}^{SM96}$/\lfir\ ratio is 
0.8. Also the luminosities ratio shows a dispersion strongly related to the dust temperature of the 
cold component. The $\rm L_{FIR}^{SM96}$/\lfir\ ratio is equal to 1 only for sources with \tc\ between 24 and 26 K.

The bottom panel of Fig.\ref{fig:sanders} shows the same comparison as the top panel 
but for \lir. The ratio between the result obtained from SM96 equation 
($\rm L^{SM96}_{IR}$) and our \lir\ has a median value of median $\rm L_{IR}^{SM96}/L_{IR} = 0.95$ with a small 
dispersion ($\sigma=0.005$). However, since the SM96's equation to obtain the total \lir\ needs 4 data points 
(fluxes at 12, 25, 60 and 100~\mum), the number of sources is very limited (e.g., only 72 sources have \lir\ but 1599 have \lfir, 
as the SM96's equations need only two fluxes in order to estimate \lfir.)

\begin{figure}[ht]
\begin{center}
  \includegraphics[bb=49 153 480 422,width=0.35\textwidth,clip]{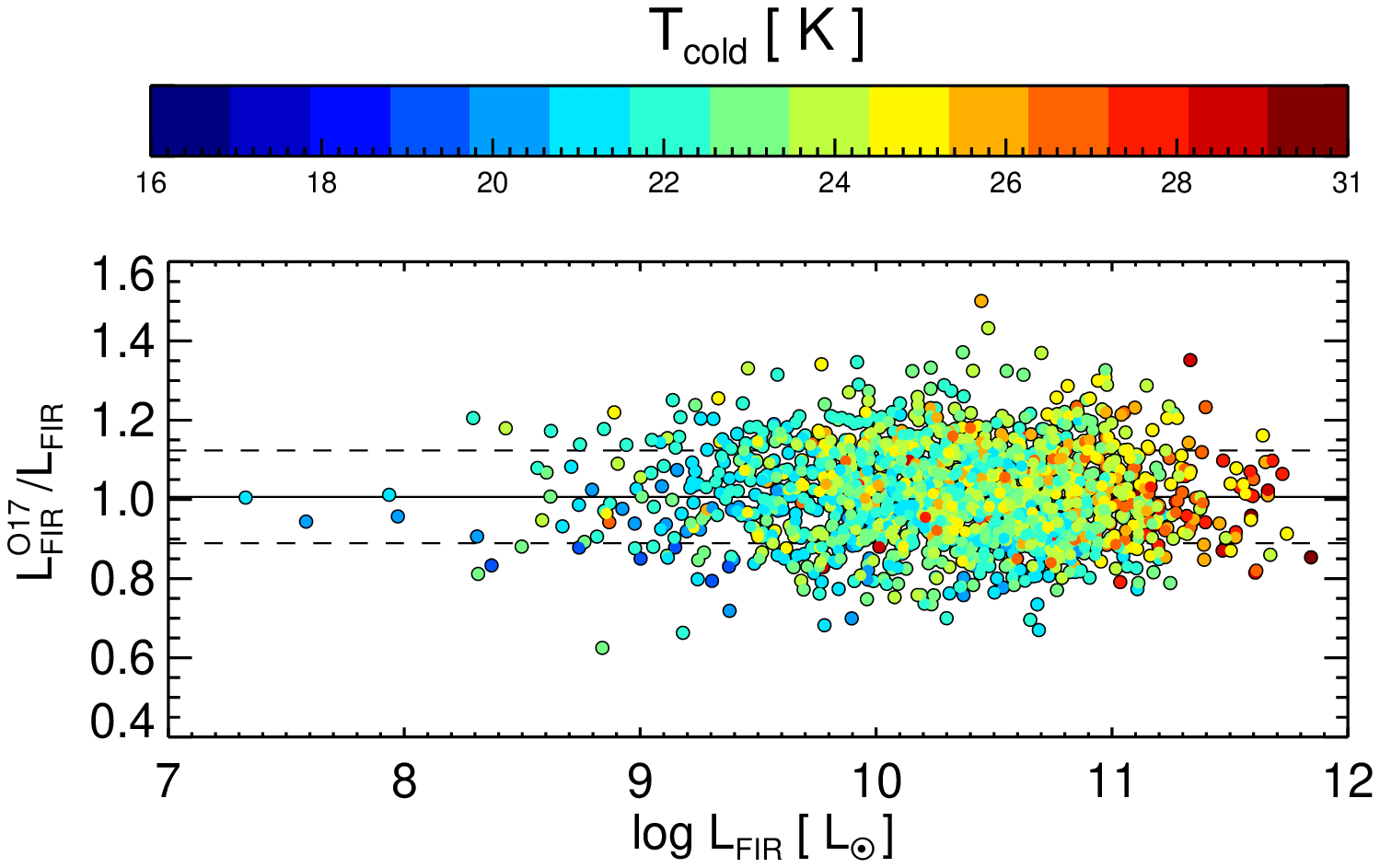}
  \includegraphics[bb=49 153 480 350,width=0.35\textwidth,clip]{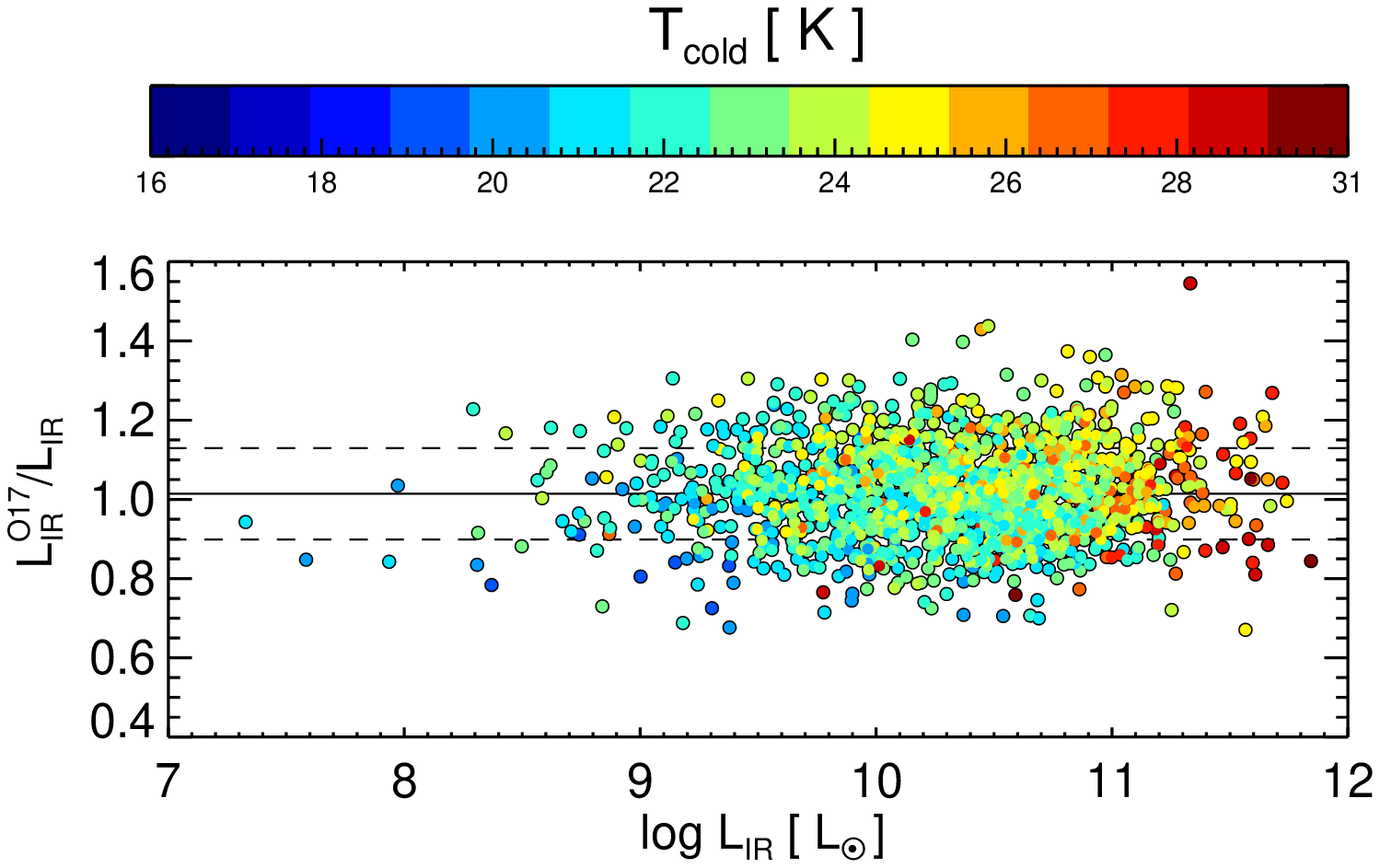}
   \caption{ Ratio between the  \lfir\ (top) and \lir\ (bottom) from our calibration 
   (O17) using the respective values obtained by the direct integration.
   The solid line is the median value and colors represent the dust temperature
   of the cold component. Dashed lines represent the $1 \sigma$ dispersion.
}  
\label{fig:sanders2}
\end{center}
\end{figure}

 The SM96 calibration is based on the use of only IRAS fluxes (mainly at 60 and 100~\mum), and a modified 
gray-body with $\beta$ between 1 and 2. 
The problem is that this calibration use the simplest dust emission model with no information at wavelengths 
greater than 100~\mum. 
Thus this model has the problem that the peak emission near 100\mum\ is not well defined and, therefore, it  
is not possible to identify the position of the peak of the IR emission. 
Additionally, it is expected that the flux between 50 to 1000 \mum\ contributes $\sim$65\% of \lir, implying  that 
it is important to have information from the sub-mm.
However, in general terms, the calibration showed by SM96 is an excellent first approximation of \lir\ ($\lambda=$ 8-1000~\mum) 
and for the far infrared luminosity ($\lambda=$ 40-500~\mum). 
However, now we have access to a greater range of photometric measurements (with more points within this range) 
and the access to more detailed dust emission models, with deeper physical constrains (e.g., the DL07 models).
With this idea in mind, we re-calibrate the SM96 equations and furthermore, we propose an easier and simpler 
linear combination of the \iras\ filters in order to estimate the \lir\ and \lfir\ using only two filters 
(at 60~\mum\ and 100~\mum ).

Our calibration is expressed by the following equations:

\begin{equation}
\mbox{ For FIR luminosity $(40-500~\mu m)$}
\nonumber
\end{equation}
\begin{equation}
\rm \frac{F_{FIR}}{\left[Wm^{-2}\right]} = 1.26 \times 10^{-14} \left(0.752\frac{f_{60}}{\left[Jy\right]}+3.236\frac{f_{100}}{\left[Jy\right]}\right)
\end{equation}

\begin{equation}
        \mbox{ For total IR luminosity $(8-1000~\mu m)$}
\nonumber
\end{equation}
\begin{equation}         
\rm \frac{F_{IR}}{\left[Wm^{-2}\right]} = 1.8\times10^{-14}\left(1.439\frac{f_{60}}{\left[Jy\right]}+2.450\frac{f_{100}}{\left[Jy\right]}\right) 
\end{equation}
where the luminosity is obtained from the conventional equation: 
\begin{equation}
\rm \frac{L}{\left[\lsun\right]} =\rm 3.13\times10^{19} \left(\frac{D_L}{\left[Mpc\right]}\right)^2 \frac{F}{\left[Wm^{-2}\right]}
\end{equation}

Figure \ref{fig:sanders2} top (bottom) panel shows the ratio between the \lfir\ 
( \lir) obtained with our calibration using the values derived the direct integration of the SED templates,
as a function of the \lfir\ (\lir) also obtained for the integration of the SED templates. 

We use 1,599 sources and find that the median value of the $\rm L_{FIR}^{OR17}/\lfir$ and $\rm L_{IR}^{OR17}/\lir$ ratios 
$\sim$ 1.001 with standard deviation $\sim$ 0.13.

One other advantages of our calibration is that these is no correlation between the 
dispersion and the \tc. Furthermore, our equations do not need a temperature correction 
(as shown the SM96 with the C factor) to compensate for over - or  under - estimations.

\begin{figure}[ht]
\begin{center}
  \includegraphics[bb=41 152 472 562,width=0.35\textwidth,clip]{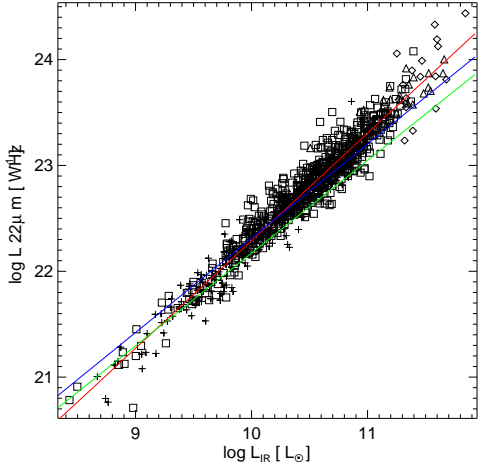}
    \caption{ Correlation between the luminosity at 22~\mum\ and the 
    total IR luminosity. 
    The red line shows the best fit, blue and green lines are the best-fit lines 
    obtained by \citet{rieet09} and \cite{jaret13}, respectively,
    from the SFR and the luminosity at 22\mum.
    Symbols represent the same sSFR as in Fig. \ref{fig:mass comp}
    }
\label{fig:24mum Lir}
\end{center}
\end{figure}  

Another interesting way to obtain the total \lir\ from a single measurement comes from the use of the luminosity at 22~\mum (WISE W4 filter). 
Using a sub-sample of 924 galaxies with at least 5 photometric points for a SED fitting with $\rm \chi^2_r<1$,
we obtain that the best fit equation is expressed by:
\begin{equation}
 \rm \frac{\lir}{[L_{\sun}]} =10^{-10.60\pm0.17} \left(\frac{L_{22~\mum}}{[WHz^{-1}]}\right)^{0.926\pm0.008}
 \end{equation}
with rms=0.096. The best-fit line for our data is plotted in Figure \ref{fig:24mum Lir} as a red solid line.
Similar relations are shown in different works, where the SFR of the galaxy is derived from the flux at 24~$\mum$. 
In particular, Figure \ref{fig:24mum Lir} reports two examples, one  from \citet{jaret13} (green line) and 
the other from \citet{rieet09} (blue line). 
The point dispersion in our calibration does not show correlations with \tc,\mstar, or \mdust.

\begin{figure}[ht]
\begin{center}
  \includegraphics[bb=15 120 500 650,width=0.4\textwidth,clip]{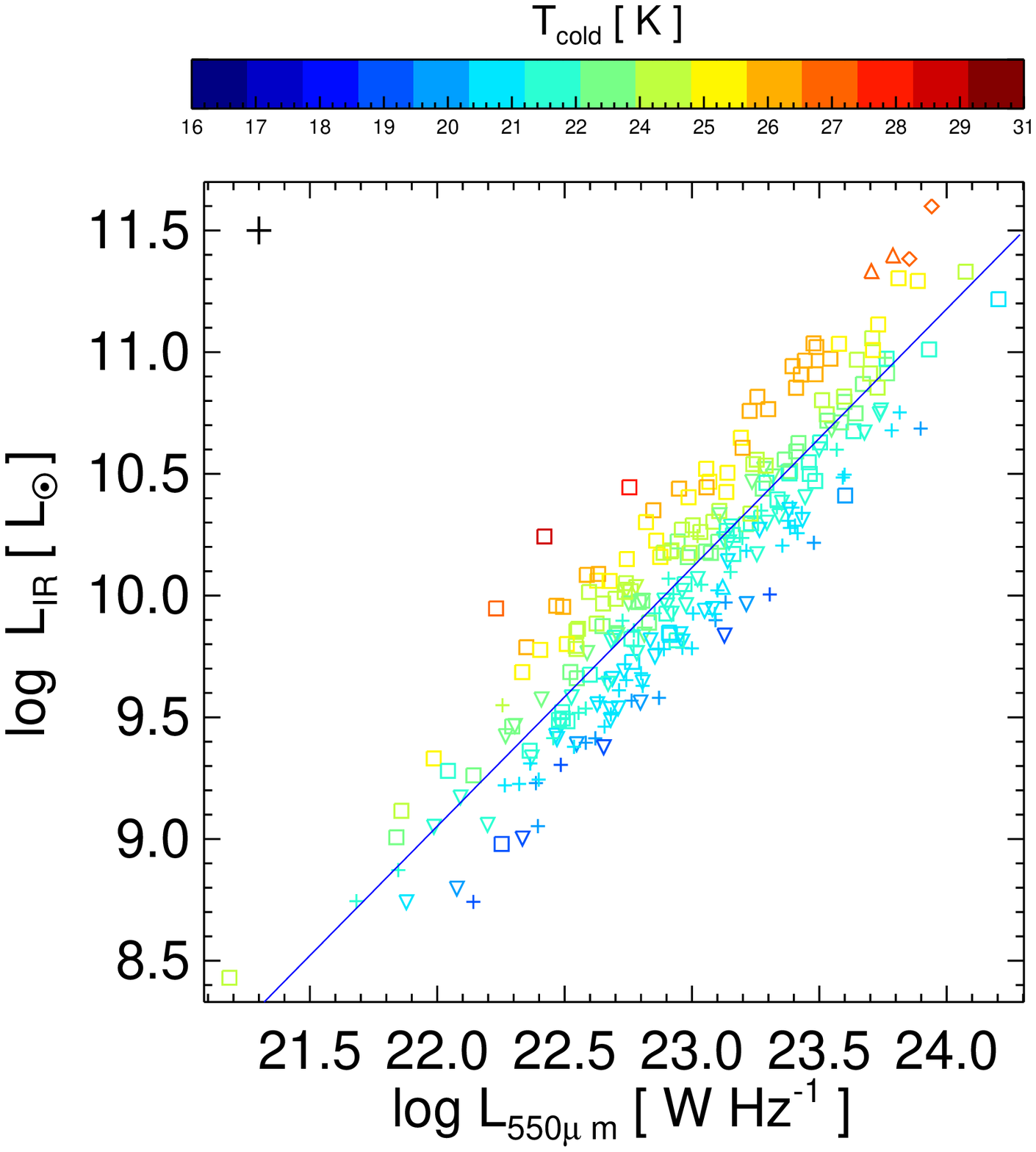}
  \includegraphics[bb=15 120 500 650,width=0.4\textwidth,clip]{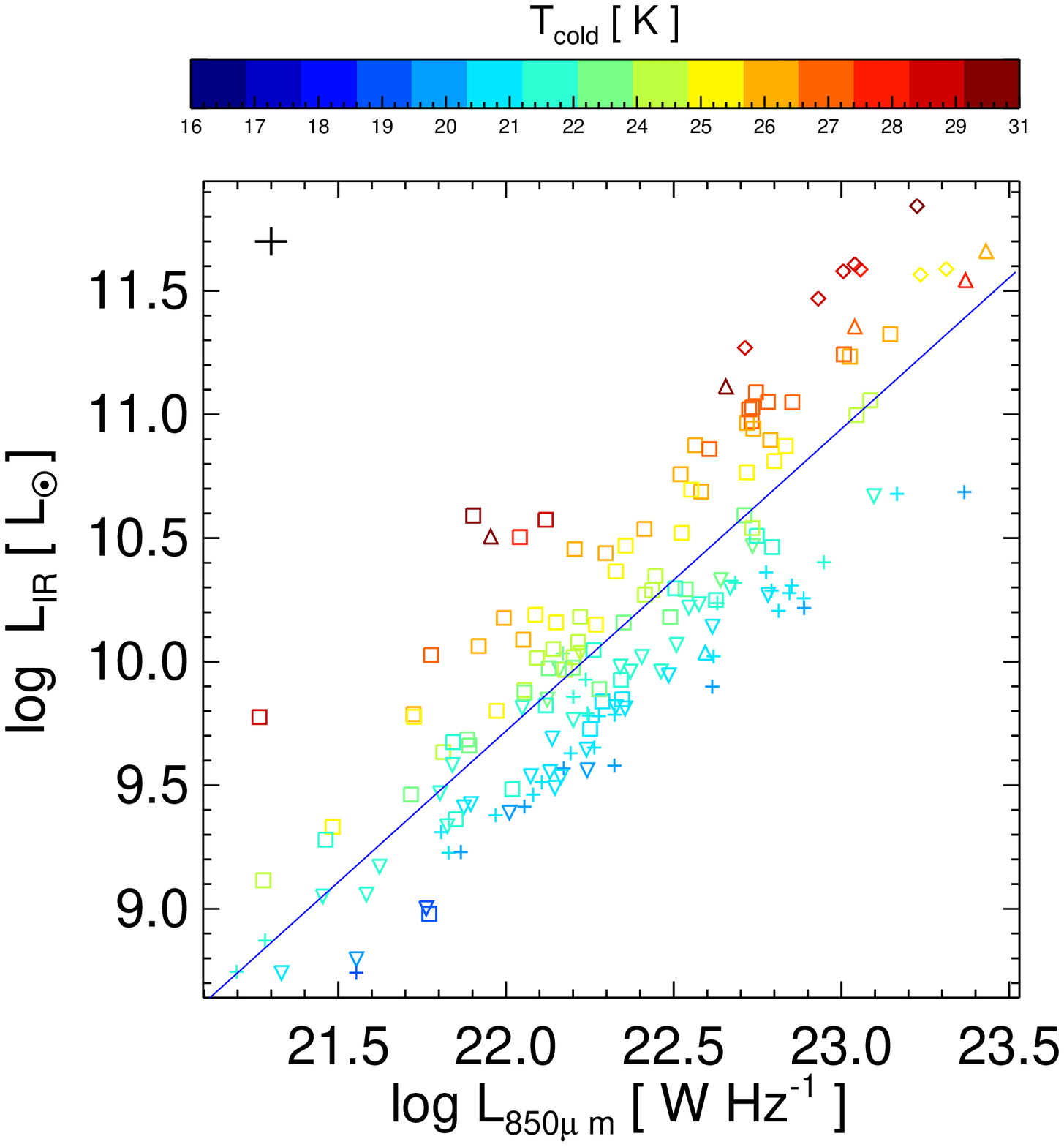}
  \caption{ Correlation between total \lir\ and the luminosities  at 550~\mum\ (top panel) 
  and 850~\mum\ (bottom panel). 
  The dust temperature (in K) of the cold component is shown by the color scale. 
  The blue line shows the best fit. 
  Symbols are as shown in Figure \ref{fig:mass comp}. 
  The black cross shows the typical error.
}  
\label{fig:L850-L550-lir}
\end{center}
\end{figure}

Finally, we find a correlation between the sub-mm fluxes (at 350~\mum\ , 550~\mum\ , 850~\mum ) 
and the total \lir. We present this correlations in Section \ref{subsec:fundamentalplane} 
for the flux at 350~\mum.

Figure \ref{fig:L850-L550-lir} shows the correlation between the \lir\ 
and the luminosity at 550~\mum\ (top panel) and 850~\mum\ (bottom panel), where the best 
fits (blue lines) are determined by the following equations, respectively:
\begin{equation}
\rm \frac{L_{IR}}{[L_{\sun}]} =  10^{-14.588\pm0.002 }\left(\frac{L_{550}}{[WHz^{-1}]}\right)^{1.074\pm0.005}                                 
\nonumber
\end{equation}
\begin{equation}
\rm \frac{L_{IR}}{[L_{\sun}]} = 10^{-17.135\pm0.002}\left(\frac{L_{850}}{[WHz^{-1}]}\right)^{1.221\pm0.0051}
\end{equation}
These relations show that the dispersion is related to the dust 
temperature of the cold component. Following the same idea of the dust plane shown 
in Section \ref{subsec:fundamentalplane}, we define a plane using the total 
IR luminosity, the luminosity at 550~\mum\ (at 850~\mum\ ), and the 
dust temperature of the cold component. The best fits are described as follow: 
\begin{equation}
\rm \log \left(\frac{L_{IR}}{ [L_{\sun}]}\right)  - 1.00\times \log \left(\frac{L_{550}}{[WHz^{-1}]}\right) 
\nonumber
\end{equation}
\begin{equation}
\rm - 0.13\times \left(\frac{T_{cold}}{[K]}\right) + 16.03 = 0 
\nonumber
\end{equation}
\begin{equation}
\rm \log \left(\frac{L_{IR}}{[L_{\sun}]}\right)  - 1.01\times \log \left(\frac{L_{850}}{[WHz^{-1}]}\right) 
\nonumber
\end{equation}
\begin{equation}
\rm - 0.15\times \left(\frac{T_{cold}}{[K]}\right) + 15.93 = 0 
\end{equation}
A similar dust plane is defined for the luminosity at 350~\mum\ , see Section
\ref{subsec:fundamentalplane} for more details.

\section{Estimating Dust masses from limited sub-mm data}
\label{apn:Mdust-RJtail}

\begin{figure}[ht]
\begin{center}
  \includegraphics[bb=20 120 500 650,width=0.4\textwidth,clip]{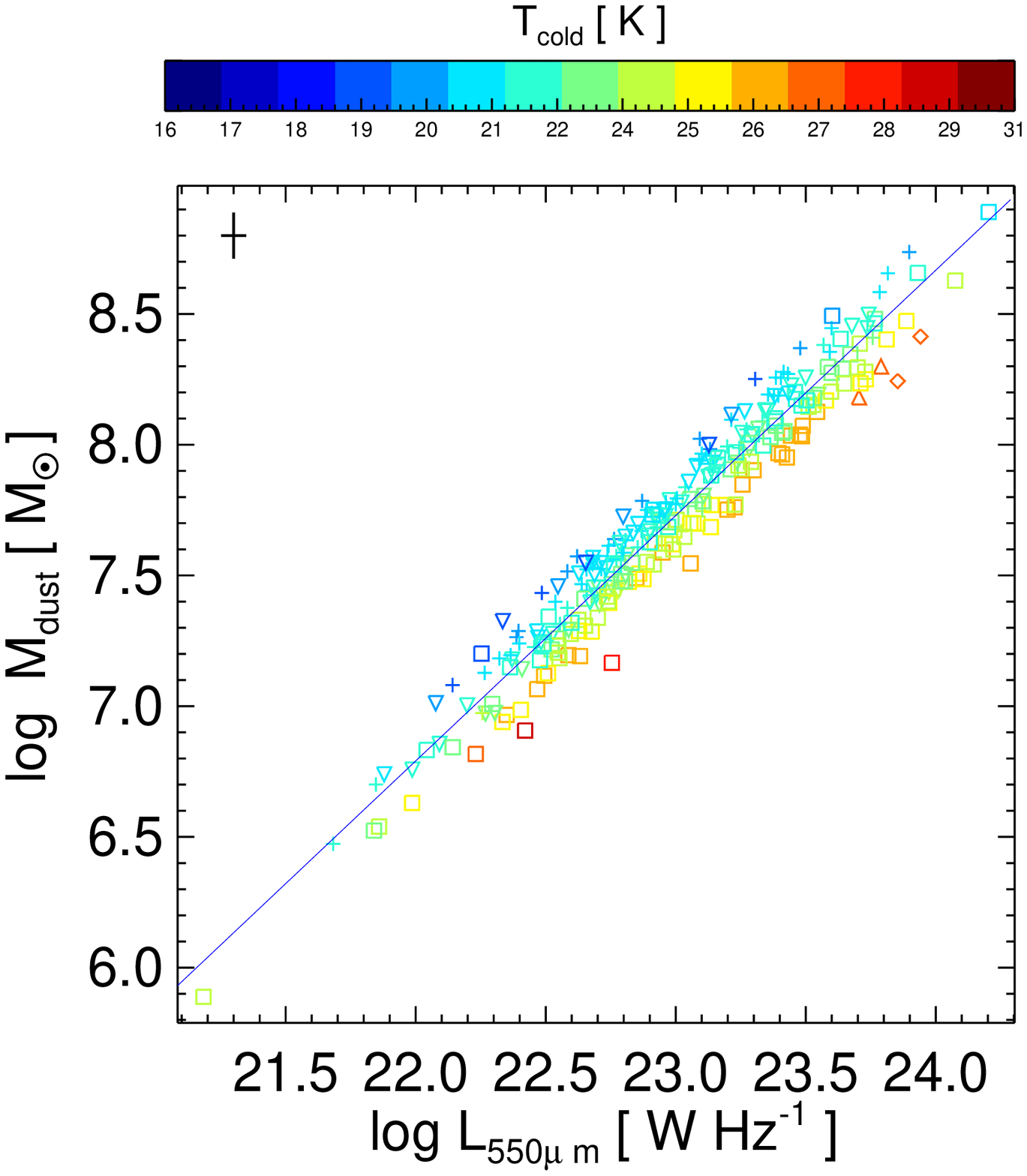}
  \includegraphics[bb=20 120 500 650,width=0.4\textwidth,clip]{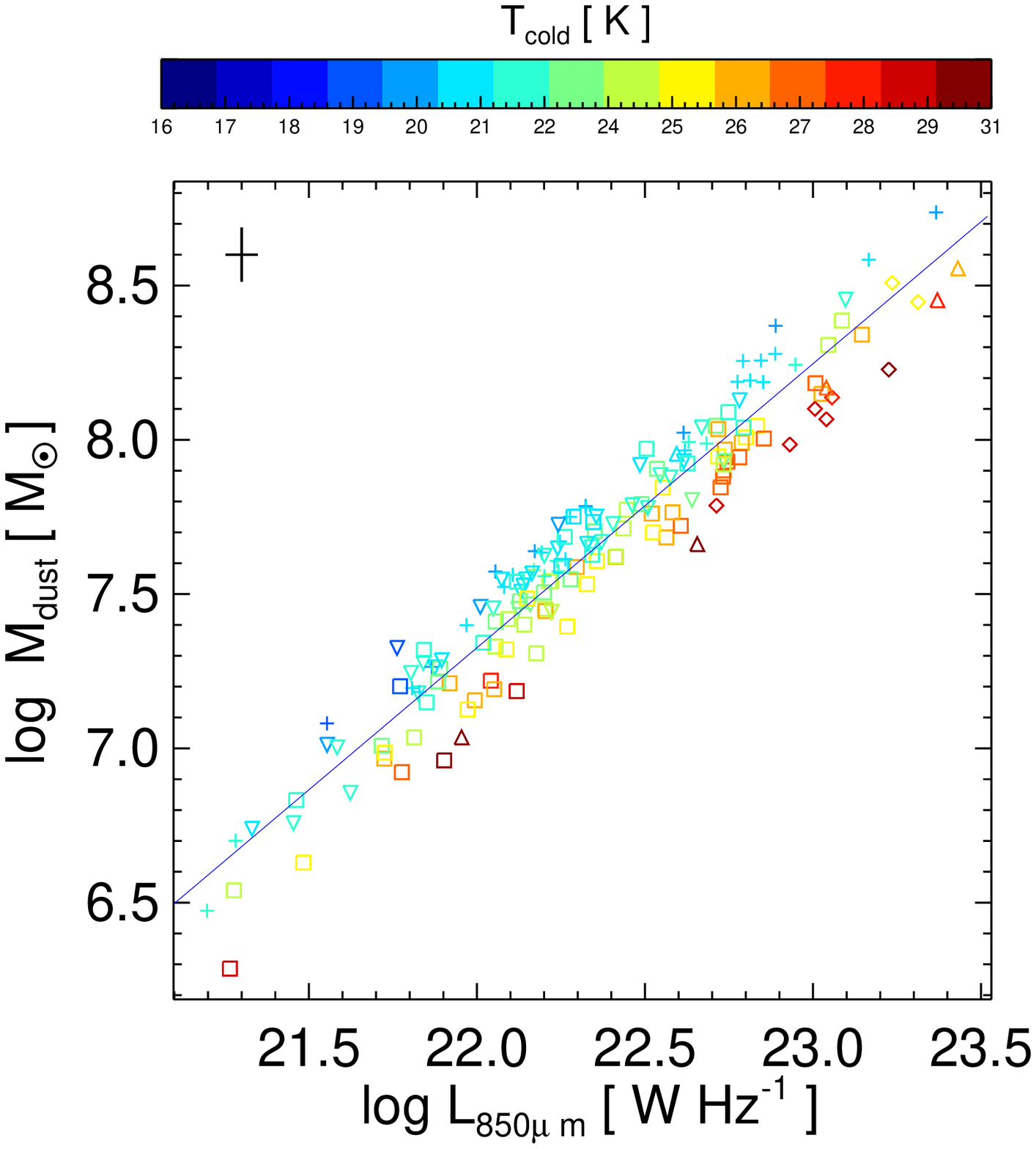}
  \caption{ The relation between dust mass and the luminosities at 550~\mum\ (top panel) 
  and at 850~\mum\ (bottom panel). 
  The dust temperature (in K) of the cold component is shown by the color scale. 
  The blue line shows the best fit. 
  Symbols are as shown in Figure \ref{fig:mass mdust-mism}.
  The typical error is shown by black the cross.
}  
\label{fig:L850-L550-mdust}
\end{center}
\end{figure}

The most common approach to estimate the dust mass in a galaxy from a single measurement 
is using the flux at 850~\mum\ (e.g., \citet{dunet00}; \citet{dunet01}).
 This is because the most common approach would be to have a set of measurements, 
and then perform a fit according to a physical dust model. 
The aforementioned results were limited by the data, since at that time there were not deep 
measurements between 100-500 \mum. 
In Section \ref{subsec:fundamentalplane} we show a correlation between the dust mass 
and the luminosity at 350~\mum. Now we extend this correlation to the luminosities 
at 550~\mum\ and 850~\mum.

Figure \ref{fig:L850-L550-mdust} shows the correlations 
between the \mdust\ and the sub-mm luminosities at 550~\mum\ (top panel) and 850~\mum\ (bottom panel), 
where the color scale shows the dust temperature of the cold component. The best
fits (blue lines) equations are represented by:
\begin{equation}
\rm \frac{M_{dust}}{[M_{\sun}]} = 10^{-13.606\pm0.002}\left(\frac{L_{550}}{[WHz^{-1}]}\right)^{0.927\pm0.05}
\nonumber
\end{equation}
\begin{equation}
\rm \frac{M_{dust}}{[M_{\sun}]} = 10^{-12.84\pm0.002}\left(\frac{L_{850}}{[WHz^{-1}]}\right)^{0.920\pm0.005}
\end{equation}

Analogously to the dust plane (Section \ref{subsec:fundamentalplane}), we define a plane among the luminosities in the sub-mm, 
the dust mass and the dust temperature of the cold component.
The equations for the best fits are:
\begin{equation}
\rm \log \left(\frac{M_{dust}}{[M_{\sun}]}\right) - 0.961\times \log \left(\frac{L_{550}}{[WHz^{-1}]}\right)
\nonumber
\end{equation}
\begin{equation}
~~~~~\rm + 0.064\times \left(\frac{T_{cold}}{[K]}\right) + 12.933 = 0  
\nonumber
\end{equation}
\begin{equation}
\rm \log \left(\frac{M_{dust}}{[M_{\sun}]}\right) - 0.993\times\log \left(\frac{L_{850}}{[WHz^{-1}]}\right)
\nonumber
\end{equation}
\begin{equation}
~~~~~\rm + 0.054\times \left(\frac{T_{cold}}{[K]}\right) + 13.310 = 0 
\end{equation}
These relations are independent to the sSFR classification.

\end{document}